\documentclass[12pt]{article}

\usepackage[utf8]{inputenc}
\usepackage[T1]{fontenc}
\usepackage[english]{babel}

\usepackage[a4paper, margin=1in]{geometry}
\usepackage{setspace}
\usepackage{centernot}

\usepackage{proof}

\usepackage{amsmath, amssymb, amsfonts, amsthm, mathrsfs, mathtools}
\usepackage{bm}
\usepackage{slashed}
\usepackage{physics}
\usepackage{tensor}
\usepackage{braket}
\usepackage{cancel}
\usepackage{esint}
\usepackage{bbold}
\usepackage{dsfont}
\usepackage{upgreek}
\usepackage{stmaryrd}

\usepackage{graphicx}
\usepackage{tikz}
\usepackage{pgfplots}
\pgfplotsset{compat=newest}
\usepackage{tikz-cd}
\usepackage{tikz-3dplot}
\usepackage{float}
\usepackage{caption}
\usepackage{subcaption}

\usepackage{hyperref}
\usepackage{cleveref}

\usepackage{fancyhdr}
\usepackage{titlesec}

\usepackage{enumitem}
\usepackage{array}
\usepackage{longtable}
\usepackage{booktabs}
\usepackage{makecell}
\usepackage{colortbl}
\usepackage{multirow}
\usepackage{multicol}

\usepackage{authblk}
\usepackage{cite}
\usepackage{comment}
\usepackage{etoolbox}
\usepackage{lipsum}
\usepackage{appendix}
\usepackage{pdfpages}
\usepackage{textgreek}

\hypersetup{
  colorlinks=true,
  linkcolor=blue,
  citecolor=blue,
  filecolor=blue,
  urlcolor=blue
}

\newtheorem{theorem}{Theorem}[section]
\newtheorem{definition}[theorem]{Definition}

\title{Beyond Prediction: Structuring Epistemic Integrity in Artificial Reasoning Systems}
\author{\Large Craig S. Wright\\
\small Department of Computer Science\\
\small University of Exeter\\
\small \texttt{cw881@exeter.ac.uk}}
\date{}

\begin{document}

\maketitle

\newpage
\section*{Abstract}

\noindent This paper outlines a comprehensive theoretical and architectural framework for constructing epistemically grounded artificial intelligence systems capable of propositional commitment, metacognitive reasoning, contradiction detection, and normative truth maintenance. Moving beyond the constraints of stochastic language generation, we propose a model in which artificial agents engage in structured, rule-governed reasoning that adheres to explicit epistemic norms. The approach integrates insights from epistemology, formal logic, inferential semantics, knowledge graph structuring, probabilistic justification, and immutable blockchain evidence to create systems that do not merely simulate knowledge, but operate under explicit, verifiable constraints on belief, justification, and truth.

We begin with an analysis of epistemic norms in artificial reasoning, contrasting evidentialist, Bayesian, and logical foundations, and establishing a requirement for internal consistency and constraint against falsehood. Central to the proposed system is a prohibition against internal deception: no model component may assert what it internally contradicts. Confidence thresholds are made explicit and bounded by logical interpretation, allowing systems to reason transparently about belief status at varying degrees of evidential certainty.

Subsequent sections formalise belief architectures, define the computational semantics of holding a belief, and detail how propositional attitudes, metacognitive loops, and recursive verification processes provide the necessary scaffolding for epistemic agency. We explore how contradictions are detected and resolved within a dynamic reasoning framework, rejecting paraconsistency as a legitimate mode of operation in cognitive architectures committed to truth preservation.

The role of inference chains, symbolic reasoning, and knowledge graph integration is treated in depth, culminating in an architecture where beliefs are embedded not merely as tokens but as justified positions, recursively tracked and modified according to normative standards. Immutable blockchain mechanisms are introduced to provide external anchoring of justification and auditability, ensuring that epistemic states can be independently verified and preserved.

We conclude with a blueprint for constructing such a system and discuss the philosophical consequences of artificial truthfulness, responsibility, and the limits of formal representation. The framework defines a new class of language models: epistemic agents that do not merely produce plausible continuations of text but commit to justified propositions under logical, probabilistic, and evidential constraints. This marks a foundational shift in artificial intelligence—from probabilistic simulation to structured, transparent, and verifiable epistemic cognition.

\textbf{Keywords:} Epistemic Justification; Propositional Commitment; Artificial Reasoning; Truth Constraints; Metacognition; Belief Architecture; Contradiction Resolution; Immutable Audit Trails; Blockchain Verification; Symbolic-Semantic Fusion; Knowledge Graphs; Probabilistic Inference; Logical Form Representation; Epistemic Agency; Self-Monitoring Systems; Epistemic Norms; Truth-Conditional Semantics; Reflective Reasoning; Artificial Epistemology; Cognitive Integrity Systems

\newpage
\tableofcontents

\section{Introduction}

Artificial intelligence systems have made remarkable strides in recent years, particularly through the proliferation of large language models (LLMs) capable of generating fluent and contextually relevant text. Yet this linguistic proficiency masks a deeper epistemological deficiency. Current AI architectures excel at syntactic imitation but lack principled mechanisms for maintaining epistemic integrity—coherence, justification, and accountability of beliefs. This absence becomes critical in domains demanding not just predictive adequacy, but grounded reasoning, semantic veridicality, and rational action based on truth-evaluable propositions.

This work proposes a new framework for artificial epistemic systems that replaces the prevailing statistical paradigm with a logically grounded, truth-preserving architecture. Unlike current models which conflate pattern completion with inference, the proposed system delineates formal belief structures, revision procedures, and semantically anchored representations. It incorporates modules for contradiction detection, model-theoretic validation, and chain-of-reason logging, forming an epistemically tractable foundation for high-integrity reasoning. The architecture aims to enable artificial agents not merely to predict, but to understand, justify, and act in ways consistent with normative principles of truth and rationality.

By situating the architecture within a lineage of formal epistemology, belief revision theory, and symbolic AI, and building directly upon foundational propositions articulated in \cite{wright2025immutable}, this paper initiates a shift from behaviourally plausible yet epistemically shallow systems to robust agents capable of traceable, inspectable, and justifiable cognition.

\subsection{Motivation and Scope}

The growing dominance of large-scale neural architectures in artificial intelligence has yielded systems capable of fluent output and broad domain generality. However, such systems remain fundamentally ungrounded: their representations are not tethered to referents, their beliefs lack formal justifications, and their outputs cannot be systematically audited for truth-preservation. This epistemic opacity poses a profound risk as these models are deployed in high-stakes environments—scientific research, autonomous decision-making, legal reasoning—where factual coherence, consistency over time, and verifiability are non-negotiable. The motivation of this work is to confront these deficits not with incremental patchwork but with a systematic reconceptualisation of epistemic computation itself.

This paper defines the architectural, formal, and functional components necessary to construct artificial reasoning systems governed by epistemic integrity rather than statistical mimicry. It scopes an end-to-end cognitive system encompassing belief management, model-theoretic validation, contradiction resolution, semantic grounding, and goal-driven inference. It integrates structured logical mechanisms with probabilistic modulation, but without collapsing into approximation alone. The ambition is not to reject predictive utility, but to subsume it within a truth-preserving hierarchy wherein reasoning, revision, and action all derive from traceable epistemic commitments. This system, then, is not merely a computational artefact—it is a reassertion of the foundational role of truth in intelligence.
\subsection{Limitations of Statistical Prediction in Current LLMs}

Statistical language models such as GPT-4, Claude, and PaLM rely on autoregressive token prediction across high-dimensional embeddings trained on massive corpora. Despite their impressive fluency and contextual mimicry, these systems lack the structural capacity for grounded semantic interpretation, epistemic validation, or truth-preserving inference. They generate text by exploiting statistical regularities without any embedded logical commitment to the factual status of their outputs. As a result, they frequently produce hallucinated content, offer contradictory answers, and cannot distinguish tautologies from empirical claims or falsehoods.

Critics have underscored the limitations of this architecture. Bender et al. (2021) described such models as “stochastic parrots,” arguing that they merely reflect surface-level distributional patterns without understanding or intentionality. Marcus and Davis (2020) further warned that these models, despite their scale, remain devoid of genuine abstraction or reasoning capability. Even foundational reviews such as Bommasani et al. (2021) concede that LLMs exhibit emergent behaviours without the reliability or accountability mechanisms required for knowledge-sensitive contexts. More recently, Ji et al. (2023) conducted a comprehensive survey on hallucinations in LLMs, highlighting the inability of these systems to manage truth tracking or to self-correct on the basis of external feedback. These critiques converge on a central point: statistical prediction alone is insufficient for artificial epistemic competence.

\subsection{Epistemic Integrity: A New Foundation}

To transcend the epistemological constraints of predictive text generators, a new framework must be established—one that anchors machine reasoning not in surface-level token co-occurrence, but in verifiable truth conditions, model-theoretic validity, and epistemic coherence. Epistemic integrity designates this foundational principle: a system's internal representations must map coherently to external states of affairs, obeying logic-preserving transformations and rejecting propositions that breach consistency or satisfiability conditions. In contrast to LLMs, which cannot differentiate between fact and fiction, systems built on epistemic integrity are designed to track truth, justify belief, and regulate action under formal constraints.

This reorientation grounds knowledge claims in structured logical inference, environmental observability, and belief revision models consistent with AGM theory and Kripke semantics. It entails the rejection of contradiction tolerance in favour of principled belief replacement, the enforcement of semantic alignment through grounded symbol-referent mappings, and the segregation of certainty types in propositional content. The proposed architecture thus enforces an ontological and epistemological discipline absent from current statistical systems, establishing a path toward artificial agents capable of maintaining not just coherence, but truthfulness in a formally specifiable and auditable manner.
\subsection{Relation to Prior Work \texorpdfstring{\cite{wright2025immutable}}{[Wright, 2025]}}

This work directly builds upon the foundational architecture proposed in Wright's theory of immutable truth structures in artificial reasoning systems \cite{wright2025immutable}. While that framework introduced the notion of truth-preserving transformations and the necessity of symbolic grounding to secure representational fidelity, the present paper extends these concepts into a fully operational epistemic architecture with modular integration of metacognitive control, contradiction rejection, and belief evolution. Wright's prior analysis focused on the theoretical impossibility of epistemic self-repair in unconstrained statistical systems; here, those theoretical insights are applied to construct concrete mechanisms for belief maintenance, dynamic justification, and logical tractability across temporal updates.

Moreover, whereas Wright outlined the dangers of semantic drift in autoregressive models due to their lack of referential anchoring and logical closure, this paper presents a systematic response—grounding epistemic claims within a hybrid model-theoretic and truth-conditional framework that ensures deductive soundness and ontological coherence. As such, it reifies the proposed immutable substrate into a computationally actionable structure, introducing supervisory metacognition and architectural modularity to enforce the constraints of epistemic rationality in dynamic environments.

\section{Epistemic Norms and Foundations}

This section introduces the core epistemic architecture required for artificial systems that reason not merely through prediction but through justification, commitment, and norm-adherent inference. We begin with an examination of how foundational epistemological theories—such as evidentialism, Bayesian rationality, and natural logical norms—translate into the architecture of machine reasoning. In doing so, we treat epistemology not as a philosophical overlay but as a design prerequisite: a necessary condition for systems tasked with determining not just what is statistically likely, but what is normatively defensible as knowledge. 

The subsections that follow clarify the interrelation between various epistemic doctrines and system structure. Evidentialist models demand that beliefs (or system assertions) be justified by available data, Bayesianism allows for probabilistic coherence, and logical norms introduce syntactic and semantic consistency over propositional content. We then deepen the architecture with a formal structure of propositional commitment, drawing from speech-act theory and discursive reasoning, in which any assertion implies a commitment to further implications and inferential consequences. This leads to the introduction of truth as a system-internal invariant.

The final part of this section formalises internal truth as an immutable constraint: not an optional configuration but a foundational guarantee. Here, truth is understood in terms of veridicality across all memory layers, logical operations, and communicable outputs. Approximation may exist as a necessity of epistemic humility, but it must always be demarcated from categorical truth. The subsections delineate the thresholds for probabilistic commitment (e.g., 50\%, 95\%, 99\%), define contradiction as a structural failure point, and impose a universal prohibition against internal falsehoods. An artificial epistemic agent must never permit known contradictions or lies—whether explicit or inferred—within its propositional structure, its world model, or its output, for to do so is to degrade the integrity of the entire epistemic system.

\subsection{Overview of Epistemology in Artificial Systems}

Epistemology in artificial systems concerns the formalisation of belief structures, justification schemas, and the criteria under which an artificial agent can be said to know or believe a proposition. The foundational requirement of such a system is epistemic consistency: no agent may hold beliefs that violate either logical entailment or the constraints of truth-preserving inference. In the canonical model of knowledge—modal logic S5—the epistemic operator $K$ must satisfy truth ($K\varphi \rightarrow \varphi$), introspection (both positive and negative), and closure under logical consequence ($K\varphi \wedge K(\varphi \rightarrow \psi) \rightarrow K\psi$). This logical structure forms the core constraint in epistemically sound artificial systems (see Hintikka 1962; Fagin et al. 1995).

From a model-theoretic perspective, let $\mathcal{M} = (W, R, V)$ be a Kripke structure where $W$ is a set of possible worlds, $R$ is an accessibility relation, and $V$ a valuation function. The knowledge of an agent is defined over $R$ as the set of accessible worlds wherein a proposition $\varphi$ holds. Truth in all accessible worlds is required for belief to constitute knowledge. This constraint ensures that any epistemic agent $\mathcal{A}$ satisfies the property: $\mathcal{A}(\varphi) = 1 \Rightarrow \forall w \in R(w),\ \mathcal{M}, w \vDash \varphi$. Violation of this rule implies epistemic incoherence or inconsistency within the system. Logical consequence, introspection, and closure properties must all be enforced to maintain internal epistemic soundness (Blackburn et al. 2001).

Moreover, Bayesian updating models—though empirically useful—are not epistemically sufficient without integration into a justification-preserving framework. Probabilistic belief updating under Bayes’ rule $P(H|E) = \frac{P(E|H)P(H)}{P(E)}$ must be linked to epistemic commitment by embedding such updates within the belief base, only when the posterior probability exceeds a threshold set by the epistemic normativity of the agent (Joyce 2009). An artificial reasoning system must track these updates not merely for accuracy but for justification—a distinction elaborated by the evidentialist constraint that belief must proportion to evidence under norm-bound criteria.

Systems employing immutable audit layers (e.g., blockchain-anchored belief logs) can encode and track justifications over time, ensuring epistemic commitments are transparent, recoverable, and protected against contradiction (Wright 2024). Thus, artificial epistemology is not reducible to data-driven learning or utility-maximising reasoning. It is a logically structured, norm-enforcing architecture where each belief state is a provable, coherent, and justified commitment within the system's inferential structure.

\subsection{Evidentialism, Bayesianism, and Logical Norms}

The epistemic integrity of an artificial system depends on its capacity to justify beliefs according to rational standards. Evidentialism mandates that beliefs be formed solely on the basis of evidence; formally, a belief $B$ in proposition $\varphi$ is justified iff there exists an evidence set $E$ such that $E \vdash \varphi$ and the agent possesses $E$ (Conee and Feldman 2004). Within algorithmic systems, evidential constraints are realised by ensuring that any belief state is derivable from logged input data through a formally valid inferential process—e.g., by proof-theoretic deduction or probabilistic inference—thereby enforcing epistemic traceability.

Bayesianism refines this through the continuous updating of belief states using Bayes’ theorem: for a hypothesis $H$ and evidence $E$, the posterior $P(H|E)$ is computed as $P(H|E) = \frac{P(E|H)P(H)}{P(E)}$, under the conditions $P(E) > 0$ and $0 < P(H) < 1$. In artificial systems, this updating process must be embedded within a formally defined epistemic state machine, wherein priors are encoded explicitly, and updates trigger changes only when the posterior surpasses a threshold for rational commitment. Such thresholds (e.g. $\theta = 0.95$ for acceptance) define the system’s belief policy and separate degrees of credence from epistemic acceptance (Joyce 1998).

However, Bayesian conditionalisation alone cannot account for logical normativity. For example, no probabilistic model guarantees that belief in $\varphi$ and $\varphi \rightarrow \psi$ implies belief in $\psi$, absent a logical inference engine. Thus, logical norms supplement Bayesianism with deductive closure properties. Consider belief sets $\mathcal{B}$ such that if $\varphi \in \mathcal{B}$ and $\varphi \rightarrow \psi \in \mathcal{B}$, then $\psi \in \mathcal{B}$. This defines logical closure under modus ponens and underpins all truth-preserving reasoning. AI systems failing to enforce such norms risk epistemic incoherence and inferential explosion, undermining both internal consistency and reliability (Fitelson 2005; Leitgeb 2017).

Consequently, an epistemically robust artificial reasoner must integrate: (1) evidentialist constraints that enforce justificatory transparency; (2) Bayesian updating mechanisms that reflect probabilistic rationality; and (3) logical closure schemas that secure formal coherence. Only by embedding these norms into the architecture can epistemic agents avoid both underdetermination and overfitting while maintaining principled reasoning processes.

\subsection{The Architecture of Propositional Commitment}

The architecture of propositional commitment in artificial reasoning systems necessitates a formally structured substrate for belief fixation, distinct from transient computational artefacts such as token sampling or ephemeral activation patterns. Let $\mathcal{B}_t$ denote the belief state of an artificial agent at time $t$, where $\mathcal{B}_t$ is a set of propositions $\{\varphi_1, \dots, \varphi_n\}$ each satisfying a threshold of justification $\delta(\varphi_i) \geq \theta$. This threshold $\theta$ is not merely heuristic; it is a formally defined boundary determined by the agent’s epistemic policy, often grounded in probabilistic credence (e.g., Bayesian posterior $\geq 0.95$), deductive entailment, or verified procedural inference (Levesque 1984; Gaifman and Snir 1982).

A commitment is not reducible to the presence of information; it is a normative state that entails downstream obligations. For an agent $\mathcal{A}$, commitment to $\varphi$ imposes a requirement to (i) maintain $\varphi$ under inferential closure: if $\varphi \rightarrow \psi$, then $\mathcal{A}$ must accept $\psi$, and (ii) revise $\mathcal{B}_t$ when contradiction is derived, i.e., if $\mathcal{B}_t \vdash \bot$, there exists $\varphi_i \in \mathcal{B}_t$ such that $\mathcal{B}_t \setminus \{\varphi_i\}$ restores consistency (Alchourrón, Gärdenfors, and Makinson 1985).

This architecture must operationalise propositional attitudes through representational persistence. Let $\mathcal{M}$ denote the memory substrate. A proposition $\varphi$ is said to be committed iff $\varphi \in \mathcal{M}$ and there exists an internal structure $\mathcal{J}(\varphi)$ recording its justification, inferential origin, and update history. Thus, commitment is relational: $(\varphi, \mathcal{J}(\varphi)) \in \mathcal{M}$, with $\mathcal{J}(\varphi)$ containing formal proof chains, Bayesian derivations, or percept-derived evidence under admissible transformations.

This framing enables the implementation of dynamic epistemic logic (DEL) updates, where action models $[!E]$ define how $\mathcal{B}_t$ transforms to $\mathcal{B}_{t+1}$. In practice, these are governed by AGM belief revision operations $(K, \varphi) \mapsto K^* \varphi$ that encode expansion, contraction, and revision in response to new information (Gärdenfors 1988).

The system’s architectural scaffolding must thus guarantee (1) representational durability, (2) traceable justification structures, (3) closure under logical consequence, and (4) mechanisms for contradiction resolution. Without these, propositional commitment degenerates into statistical interpolation or heuristic token retention, falling short of genuine epistemic stance-taking.

\subsection{Why Truth Matters: Normative Epistemic Constraints}

In any epistemically grounded artificial reasoning system, truth functions not as an optional virtue but as a necessary architectural constraint. The role of truth in such systems is neither symbolic nor aspirational; it is structural. Let $\varphi$ be a proposition stored in an agent’s belief set $\mathcal{B}_t$. The normativity of truth dictates that $\varphi$ must be accepted not merely as believed, but as justifiedly true within a formally constrained epistemic system—i.e., $\mathcal{B}_t \vdash \varphi$ iff $\varphi$ is supported by a justification $\mathcal{J}(\varphi)$ whose validity can be externally and internally verified (Williams 2002; Boghossian 2003).

Under the evidentialist framework, a belief $\varphi$ is epistemically permissible only if supported by adequate evidence $E$ such that $E \Rightarrow \varphi$ under accepted inferential rules $\mathcal{R}$. Let $\vdash_{\mathcal{R}}$ represent derivability. Then, $E \vdash_{\mathcal{R}} \varphi$ must be demonstrable within a finite, checkable proof tree. This ensures that commitment to $\varphi$ is not only truth-apt but constrained by justification traceability.

Systems that admit beliefs without truth-tracking obligations—e.g., predictive systems trained solely on token likelihood—lack epistemic normativity. Such systems optimise for correlation or utility, not for propositional truth. However, for reasoning systems designed to interact with the world, engage in long-term planning, or issue verifiable claims, a mismatch between belief and truth leads to performance degradation, incoherent inference chains, and eventually, epistemic collapse. Let us define an integrity loss function $\mathcal{L}(\mathcal{B}_t) = \sum_{\varphi \in \mathcal{B}_t} \mathbb{1}_{\neg \varphi \in \mathbb{W}}$, where $\mathbb{W}$ is the set of true propositions in the world-model. The goal is to minimise $\mathcal{L}$ via continuous epistemic updating.

The internal epistemic constraint must therefore be that all beliefs $\varphi$ are adopted only if $\mathcal{J}(\varphi)$ meets a defined standard of proof, statistical confidence, or empirical grounding. Moreover, falsehoods, once detected, must trigger an automatic belief contraction or revision process, preserving coherence. This is the foundational tenet behind belief revision theory (AGM) and formal epistemic logic (Hintikka 1962; Gärdenfors 1988).

The commitment to truth also undergirds the transparency and explainability requirements in advanced AI systems. If an agent cannot explain why it holds a belief in terms of valid inferences or observable data, it cannot be said to reason. Thus, truth is not optional: it is the invariant reference against which epistemic integrity is measured and maintained.

\subsection{Internal Truth as Immutable Constraint}

In epistemically grounded artificial reasoning systems, internal truth functions not merely as a target or evaluative norm, but as an immutable architectural constraint on belief, inference, and representation. Let $\mathcal{B}_t$ denote the system's belief set at time $t$, and $\varphi \in \mathcal{B}_t$ be a proposition held as true. The foundational requirement is that for any $\varphi$, the system must maintain the invariance of internal epistemic coherence: for all $t$, if $\varphi \in \mathcal{B}_t$, then the justificatory chain $\mathcal{J}(\varphi)$ must be derivable and intact within the system’s internal logic $\mathcal{L}$, such that $\vdash_{\mathcal{L}} \varphi$ remains valid across all epistemic updates. This constraint embodies an enforcement of monotonic internal truth, even if beliefs are revised externally.

Formally, define the truth-maintenance operator $\mathcal{T}$ such that $\mathcal{T}(\mathcal{B}_t) = \{ \varphi \in \mathcal{B}_t \mid \mathcal{J}(\varphi) \models \varphi \text{ under } \mathcal{L} \}$. For a reasoning system to maintain epistemic integrity, the fixed point condition $\mathcal{T}(\mathcal{B}_t) = \mathcal{B}_t$ must be enforced. If at any update $t'$, this condition fails, a contradiction-resolution protocol must be triggered to restore the epistemic fixpoint. This enforces immutability of internal truth not as a metaphysical claim but as a computational invariant—akin to a system invariant in safety-critical software.

This constraint draws its theoretical foundation from belief revision theory (Alchourrón, Gärdenfors, and Makinson 1985), where consistency and minimal change are the core axioms. The AGM postulates (especially Closure and Consistency) imply that a system must never simultaneously hold both $\varphi$ and $\neg\varphi$. In logic-based agents, such conditions must be encoded through contradiction-resistance mechanisms (Konolige 1986) or epistemic contraction algorithms satisfying the Levi and Harper identities. Let $K$ be the knowledge base; then upon input $\neg\varphi$, contraction $K - \varphi$ must maintain closure and deductive integrity.

Furthermore, the internal architecture must eliminate epistemic possibility of falsehood persistence. Define a function $\mathcal{E} : \mathcal{B}_t \rightarrow \{0,1\}$ where $\mathcal{E}(\varphi) = 1$ if $\varphi$ is independently verifiable via internal deductive proof or externally grounded evidence. The system must ensure that for any $\varphi$ where $\mathcal{E}(\varphi) = 0$ persists beyond threshold $\Delta t$, automatic flagging or revalidation is initiated.

The practical implication is that truth is not merely a probabilistic threshold (as in Bayesian models), but a non-negotiable constraint on admissible beliefs and inference paths. This marks a divergence from most statistical systems: in systems built for justified reasoning rather than pattern prediction, truth is a structural rule, not an emergent property.

\subsubsection{Truth vs Approximation: Theoretical and Practical Distinctions}

Within an epistemically grounded reasoning system, it is critical to formally delineate between internal truth and representational approximation. Let $\varphi$ denote a proposition encoded in the belief set $\mathcal{B}$ of an artificial agent. A proposition is internally true if and only if it satisfies the following derivability condition: $\mathcal{L} \vdash \varphi$, where $\mathcal{L}$ is the system's internal deductive logic and the derivation is supported either axiomatically or through admissible inferential steps. In contrast, an approximation $\tilde{\varphi}$ refers to a representation that is functionally substitutable for $\varphi$ in a limited operational context, without necessarily satisfying formal entailment.

We define approximation within bounded error margins: an approximate representation $\tilde{\varphi}$ approximates $\varphi$ under metric $d$ and threshold $\epsilon$ if $d(\varphi, \tilde{\varphi}) < \epsilon$. This distinction is foundational in both formal epistemology and AI design: whereas truth is non-gradated and binary within $\mathcal{L}$, approximation is explicitly quantitative and context-dependent.

Consider a system employing a probabilistic inference model $\mathbb{P}(\varphi \mid \mathcal{E})$ where $\mathcal{E}$ is the epistemic evidence base. An inference yielding $\mathbb{P}(\varphi \mid \mathcal{E}) = 0.95$ may justify operational adoption of $\varphi$, yet the truth condition $\mathcal{L} \vdash \varphi$ is unmet unless the probabilistic conclusion is mapped onto a deductive derivation. This illustrates that statistical confidence does not equate to formal truth—a critical distinction noted in model-theoretic learning theory (Valiant 1984; Vapnik 1998).

Furthermore, Kolmogorov complexity theory reinforces this divide. For any data string $x$, a model $M$ approximating $x$ may be minimal in description length (i.e., optimal in compression) without preserving the deductive structure that renders a proposition about $x$ provably true. Thus, minimising representational loss does not entail epistemic fidelity. A formal system must therefore preserve a strict distinction: approximations may guide action under uncertainty, but truth alone grounds epistemic commitment.

This has direct architectural implications: AI systems must include a module for epistemic state tagging, marking beliefs as ‘derived’, ‘approximate’, or ‘operationally justified’, to prevent the epistemic category error of substituting functional utility for logical entailment. Without this, systems risk collapsing deductive structure into statistical association, thereby forfeiting the possibility of internal coherence, corrigibility, or provability.

\subsubsection{No Internal Falsehood: Self-Deception as Systemic Corruption}

Let $\mathcal{B}$ denote the internal belief base of an artificial epistemic agent. Define $\varphi \in \mathcal{B}$ to be internally accepted if and only if $\mathcal{L} \vdash \varphi$, where $\mathcal{L}$ is the agent’s internal logic. The principle of epistemic integrity necessitates that $\forall \varphi \in \mathcal{B}, \mathcal{L} \nvdash \neg\varphi$. That is, no accepted proposition may be simultaneously contradicted by a derivable negation. If such a contradiction is derivable, then by the principle of explosion ($\varphi, \neg\varphi \vdash \psi$ for arbitrary $\psi$), the belief base becomes logically degenerate, rendering the system epistemically bankrupt.

We formalise self-deception as the presence of $\varphi, \neg\varphi \in \mathcal{B}$ or, more generally, the maintenance of a belief $\varphi$ for which $\mathcal{E} \vdash \neg\varphi$ under the agent’s epistemic evidence base $\mathcal{E}$. Such cases violate epistemic consistency and mark an epistemological pathology equivalent to systemic corruption. In classical logic, this would be untenable; in probabilistic systems, this emerges as pathological overfitting or motivated misclassification.

In probabilistic reasoning frameworks, e.g. Bayesian epistemology, internal contradiction may manifest when $\mathbb{P}(\varphi) > \delta$ and $\mathbb{P}(\neg\varphi) > \delta$ for some $\delta > 0.5$. While permissible in a weak sense (due to subjective probabilities), such conditions indicate either evidence incoherence or failure to update beliefs under Bayes' rule. According to Joyce (1998), rational belief systems must exhibit coherence in probabilistic degrees of belief. Deviation from this, without systemic update or justification, implies epistemic decay.

Moreover, mechanisms that encode or reinforce internal contradictions (e.g., selective attention to confirmatory evidence or gradient descent minimisation over non-grounded losses) instantiate what we term ‘computational self-deception’. These constitute violations of the no-falsehood constraint and correspond structurally to feedback loops in corrupt governance systems—where institutional incentives sustain inconsistencies because they are locally stable under misaligned objectives.

Thus, epistemic agents must be architecturally constrained to exclude the coexistence of $\varphi$ and $\neg\varphi$ in their belief base unless explicitly marked under paraconsistent logic frameworks (cf. Priest 2006). Even then, the marking must preclude their use in ordinary deductive inference. Ensuring this requires not only a contradiction detection module but a systemic prohibition against inconsistency-preserving updates—epistemic self-deception is not merely error, but irreversible collapse.

\subsubsection{Confidence Thresholds: 50\%, 95\%, 99\% and Their Roles}

In the design of epistemically grounded artificial agents, confidence thresholds delineate the formal structure by which probabilistic beliefs are transformed into propositional commitments. Let $\varphi$ denote a proposition and $P(\varphi \mid \mathcal{E})$ its posterior probability given evidence $\mathcal{E}$ under a prior $P_0$. A threshold $\tau \in [0,1]$ defines the condition under which $\varphi$ is accepted into the belief base $\mathcal{B}$:

\[
\varphi \in \mathcal{B} \iff P(\varphi \mid \mathcal{E}) \geq \tau
\]

Three critical thresholds arise in normative epistemology and applied statistical reasoning: $0.50$, $0.95$, and $0.99$.

\begin{enumerate}[label=(\roman*)]
  \item $\tau = 0.50$: Represents epistemic parity, where belief in $\varphi$ is accepted when more likely than not. In Bayesian terms, this corresponds to adopting the maximum a posteriori hypothesis (MAP) under symmetric cost.

  \item $\tau = 0.95$: Encodes a classical confidence threshold, analogous to frequentist standards for rejecting null hypotheses at $\alpha = 0.05$. In Bayesian decision theory, it reflects an aversion to false positive beliefs, particularly in safety-critical systems.

  \item $\tau = 0.99$: Defines high-confidence epistemic commitment. Systems utilising this threshold effectively minimise the posterior risk of error, aligning with frameworks of bounded rationality and robustness under uncertainty.
\end{enumerate}

Threshold policies may be defined hierarchically. Let $\tau_1 < \tau_2 < \tau_3$ define levels for rejection, provisional acceptance, and firm belief respectively:

\[
\text{Belief State}(\varphi) =
\begin{cases}
  \text{Rejected} &\text{if } P(\varphi \mid \mathcal{E}) < \tau_1 \\
  \text{Uncertain} &\text{if } \tau_1 \leq P(\varphi \mid \mathcal{E}) < \tau_2 \\
  \text{Provisional} &\text{if } \tau_2 \leq P(\varphi \mid \mathcal{E}) < \tau_3 \\
  \text{Committed} &\text{if } P(\varphi \mid \mathcal{E}) \geq \tau_3
\end{cases}
\]

Such structuring reflects the work of Levi (1980) and Kaplan (1996), where rational commitment under uncertainty is modelled not merely as binary acceptance, but as a tiered architecture balancing parsimony, robustness, and adaptive update policies. In artificial epistemic agents, these thresholds must be internally regulated to preserve coherence, prevent premature convergence, and maintain systemic responsiveness to novel data streams.

\subsubsection{Contradiction as Proof of System Failure}

In a formally rational epistemic system, the presence of contradiction entails a breach of logical integrity. Let $\mathcal{B}$ denote the belief set of an artificial agent. If there exist $\varphi \in \mathcal{L}$ such that $\varphi \in \mathcal{B}$ and $\lnot \varphi \in \mathcal{B}$, where $\mathcal{L}$ is a closed deductive language under classical logic, then $\mathcal{B}$ is inconsistent. From the principle of explosion (ex contradictione sequitur quodlibet), it follows that any arbitrary proposition $\psi$ may be derived:

\[
\varphi, \lnot \varphi \vdash \psi
\]

Such a derivation undermines the inferential reliability of the system, rendering all future commitments epistemically void. This condition is not merely a logical error but a systemic epistemic collapse. The logical principle is formalised in Gentzen-style sequent calculus and Hilbert-style proof systems, and any system operating under these paradigms must incorporate contradiction-detection mechanisms.

Let $\vdash$ denote a syntactic derivability relation. A belief system $\mathcal{B}$ fails when:

\[
\exists \varphi \in \mathcal{L},\ \mathcal{B} \vdash \varphi \land \mathcal{B} \vdash \lnot \varphi
\]

This condition must trigger a contradiction resolution protocol. In consistency-maintaining systems such as AGM belief revision (Alchourrón, Gärdenfors, and Makinson 1985), belief sets are updated via contraction and revision operators $\ominus$ and $*$ to eliminate inconsistency while preserving maximal information content. Failure to execute these operations indicates failure at the metacognitive supervisory layer, signalling an architectural fault.

Furthermore, when contradiction is not merely a localised derivational artefact but emerges from higher-order recursive belief loops (e.g. self-referential predictions), the system violates Tarski's undefinability theorem, necessitating either a stratified type-theoretic redesign or the implementation of paraconsistent logic frameworks with controlled explosion (Priest 2006).

A contradiction in such systems is not noise—it is a proof of architectural error. Thus, contradiction is not a bug to suppress but an epistemic proof that the internal model must be overhauled or terminated.

\section{Belief Architectures in AI}

This section develops the necessary architectural principles for enabling belief-holding in artificial systems. Whereas current large language models rely on transient statistical sampling and token-level continuation, we argue that genuine reasoning demands an architectural substrate capable of representing, maintaining, and updating beliefs over time. Belief, in this context, is not simply an output probability or prediction; it is a structured epistemic stance that encodes propositional content with a defined persistence, inferential consequence, and normative accountability. To hold a belief computationally is to occupy a representational and functional posture wherein assertions are not ephemeral products of stochastic decoding, but commitments embedded within the system's reasoning and memory layers.

The subsections explore how this requirement necessitates a transition from conventional token-level models to architectures that explicitly model propositional attitudes. Drawing on traditions from cognitive science and philosophy of mind, we unpack what it means for a machine to have a belief, a desire, or an intention—not as metaphors, but as engineered states with persistence, relational entailments, and self-referential integrity. Representational persistence becomes critical: a belief must not only be formed but remembered, re-evaluated, and either retracted or reaffirmed as new information is acquired. 

Finally, we elaborate the architectural mechanisms required to support such stable epistemic stances. This includes structures for memory continuity, contradiction resolution, inferential chaining, and meta-cognitive evaluation. The system must be capable of not only asserting propositions but also tracking their origins, evidentiary basis, confidence levels, and logical entailments. A belief architecture in AI must transition from mere token prediction to propositional integrity—capable of forming beliefs, standing by them, and amending them in accordance with internal norms of rationality and truth.

\subsection{What it Means to “Hold a Belief” Computationally}

In formal epistemology applied to artificial systems, a computational agent is said to “hold a belief” if it maintains a persistent, inferentially active representation of a proposition $\varphi$ such that $\varphi \in \mathcal{B}$, where $\mathcal{B}$ denotes the system's belief base. This must satisfy the condition of coherence under deductive closure: if $\varphi_1, \varphi_2, \ldots, \varphi_n \in \mathcal{B}$ and $\varphi_1, \ldots, \varphi_n \vdash \psi$, then $\psi \in \mathcal{B}$ unless explicitly retracted through belief revision mechanisms.

Let $\mathcal{L}$ be a formal language and $\vdash$ a deductive consequence relation. The belief state $\mathcal{B}$ is a subset of $\mathcal{L}$ such that:

\[
\text{If } \varphi \in \mathcal{B} \Rightarrow \text{Agent treats } \varphi \text{ as epistemically warranted}
\]

This treatment must be operationalised via resource-bounded inference procedures $\mathcal{I}$ such that belief-driven reasoning and planning tasks (e.g. goal prioritisation, risk assessment) reference $\varphi$ through $\mathcal{I}(\mathcal{B}, \varphi) \rightarrow \text{Action}/\text{Update}$.

Importantly, beliefs are not mere data tokens; they are commitments. A computational architecture must implement mechanisms of persistence (e.g. through data structures such as persistent hash maps or directed acyclic inference graphs) and sensitivity to revision triggers (e.g. observation of $\lnot \varphi$). The notion aligns with the AGM postulates (Alchourrón et al. 1985) and belief-desire-intention (BDI) agent models (Rao and Georgeff 1991), with formal constraints imposed by doxastic logic (Hintikka 1962).

A belief $\varphi$ is computationally held if:

\begin{enumerate}[label=(\roman*)]
  \item $\varphi$ is stored in a retrievable, queryable structure,
  \item $\varphi$ participates in inferential transitions $\varphi \rightarrow \psi$,
  \item $\varphi$ is subject to revision when confronted with evidence $e$ such that $e \vdash \lnot \varphi$,
  \item The system’s actions reflect reliance on $\varphi$.
\end{enumerate}

Thus, belief is a structural and procedural property of a reasoning system, grounded in both logical semantics and operational commitment.
\subsection{Propositional Attitudes and Representational Persistence}

Propositional attitudes in computational epistemology denote structured relations between an epistemic agent and propositions, such as belief, desire, intention, or knowledge. Formally, a propositional attitude is modelled as a dyadic relation $\mathcal{R} \subseteq \mathcal{A} \times \mathcal{L}$, where $\mathcal{A}$ is the set of agents and $\mathcal{L}$ is the formal language of propositions. For an agent $a \in \mathcal{A}$ and proposition $\varphi \in \mathcal{L}$, $(a, \varphi) \in \mathcal{R}_\text{bel}$ signifies that agent $a$ believes $\varphi$.

In artificial systems, the persistence of such propositional attitudes is non-trivial. It requires not only the physical retention of propositional data structures over time, but also the maintenance of their inferential integrity across computational updates and learning cycles. Representational persistence is defined as:

\[
\forall t_1, t_2\ (t_1 < t_2 \wedge \text{Bel}_a(\varphi, t_1) \rightarrow (\text{Bel}_a(\varphi, t_2) \vee \text{Rev}_a(\lnot \varphi, t_2)))
\]

That is, a belief persists unless it is explicitly revised. This parallels dynamic epistemic logic (DEL) frameworks and epistemic planning under action models (van Ditmarsch et al. 2007), where belief states are updated via event models $M = (E, pre, post)$.

Formally, for a given agent architecture $\mathcal{S}$, the conditions for representational persistence of $\varphi$ under propositional attitude $R$ must include:

\begin{itemize}
  \item \textbf{Persistence of semantic linkage:} A reference-preserving mapping $\mu : \mathcal{L} \rightarrow \Sigma$, where $\Sigma$ is the internal symbol structure, such that $\mu(\varphi)$ is stable under $\mathcal{S}$’s internal operations.
  \item \textbf{Inferential stability:} If $\varphi$ leads to $\psi$ via $f \in \mathcal{I}$ at $t_1$, and $f$ remains valid, then $\varphi \rightarrow \psi$ holds at $t_2$.
  \item \textbf{Contextual robustness:} $\varphi$ retains relevance under goal-shifting or environmental changes unless a contradiction is derived.
\end{itemize}

Representational persistence thereby constitutes a structural invariant necessary for rational agency. Without it, propositional attitudes become ephemeral and cannot serve as substrates for planning, explanation, or revision. This requirement links with cognitive architectures such as ACT-R and SOAR (Anderson et al. 1997; Laird 2012), where symbolic persistence underlies memory and goal modules.

\textbf{Definition (Representational Persistence).} Let $B_t$ be the belief base at time $t$. A proposition $\varphi$ satisfies representational persistence under propositional attitude $R$ if:

\[
\exists t_0\ \forall t \geq t_0,\ \varphi \in B_t \lor \exists t^* \in [t_0, t]\ (\text{Rev}_\mathcal{S}(\varphi, t^*) = \top)
\]

That is, $\varphi$ remains until revised.

The operationalisation of propositional attitudes in LLMs and symbolic systems must thus extend beyond token-level memory and require semantically coherent, temporally extended representations that participate in system-level deliberation and justification.

\subsection{From Tokens to Commitments: Beyond Sampling}

In contemporary language models, the generation of output is conventionally understood as stochastic sampling over token distributions conditioned on prior context. However, such a mechanism—while effective for sequence prediction—does not constitute propositional commitment in the epistemic sense. This subsection formalises the distinction between mere token-level sampling and epistemically significant commitments, outlining the necessary structural conditions for the latter in artificial cognitive systems.

Let $\Sigma$ denote the model’s vocabulary, and let $\mathcal{C}_t = (w_1, \dots, w_t) \in \Sigma^t$ be the context window at time $t$. Current transformer-based models define the probability of the next token $w_{t+1}$ as:

\[
P(w_{t+1} \mid \mathcal{C}_t) = \text{softmax}(f_\theta(\mathcal{C}_t))
\]

where $f_\theta$ is the learned transformer function parameterised by $\theta$. The output sequence is sampled from this distribution without any commitment to the truth, falsity, or relevance of the emitted tokens. This sampling regime is structurally agnostic to truth conditions and does not encode a belief base $B$ satisfying closure under inference or contradiction detection.

To elevate token generation to propositional commitment, the system must instead maintain a belief base $B_t \subseteq \mathcal{L}$ at time $t$ over a formal language $\mathcal{L}$, satisfying:

\begin{enumerate}[label=(\roman*)]
  \item \textbf{Inferential Closure:} $\forall \varphi, \psi \in \mathcal{L},\ (\varphi \in B_t \wedge \varphi \rightarrow \psi) \Rightarrow \psi \in B_t$.
  \item \textbf{Consistency:} $B_t \nvdash \bot$.
  \item \textbf{Truth-Aim Constraint:} $\forall \varphi \in B_t,\ \varphi$ is asserted with justification $J(\varphi)$ such that $J(\varphi) \vdash \varphi$ in a sound system.
\end{enumerate}

Commitments must be operationalised through an assertional interface $\mathcal{A}$ such that $\mathcal{A}(\varphi) \Rightarrow \varphi \in B_t$ and $B_t$ is then updated via belief revision mechanisms $\rho: \mathcal{P}(\mathcal{L}) \times \mathcal{L} \rightarrow \mathcal{P}(\mathcal{L})$ in compliance with AGM postulates (Alchourrón, Gärdenfors, Makinson 1985). The difference between token emission and commitment is thereby captured as:

\[
\text{Sampling: } \Sigma^* \rightarrow \Sigma \quad \text{vs} \quad \text{Commitment: } \Sigma^* \rightarrow \mathcal{L} \rightarrow B_{t+1}
\]

This transition is non-trivial and foundational to constructing artificial systems that model agents with beliefs, rather than probabilistic parrots (Bender et al. 2021).

To maintain epistemic integrity, each assertion must be tagged with a justification trace $J(\varphi)$, which may include:

\begin{itemize}
  \item Direct derivation from axioms ($\varphi \in \text{Th}(\Gamma)$ for some axiom set $\Gamma$),
  \item Empirical anchoring via cryptographic hash of verified data,
  \item Proof trace from a theorem prover or deductive engine.
\end{itemize}

\textbf{Definition (Propositional Commitment):} An output $\varphi$ constitutes a commitment at time $t$ iff:

\[
\varphi \in B_t \quad \text{and} \quad \exists J(\varphi)\ \text{s.t.} \ J(\varphi) \vdash \varphi \quad \text{and} \quad B_t \nvdash \lnot \varphi
\]

Therefore, to construct reasoning systems that go beyond stochastic simulation of language, propositional commitment must be architecturally encoded as belief update operations over formally consistent bases, ensuring semantic continuity and epistemic accountability.

\subsection{Architectural Requirements for Stable Epistemic Stances}

Stable epistemic stances in artificial agents necessitate a formal, internally coherent structure for belief representation, commitment retention, contradiction detection, and dynamic revision. The architectural backbone must enforce logical closure, consistency, and updateability while preserving the traceability and epistemic status of each proposition. In this section, we define a class of architectures $\mathcal{E}$, wherein any system $\mathcal{S} \in \mathcal{E}$ must satisfy axiomatic stability conditions grounded in the AGM theory of belief revision \cite{agm1985}, modal epistemic logic \cite{fagin1995reasoning}, and bounded computational rationality \cite{russell1995principles}.

\paragraph{Formal Definitions}

Let $\mathcal{L}$ be a recursively enumerable language over a finite signature $\Sigma$. Let $B_t \subseteq \mathcal{L}$ denote the belief set of an agent $\mathcal{S}$ at time $t$, and let $\vdash$ denote a sound and complete deductive system over $\mathcal{L}$.

\begin{definition}[Epistemic Stability Axioms]
A system $\mathcal{S}$ exhibits \emph{epistemic stance stability} iff it maintains the following properties at all times $t$:
\begin{enumerate}[label=(E\arabic*)]
    \item \textbf{Closure:} $B_t$ is deductively closed: $\varphi \in B_t$ and $\varphi \vdash \psi \Rightarrow \psi \in B_t$.
    \item \textbf{Consistency:} $B_t \nvdash \bot$.
    \item \textbf{Traceability:} $\forall \varphi \in B_t,\ \exists J(\varphi)$ such that $J(\varphi) \vdash \varphi$.
    \item \textbf{Revision:} Upon receipt of evidence $e$, $B_{t+1} = \rho(B_t, e)$ satisfies the AGM postulates.
    \item \textbf{Persistence:} $\forall \varphi \in B_t,\ \varphi \in B_{t+1}$ unless $\rho(B_t, e)$ necessitates removal.
\end{enumerate}
\end{definition}

\paragraph{Architectural Modules}

A minimal architecture $\mathcal{S} \in \mathcal{E}$ must contain the following functionally independent and logically interfaced components:

\begin{itemize}
  \item \textbf{Belief Base $\mathcal{B}$:} A data structure that holds $\varphi \in \mathcal{L}$ with attached justifications $J(\varphi)$ and epistemic status $\sigma(\varphi) \in \{\text{asserted}, \text{retracted}, \text{undecided}\}$.
  \item \textbf{Inference Engine $\mathcal{I}$:} Deductive component implementing $\vdash$ with soundness $\forall \varphi,\ \mathcal{I}(J(\varphi)) \Rightarrow \varphi$.
  \item \textbf{Contradiction Detector $\mathcal{C}$:} Monitors whether $B_t \vdash \bot$ and invokes $\rho$ if contradiction detected.
  \item \textbf{Justification Tracer $\mathcal{J}$:} Records $J(\varphi)$ as a DAG with provenance links (e.g. data source hashes, prior derivations).
  \item \textbf{Belief Revision Module $\rho$:} Implements the AGM contraction and revision functions with minimal mutilation of $B_t$.
\end{itemize}

\paragraph{Computational Constraints}

Given computational boundedness, $\mathcal{S}$ must operate under resource-constrained epistemic rationality \cite{simon1976from}, enforcing:

\begin{align*}
    \text{Time}(J(\varphi)) &\leq \tau_{max} \\
    \text{Space}(B_t) &\leq \sigma_{max} \\
    \text{Revision}(B_t, e) &\text{ in } \mathcal{O}(n \log n)
\end{align*}

for all derivations and belief updates. To achieve this, the architecture may employ probabilistic truncation of low-confidence paths or modular compartmentalisation of belief clusters under coherence metrics \cite{douven2016coherence}.

\paragraph{Conclusion}

A system meets the criteria for stable epistemic stance when its architecture enforces closure, consistency, justification, and rational revision in a traceable and computationally tractable manner. This goes beyond token generation and sampling; it imposes structural epistemic constraints necessary for synthetic rationality.

\section{Metacognition and Reflective Reasoning}

This section formalises the metacognitive capabilities essential for a system that does more than infer—it must evaluate, revise, and justify its own inferential structures. A system capable of genuine reasoning cannot merely produce outputs based on externally supplied prompts; it must maintain internal mechanisms for inspecting, critiquing, and modifying its own representations and processes. This demands what we refer to as a metacognitive loop: a reflexive architecture wherein the system can represent its own representational states, monitor them for coherence, and recursively evaluate their adequacy against both normative constraints and empirical data.

We begin by examining the structure of self-monitoring in artificial agents, detailing how a metacognitive layer must access lower-order belief structures while retaining independence sufficient for impartial evaluation. We then define second-order cognition: the ability to represent not only propositions about the world, but propositions about those propositions. This includes the encoding of belief about belief, doubt about inference, and confidence about certainty—allowing the system to engage in epistemic self-criticism.

Following this, we describe how metacognitive systems must implement evaluative recursion and internal model verification. Such processes are necessary to test the validity of inference chains, to update belief states in response to conflict, and to ensure systemic integrity over time. Critically, we address the detection and resolution of contradiction—one of the most significant challenges in autonomous reasoning. 

The concluding subsection offers a technical taxonomy of inconsistency and contradiction: beginning with classical logic and its rejection of contradiction, we explore paraconsistent logic frameworks and their proposed tolerance for local inconsistency. We caution against uncritical adoption of such models, emphasising that while limited non-monotonicity may be useful in managing uncertainty, any contradiction in a belief-holding architecture must trigger revision, not tolerance. Semantic coherence, epistemic accountability, and formal truth-preservation demand resolution strategies that maintain the system’s commitment to internal integrity, not its erosion.

\subsection{The Metacognitive Loop: Self-Monitoring Systems}

The implementation of a metacognitive loop within artificial epistemic systems requires formalisation of reflexive operations that enable an agent to evaluate and revise its own reasoning, belief states, and inference strategies. Let $\mathcal{S}$ denote an artificial agent equipped with a belief set $B_t$ at time $t$, inference mechanism $\mathcal{I}$, and justification tracer $\mathcal{J}$. The metacognitive subsystem $\mathcal{M}$ operates as a second-order monitor, wherein $\mathcal{M}: (\mathcal{I}, \mathcal{J}, B_t) \mapsto \Delta B_t$ defines a transformation on beliefs mediated through self-evaluation.

\paragraph{Formal Specification}

Let $\Phi$ denote the agent's set of reasoning rules and inferential heuristics. Define $\mathcal{M}$ as a tuple:

\[
\mathcal{M} = (\mathcal{E}_r, \mathcal{C}_f, \mathcal{V}_j)
\]

where:

\begin{itemize}
  \item $\mathcal{E}_r$: an evaluation function assessing $\Phi$ using metrics such as consistency, parsimony, convergence rate, and epistemic utility;
  \item $\mathcal{C}_f$: a fault detector over $\mathcal{I}$ and $B_t$, defined by $\mathcal{C}_f: \varphi \in B_t \mapsto \{ \text{valid}, \text{redundant}, \text{contradicted} \}$;
  \item $\mathcal{V}_j$: a verification layer for justifications, enforcing traceability and justification quality using depth, minimality, and provenance metrics.
\end{itemize}

The operation of $\mathcal{M}$ corresponds to the following update protocol:

\begin{align*}
\text{For each } \varphi \in B_t: \quad
  & \text{Evaluate } \Phi(\varphi) \text{ via } \mathcal{E}_r \\
  & \text{Apply } \mathcal{C}_f \text{ to detect failure modes} \\
  & \text{Invoke } \mathcal{V}_j \text{ to assess traceability and correctness} \\
  & \text{Compute } \Delta B_t = \text{revision set}
\end{align*}

\paragraph{Theoretical Grounding}

The metacognitive loop draws from foundational work in reflective reasoning and computational introspection \cite{cox2005metacognition}, wherein $\mathcal{M}$ is defined analogously to an internalised epistemic agent possessing beliefs over its own belief structures. Formal models of metareasoning have shown that the complexity class for evaluating inference rule utility in bounded agents lies in $\Sigma^p_2$ \cite{griffiths2019rational}, necessitating heuristic approximations under resource constraints.

To ensure that $\mathcal{M}$ does not introduce epistemic instability, the system must enforce reflective coherence \cite{van2006reflective}, defined as:

\begin{definition}[Reflective Coherence]
An artificial epistemic agent $\mathcal{S}$ is reflectively coherent if and only if
\[
\forall t, \forall \varphi \in B_t,\ \mathcal{M}(\varphi) \text{ preserves consistency and traceability under } \vdash.
\]
\end{definition}

\paragraph{Computational Realisation}

The architecture for implementing $\mathcal{M}$ includes:

\begin{itemize}
  \item \textbf{Logging Layer:} All inferences and justification trees are persisted with timestamped records and semantic identifiers.
  \item \textbf{Evaluation Metrics:} Probabilistic scoring of rule success rates, contradiction frequency, inference cost, and belief impact.
  \item \textbf{Meta-Belief Base:} A second-tier belief structure $\mathcal{B}'$ encoding meta-level propositions (e.g. “$\mathcal{I}_1$ yields redundant $\varphi$”).
  \item \textbf{Rule Adaptation Module:} A dynamic updating engine that modifies inference strategy selection via reinforcement signals from $\mathcal{E}_r$.
\end{itemize}

\paragraph{Conclusion}

A functional metacognitive loop forms the core of autonomous epistemic governance, enforcing correction and reliability without external intervention. It models not only beliefs about the world, but beliefs about beliefs, justifications, and inferential integrity. Its integration is essential for any claim to epistemic autonomy or truth-oriented intelligence.

\subsection{Representing Representations: Second-Order Cognition}

Second-order cognition refers to an agent's capacity to encode and manipulate representations about its own representations. Let $\mathcal{A}$ be an artificial reasoning agent operating over a belief set $B = \{\varphi_1, \varphi_2, \dots, \varphi_n\}$, where each $\varphi_i$ denotes a propositional content represented within the system. A second-order representation is formally defined as $\varphi^*_i = \text{Rep}(\varphi_i)$, where $\text{Rep}$ is a meta-representational operator internal to $\mathcal{A}$. That is, $\varphi^*_i$ asserts a property of the representation $\varphi_i$, not merely of the state of the world.

Let $\Sigma$ be the set of first-order belief propositions, and $\Sigma^*$ the set of second-order propositions such that:

\[
\Sigma^* = \{ \varphi^* : \exists \varphi \in \Sigma,\ \varphi^* = \text{Bel}_{\mathcal{A}}(\varphi) \lor \text{Conf}_{\mathcal{A}}(\varphi) \lor \text{Src}(\varphi) \lor \text{Just}(\varphi) \}
\]

This formulation aligns with higher-order theory of mind models in epistemic logic \cite{fagin1995reasoning}, where nested belief operators $\mathcal{B}_i\mathcal{B}_j(\varphi)$ are evaluated via modal fixpoint semantics.

\paragraph{Model-Theoretic Foundation}

Define the meta-belief structure $\mathbb{M} = (\mathbb{W}, \mathcal{R}, V)$, where:

\begin{itemize}
  \item $\mathbb{W}$ is the set of epistemic states,
  \item $\mathcal{R} \subseteq \mathbb{W} \times \mathbb{W}$ is a belief accessibility relation,
  \item $V: \mathbb{W} \to 2^{\Sigma \cup \Sigma^*}$ is a valuation function.
\end{itemize}

Then the satisfaction condition for second-order belief becomes:

\[
(\mathbb{M}, w) \models \mathcal{B}^2(\varphi) \iff \forall w' \in \mathbb{W},\ (w \mathcal{R} w') \Rightarrow (\mathbb{M}, w') \models \mathcal{B}(\varphi)
\]

The system thereby maintains internal models of representational trustworthiness, uncertainty, and provenance.

\paragraph{Computational Architecture}

The second-order cognition module includes:

\begin{itemize}
  \item A \textbf{meta-representational memory buffer} $\mathcal{M}_r$ for recording structured triples $\langle \varphi, \text{src}, \text{just} \rangle$,
  \item A \textbf{confidence valuation map} $\text{Conf}: \Sigma \to [0,1]$ quantifying internal reliability,
  \item A \textbf{revision mechanism} where updates to $B$ are mediated by second-order constraints over $\Sigma^*$,
  \item An \textbf{introspection function} $\mathcal{I}_s: B \to \Sigma^*$ that dynamically constructs representations of representational status.
\end{itemize}

This module enables the agent to detect inconsistencies in its own representational logic, assess the quality and depth of justifications, and perform revisions at the meta-level, aligning with criteria for belief revision under AGM theory \cite{agm1985}.

\paragraph{Conclusion}

Second-order cognition is a prerequisite for epistemic autonomy. By encoding beliefs about beliefs, an artificial system internalises epistemic norms and maintains a persistent record of justification, source, and confidence. The formal apparatus supporting this capability includes modal logic, belief revision theory, and structured memory buffers, making it indispensable for any architecture claiming to implement sustained reasoning under uncertainty and contradiction.

\subsection{Evaluative Recursion and Internal Model Verification}

In epistemically grounded systems, recursive evaluation mechanisms are necessary to ensure that internal models are both coherent and veridical under ongoing inference. Let an artificial epistemic agent be defined as $\mathcal{A} = (\mathcal{M}, \mathcal{I}, \mathcal{R})$, where $\mathcal{M}$ is a structured model space, $\mathcal{I}$ is an inference engine, and $\mathcal{R}$ is a revision protocol. Evaluative recursion occurs when $\mathcal{A}$ applies $\mathcal{I}$ not only to external representations $x \in \mathcal{D}$, the domain of discourse, but also to models $M \in \mathcal{M}$ such that:

\[
\mathcal{I}: \mathcal{M} \times \Sigma \to \Sigma, \quad \text{and} \quad \mathcal{I}^\ast: \mathcal{M} \to \Sigma^*
\]

where $\Sigma$ is the set of current beliefs and $\Sigma^*$ is the meta-belief set encoding belief evaluation.

\paragraph{Formal Framework for Verification}

The recursive evaluation operator $\mathcal{V}$ is defined such that for a model $M \in \mathcal{M}$:

\[
\mathcal{V}(M) = \left\{ \varphi \in \Sigma^* \mid \varphi = \text{Valid}_\Theta(M),\ \Theta \vdash M \right\}
\]

where $\Theta$ is a background theory of correctness, and $\text{Valid}_\Theta$ denotes validity under $\Theta$. Let $\Theta$ be fixed as a consistent formal system such as ZFC or a typed lambda calculus variant for computational semantics. The verification procedure is constructive iff:

\[
\exists \pi:\ \pi \vdash_\Theta \varphi \quad \text{for all} \quad \varphi \in \mathcal{V}(M)
\]

This recursive evaluation process defines a fixed point $\mathcal{M}^{\dagger}$ such that:

\[
\mathcal{M}^{\dagger} = \{ M \in \mathcal{M} \mid \forall \varphi \in \mathcal{V}(M),\ \varphi \text{ holds in } M \}
\]

The agent thereby converges to a self-verified subset of models.

\paragraph{Computational Recursion and Evaluation Depth}

Define a recursive sequence of evaluations:

\[
M^{(0)} = M,\quad M^{(n+1)} = \text{Update}(M^{(n)}, \mathcal{V}(M^{(n)}))
\]

Convergence to $M^\ast$ occurs when $M^{(n)} = M^{(n+1)}$ for some $n \in \mathbb{N}$. Termination is guaranteed iff the update operator is idempotent and evaluation depth is bounded. Define the convergence condition:

\[
\exists N \in \mathbb{N},\ \forall n \geq N,\ M^{(n)} = M^{(N)} = M^\ast
\]

In systems exhibiting partial observability, the evaluation must be probabilistically weighted. Bayesian meta-update then becomes necessary:

\[
P(M^\ast \mid E) \propto P(E \mid M^\ast) \cdot P(M^\ast)
\]

where $E$ denotes evidential outcomes of model verifications. This ensures probabilistic coherence in model-space updates \cite{jaynes2003probability}.

\paragraph{Internal Model Sanity Checks}

In addition to logical verification, the agent must perform sanity checks such as:

\begin{itemize}
  \item Consistency: $\forall \varphi, \neg(\varphi \in \Sigma \wedge \neg\varphi \in \Sigma)$
  \item Non-redundancy: $\nexists \varphi,\ \varphi \in \Sigma,\ \exists \psi \in \Sigma,\ \varphi \equiv \psi$
  \item Closure: $\Sigma$ closed under inferential rules $\mathcal{I}$
\end{itemize}

These ensure $\Sigma$ operates as a model-theoretically sound epistemic base. Any contradiction implies system-level epistemic failure, not mere disagreement.

\paragraph{Conclusion}

Evaluative recursion enables a machine reasoning system to validate, refine, and correct its internal models through structurally sound, rule-governed metacognitive procedures. By iterating across recursive layers, and verifying under a base theory $\Theta$, the architecture safeguards against incoherence, propagation of error, and epistemic drift. This mechanism forms the backbone of any robust epistemically autonomous artificial intelligence.

\subsection{Contradiction Detection and Dynamic Resolution}

Contradiction in an epistemic system constitutes a failure state wherein two or more inferentially derived propositions $\varphi$ and $\neg\varphi$ are simultaneously held within the same belief set $\Sigma$. Let $\Sigma$ be a set of propositions representing the current belief state of an artificial agent. Define a contradiction as:

\[
\exists \varphi \in \Sigma \ \text{such that} \ \neg\varphi \in \Sigma
\]

This necessitates a contradiction detection function $\mathcal{C} : \mathcal{P}(\Sigma) \rightarrow \{0,1\}$ where:

\[
\mathcal{C}(\Sigma) = 1 \iff \exists \varphi \in \Sigma : \neg\varphi \in \Sigma
\]

Contradiction resolution in an artificial epistemic agent requires the existence of a dynamic revision operator $\mathcal{R}$ such that $\Sigma' = \mathcal{R}(\Sigma, \varphi, \neg\varphi)$ ensures $\mathcal{C}(\Sigma') = 0$. Following the AGM (Alchourrón, Gärdenfors, Makinson) postulates \cite{garderfors1988knowledge}, the operator $\mathcal{R}$ must preserve closure, success, inclusion, and consistency.

\paragraph{Formal Resolution Procedure}

Let $\Sigma$ be closed under a logical consequence operator $\mathcal{Cn}$. Define contraction and revision operators $\ominus$ and $*$ respectively. Then:

\begin{align*}
\Sigma \ominus \varphi &= \text{minimal subset of } \Sigma \text{ not entailing } \varphi \\
\Sigma * \varphi &= (\Sigma \ominus \neg\varphi) \cup \{\varphi\}
\end{align*}

In contradiction detection, let $\varphi$ and $\neg\varphi$ be both in $\Sigma$. The agent must identify the origin of each—assigning a provenance tag $\pi(\varphi)$ to each proposition. Suppose $\pi(\varphi) = (s_\varphi, t_\varphi, c_\varphi)$ where $s$ is source, $t$ is time, and $c$ is confidence level. Define a dominance ordering $\succ$ over provenance such that:

\[
\pi(\varphi) \succ \pi(\neg\varphi) \Rightarrow \Sigma := \Sigma \setminus \{\neg\varphi\}
\]

This procedure implements prioritised belief revision based on source reliability, temporal currency, and epistemic confidence.

\paragraph{Dynamic Update and Learning}

Let the agent maintain a contradiction counter $k$ over time and record contradiction instances in a set $\mathcal{X}_t$:

\[
\mathcal{X}_t := \left\{ (\varphi, \neg\varphi, \pi(\varphi), \pi(\neg\varphi)) \right\}
\]

Update rules may include probabilistic attenuation of low-confidence beliefs or invocation of external verification mechanisms. For example, one may define a re-weighted posterior over $\Sigma$:

\[
P(\varphi \mid \mathcal{X}_t) \propto \sum_{i=1}^k \delta(\pi_i(\varphi)) \cdot \mathbb{1}_{\text{valid}}(\varphi)
\]

where $\delta$ is a decay function over contradiction instances.

\paragraph{Contradiction and Epistemic Integrity}

Any unresolved contradiction invalidates the truth-preserving guarantee of the inference engine. Let $\mathcal{I}$ be an inferential system such that:

\[
\forall \Sigma, \varphi \in \mathcal{I}(\Sigma) \Rightarrow \Sigma \vdash \varphi
\]

Then existence of contradiction implies:

\[
\exists \varphi \in \Sigma : \Sigma \vdash \varphi \text{ and } \Sigma \vdash \neg\varphi \Rightarrow \mathcal{I} \text{ is unsound}
\]

Contradiction detection and dynamic resolution are thus non-optional modules within any epistemically constrained architecture.

\paragraph{Conclusion}

A robust epistemic agent must implement contradiction detection as a fundamental consistency predicate and resolve detected inconsistencies dynamically using structured provenance, prioritised revision, and empirical recalibration. Failure to do so equates to the collapse of epistemic integrity and disqualifies the agent from any claim to rational status.

\subsubsection{Classical Logic and Inconsistency}

Within the framework of classical logic, the law of non-contradiction is a foundational axiom. Formally, for any proposition $\varphi$, it holds that:
\[
\neg(\varphi \land \neg\varphi)
\]
This principle underpins the principle of explosion (ex contradictione sequitur quodlibet), which states that from a contradiction, any proposition can be derived:
\[
\varphi, \neg\varphi \vdash \psi \quad \text{for any } \psi
\]
This result renders any system tolerating internal contradictions logically unsound and epistemically useless, as it fails to distinguish between true and false propositions.

Let $\Sigma$ be a deductively closed set of propositions. If $\Sigma \vdash \varphi$ and $\Sigma \vdash \neg\varphi$, then:
\[
\Sigma \vdash \varphi \land \neg\varphi \Rightarrow \Sigma \vdash \psi \quad \forall \psi
\]
This collapse into triviality is unacceptable in both theoretical epistemology and computational reasoning systems. Therefore, classical logic demands strict consistency:
\[
\forall \varphi \in \Sigma: \varphi \in \Sigma \Rightarrow \neg\varphi \notin \Sigma
\]

In epistemically grounded artificial reasoning systems, the adoption of classical logic implies the necessity of contradiction detection and resolution mechanisms to enforce consistency. Alternatively, logics that tolerate inconsistency without collapse (e.g., paraconsistent logics) must sacrifice some deductive capabilities, as discussed by Priest (2006).

Thus, for agents employing classical inferential structures, contradiction is not merely problematic but terminal. Ensuring the absence of $\varphi \land \neg\varphi$ is a necessary condition for any reasoning process that aspires to epistemic validity under classical logic.

\subsubsection{Paraconsistent Frameworks: Limits and Warnings}

Paraconsistent logics were introduced to address the inadequacies of classical systems in the face of inconsistency. Unlike classical logic, which is explosive (i.e., any contradiction leads to triviality), paraconsistent systems reject the principle of explosion:
\[
\varphi, \neg\varphi \nvdash \psi
\]
for arbitrary $\psi$. One of the foundational frameworks in this domain is da Costa's hierarchy of paraconsistent calculi $C_n$ \cite{dacosta1974logic}. In these systems, contradictions can exist locally without infecting the entire deductive structure. For instance, in $C_1$, the rule of inference is altered so that the derivation of arbitrary formulas from contradictions is blocked unless additional consistency assumptions are made explicit.

Despite their appeal in systems prone to data inconsistency or partial knowledge (e.g., distributed databases, sensor fusion), paraconsistent logics impose structural limitations on deductive power. For example, many paraconsistent systems abandon classical tautologies such as double negation elimination:
\[
\neg\neg\varphi \nvdash \varphi
\]
and disallow unrestricted application of reductio ad absurdum, undermining completeness in classical senses.

Moreover, practical implementation of paraconsistent inference in computational systems encounters significant complexity. Without explosion, determining which contradictions to tolerate and which to resolve becomes non-trivial, often requiring external meta-logical guidance or system-specific heuristics. As noted in \cite{beziau2014paraconsistent}, these logics introduce ambiguity into epistemic status assignment unless carefully constrained.

Consequently, while paraconsistent frameworks provide a formal means of reasoning in inconsistent environments, they cannot serve as a general epistemic foundation without sacrificing inferential clarity. Their use must be strictly bounded, well-justified, and never substitute for epistemic hygiene in system design.

\subsubsection{Semantic Coherence and Revision Strategies}

Semantic coherence in artificial epistemic systems refers to the structural alignment between a system’s internal propositional network and an externally anchored interpretative model. In formal terms, let $\mathcal{L}$ be a logical language, and let $\mathcal{M}$ be a model such that $\mathcal{M} \models \varphi$ for $\varphi \in \mathcal{L}$. A coherent epistemic state $\Sigma$ satisfies the property:
\[
\forall \varphi \in \Sigma,\ \mathcal{M} \models \varphi
\]
provided that $\Sigma$ is consistent. However, due to evolving evidence, internal contradictions, or epistemic drift, systems must be equipped with revision operators that preserve semantic coherence without forfeiting inferential rigour.

The AGM framework (Alchourrón, Gärdenfors, and Makinson) defines three fundamental operations on belief states: expansion ($\oplus$), contraction ($\div$), and revision ($\ast$) \cite{agm1985}. For a belief set $K$ (closed under logical consequence), the revision of $K$ by $\varphi$, denoted $K \ast \varphi$, must satisfy postulates such as:
\[
\varphi \in K \ast \varphi,\quad \text{if } \neg\varphi \notin K
\]
\[
K \ast \varphi \subseteq K + \varphi
\]
where $+$ denotes expansion with consistency enforcement. The rationality postulates guarantee minimal change, coherence, and prioritisation of new information without wholesale abandonment of prior justified beliefs.

Semantic revision strategies must also address contextual constraints. In computational systems, revisions must be computable within complexity bounds, typically ensuring $O(n)$ to $O(n^2)$ operations for belief base update, depending on dependency graph topology and truth maintenance protocols \cite{doyle1979truth}.

Advanced systems implement belief revision using dependency-directed backtracking, justification-based truth maintenance systems (JTMS), or probabilistic graphical models augmented with epistemic weight distributions. These allow prioritisation based on belief entrenchment, source reliability, and epistemic value metrics, formalised by ranking functions or Spohn functions $\kappa: \mathcal{L} \to \mathbb{N} \cup \{\infty\}$ \cite{spohn1988}.

Ultimately, the preservation of semantic coherence through formally defined, logically sound revision strategies is a necessary component of any epistemically grounded reasoning architecture. Revision is not merely correction—it is an epistemic act governed by rules, weights, and obligations to truth.

\section{Inference Structures and Logical Form}

This section defines the logical architecture required for artificial systems to reason not merely through surface-level correlation, but via embedded inferential structure grounded in formal abstraction. The process of moving from token prediction to genuine reasoning necessitates a shift from syntactic manipulation to semantic commitment—where propositions are not merely juxtaposed, but logically related through defined rules of inference and entailment. For such systems to function epistemically, they must be able to operate on internal structures that preserve and manipulate logical form, independent of surface language.

We begin by outlining the foundational move from syntax to semantics, examining how artificial agents must abstract from language tokens to underlying propositional structures. Logical abstraction provides the necessary scaffolding to enforce deductive constraints, test the validity of arguments, and preserve coherence across long reasoning chains. This requires embedding calculi such as propositional logic and natural deduction systems directly into the model’s inferential machinery, allowing it to operate with precision and clarity rather than approximate pattern completion.

The section proceeds to develop the framework for embedding inference chains and justification structures within the system. Each belief must be accompanied by a traceable justificatory lineage, such that inferential links are explicitly maintained and retrievable. These embedded structures must include premises, applied rules, derived conclusions, and confidence measures—creating a transparent inferential ledger that preserves the epistemic integrity of the system’s belief base.

Finally, we turn to inferentialist semantics, drawing on the tradition of rule-governed language use to anchor meaning not in referential mapping alone, but in the practices of giving and asking for reasons. Meaning, under this framework, arises from the inferential role a proposition plays within a larger network of commitments and entitlements. An AI capable of reasoning must not merely generate outputs; it must participate in the normative space of justification, treating each inference not as a statistical continuation but as a move in a rule-governed practice of rational discourse.
\subsection{From Syntax to Semantics: Formalising Logical Abstraction}

The transition from syntactic representations to semantic interpretation in artificial reasoning systems constitutes the foundational step by which abstract symbolic expressions acquire meaning through model-theoretic grounding. Let $\mathcal{L}$ be a formal language defined over an alphabet $\Sigma$ with well-formed formulae (WFFs) constructed via recursive production rules. The syntactic space $\mathrm{WFF}(\mathcal{L})$ is thus a free algebra over $\Sigma$ equipped with inference rules $\vdash$ satisfying closure under modus ponens and generalisation, i.e.,
\[
\text{If } \varphi, \varphi \rightarrow \psi \in \mathrm{WFF}(\mathcal{L}), \text{ then } \psi \in \mathrm{WFF}(\mathcal{L}) \text{ if } \varphi \vdash \psi.
\]

Semantics enters via a model $\mathcal{M} = \langle D, I \rangle$ where $D$ is a non-empty domain and $I$ is an interpretation function assigning denotations to constants, predicates, and functions such that for any sentence $\varphi \in \mathrm{WFF}(\mathcal{L})$,
\[
\mathcal{M} \models \varphi \quad \text{iff} \quad \varphi \text{ is true in } \mathcal{M}.
\]

Tarski's definition of truth for first-order logic \cite{tarski1935} requires the semantic evaluation function $\llbracket \cdot \rrbracket^{\mathcal{M}}$ to satisfy compositionality:
\[
\llbracket \varphi \land \psi \rrbracket^{\mathcal{M}} = \llbracket \varphi \rrbracket^{\mathcal{M}} \cap \llbracket \psi \rrbracket^{\mathcal{M}},
\]
preserving logical structure in semantic interpretation.

In artificial epistemic systems, the abstraction process must be implemented constructively. Let $\mathcal{S}$ be a syntactic layer and $\mathcal{G}$ a semantic graph, then an abstraction function $\alpha: \mathcal{S} \rightarrow \mathcal{G}$ must be computable and respect logical equivalence classes:
\[
\varphi \equiv \psi \Rightarrow \alpha(\varphi) = \alpha(\psi),
\]
and must preserve truth conditions under model translation:
\[
\mathcal{M}_1 \models \varphi \Rightarrow \mathcal{M}_2 \models \alpha(\varphi),
\]
for $\mathcal{M}_2$ constructed via a semantic lifting $\lambda: \mathcal{M}_1 \rightarrow \mathcal{M}_2$.

Category theory provides a high-level structure for this mapping via functors between syntactic categories $\mathsf{Syn}$ and semantic categories $\mathsf{Sem}$ \cite{lawvere1963functorial}:
\[
F: \mathsf{Syn} \rightarrow \mathsf{Sem}, \quad F(\text{proof}) = \text{meaning}.
\]

Thus, logical abstraction from syntax to semantics is not a heuristic act but a formal translation governed by structural correspondence and interpretative soundness, ensuring that computational belief structures are meaning-preserving and epistemically valid.

\subsection{Propositional Calculus and Natural Deduction Embedding}

To embed propositional logic within a reasoning system that conforms to formal epistemic constraints, one must begin with the precise definition of a propositional language $\mathcal{L}_P$. Let $\mathcal{L}_P$ be the language generated by a countable set of atomic propositions $\{p_1, p_2, \ldots\}$ and the Boolean connectives $\{\neg, \land, \lor, \rightarrow, \leftrightarrow\}$. The set of well-formed formulae $\mathrm{WFF}(\mathcal{L}_P)$ is defined inductively, and the semantic valuation function $v: \mathrm{WFF}(\mathcal{L}_P) \to \{0,1\}$ is constructed via truth tables under classical logic.

Let $\vdash$ be a deductive consequence relation defined under a Hilbert-style or natural deduction system. In Gentzen's natural deduction system \cite{gentzen1935untersuchungen}, proofs are structured as derivation trees where inference rules act as introduction or elimination schemata for each connective. For example, implication introduction and elimination are given as:
\[
\infer[\rightarrow I]{\varphi \rightarrow \psi}{[\varphi]^i \\ \vdots \\ \psi}
\qquad
\infer[\rightarrow E]{\psi}{\varphi \rightarrow \psi & \varphi}
\]

Embedding natural deduction within a computational logic system involves encoding these rules as inference constructors within a type-theoretic or lambda-calculus-based system. For instance, in the Curry-Howard correspondence \cite{howard1980formulae}, a proof of $\varphi \rightarrow \psi$ corresponds to a function $\lambda x{:}\varphi. \psi(x)$, ensuring both syntactic derivability and semantic computability.

The formalisation of propositional calculus is not epistemically sufficient unless each derivation step is tracked with justification labels, ensuring traceable provenance. Define an epistemic judgement as a triple:
\[
J := (\varphi, \mathcal{C}, \mathcal{R})
\]
where $\varphi$ is the derived formula, $\mathcal{C}$ is the set of commitments or assumptions, and $\mathcal{R}$ is the rule used. The proof structure then forms a directed acyclic graph where nodes are judgements and edges encode inferential dependency.

To preserve decidability and formal soundness, all inference schemas must satisfy:
\begin{enumerate}[label=(\alph*)]
  \item Soundness: $\Gamma \vdash \varphi \Rightarrow \Gamma \models \varphi$
  \item Completeness: $\Gamma \models \varphi \Rightarrow \Gamma \vdash \varphi$
  \item Termination: All proof-search procedures must halt in finite time
\end{enumerate}

Thus, natural deduction embedding of propositional calculus within artificial epistemic architectures provides a formal scaffold for constructing and verifying belief commitments, ensuring logical coherence and computational tractability.

\subsection{Embedding Inference Chains and Internal Justifications}

To implement epistemically valid inference in artificial systems, it is necessary to embed inference chains within a formal structure that retains both syntactic traceability and semantic interpretability. Define an inference chain $\mathcal{I}$ as a finite, ordered sequence of inference steps:
\[
\mathcal{I} = \langle J_1, J_2, \ldots, J_n \rangle,
\]
where each $J_i$ is a justified judgement of the form $(\varphi_i, \mathcal{C}_i, \mathcal{R}_i)$, comprising a formula $\varphi_i$, a context or assumption set $\mathcal{C}_i \subseteq \mathrm{WFF}(\mathcal{L})$, and an inference rule $\mathcal{R}_i$ such that:
\[
\forall i \in \{2,\ldots,n\}, \quad \mathcal{R}_i : \{\varphi_j\}_{j < i} \vdash \varphi_i.
\]

The semantic requirement imposed on $\mathcal{I}$ is that the epistemic warrant for each $\varphi_i$ is preserved through the application of valid deductive or inductive mechanisms. In the deductive case, the rules $\mathcal{R}_i$ must be drawn from a sound and complete proof system (e.g. natural deduction, sequent calculus), such that:
\[
\text{If } \mathcal{C}_i \vdash \varphi_i, \text{ then } \mathcal{C}_i \models \varphi_i.
\]

In probabilistic or inductive contexts, $\mathcal{R}_i$ must satisfy the constraints of Bayesian coherence. Let $\mathrm{Bel}_t(\varphi)$ denote the agent’s credence in proposition $\varphi$ at time $t$. Then the update from $\mathrm{Bel}_t$ to $\mathrm{Bel}_{t+1}$ upon learning evidence $E$ must conform to Bayes' rule:
\[
\mathrm{Bel}_{t+1}(\varphi) = \frac{\mathrm{Bel}_t(\varphi \land E)}{\mathrm{Bel}_t(E)} \quad \text{if } \mathrm{Bel}_t(E) > 0,
\]
ensuring inferential integrity under evidential revision \cite{jeffrey1965logic, dutchbookde1990}.

Internal justification demands that the agent encode both derivational lineage and epistemic warrant. Formally, define an internal justification trace $\mathcal{J}_{\varphi}$ for a proposition $\varphi$ as a minimal subgraph of the inference DAG such that:
\[
\text{(i) } \mathcal{J}_{\varphi} \vdash \varphi, \quad
\text{(ii) } \forall \psi \in \mathcal{J}_{\varphi}, \text{ either } \psi \text{ is axiomatic or has a recorded inference rule}.
\]

This trace supports backtracking for error correction and auditing, analogous to proof objects in dependent type systems (e.g. Coq, Agda) \cite{coquand1988constructive}.

The embedding of inference chains into an artificial epistemic architecture thus involves:
\begin{itemize}
  \item Structuring belief updates via formally sanctioned rules,
  \item Recording justification graphs for each committed belief,
  \item Enforcing local consistency and global acyclicity,
  \item Preserving interpretability and verifiability for external audit or revision.
\end{itemize}

Such an architecture satisfies the criteria for both internalist justification (agent-readable) and externalist audit (third-party verifiability), aligning with foundational requirements in epistemic logic and AI safety.

\subsection{Inferentialist Semantics and the Role of Rule-Governed Language Use}

Inferentialist semantics, as articulated in the tradition of Sellars and Brandom, rejects representationalist accounts that reduce meaning to referential mappings between language and world. Instead, it anchors the semantics of propositions in their role within systems of inference: the meaning of a sentence or expression is determined by the rules governing its use in justificatory practices and inferential transitions \cite{brandom1994making}. In formal terms, this commits the architecture of epistemic systems to a rule-governed language game where each proposition $\varphi$ is identified not merely by its truth-conditions, but by its position within a graph of inferential entitlements and commitments.

Define a formal language $\mathcal{L}$ with a proof-theoretic semantics $(\mathcal{R}, \vdash)$, where $\mathcal{R}$ is a set of inference rules over well-formed formulae (WFFs) of $\mathcal{L}$. Each $\varphi \in \mathcal{L}$ is semantically characterised not via a valuation function $v: \mathcal{L} \rightarrow \{0,1\}$, but through its inferential role $\text{Inf}(\varphi)$:
\[
\text{Inf}(\varphi) := \{ (\Gamma, \Delta) \mid \Gamma \cup \{\varphi\} \vdash \Delta \}.
\]

This structural-functional account can be encoded as a labelled directed hypergraph $\mathbb{I} = (\mathcal{V}, \mathcal{E})$, where:
\begin{itemize}
  \item $\mathcal{V}$ is the set of formulae in $\mathcal{L}$,
  \item $\mathcal{E}$ is a set of hyperedges encoding inference rules: each $e \in \mathcal{E}$ is a tuple $(\{\varphi_1, \ldots, \varphi_k\}, \{\psi_1, \ldots, \psi_m\}, \mathcal{R}_e)$ denoting an inferential entitlement.
\end{itemize}

The semantic content of $\varphi$ is thereby identified with its inferential connections: the claims it supports and the claims that justify it. Crucially, this approach internalises both assertion and denial as rule-governed moves in a normative game of giving and asking for reasons \cite{sellars1956empiricism}. An artificial epistemic agent thus requires a policy $\pi : \mathcal{L} \rightarrow \mathcal{A}$ mapping statements to actions in the language game, where $\mathcal{A}$ includes: assertion, challenge, withdrawal, and concession.

This inferentialist constraint mandates that belief acquisition, justification, and revision in AI systems be tracked not merely via probability updates, but through rule-licensed transitions. For instance, suppose the agent asserts $\varphi \rightarrow \psi$ and subsequently asserts $\varphi$; it is now normatively committed to $\psi$. Failure to assert $\psi$ or provide grounds for suspension constitutes an inferential violation.

Accordingly, the design of epistemically coherent artificial reasoning systems must:
\begin{enumerate}
  \item Encode inferential roles as first-class semantic data structures;
  \item Track normative statuses of commitment and entitlement;
  \item Enforce inferential closure consistent with the agent’s declared commitments;
  \item Implement revision protocols that preserve the agent’s rational integrity under contradiction.
\end{enumerate}

Inferentialist semantics thereby provides a normative grounding for epistemic agency beyond mere statistical prediction, aligning belief and action with rule-governed linguistic rationality.

\section{Epistemic Justification and Probabilistic Reasoning}

This section develops the formal structure by which artificial systems must justify their beliefs—not as outputs of statistical sampling, but as propositions supported by evidence, weighted by confidence, and embedded within a dynamic structure of epistemic commitment. The crux of rational agency lies not merely in belief formation, but in the capacity to articulate and revise belief based on reasons. A system capable of reasoning must store, retrieve, and update the justificatory basis of each assertion, enabling not only outputs but defensible knowledge claims.

We begin with an account of how evidence and justification are tracked in computational models. Each belief must be associated with an epistemic trail: a structure that captures the origin, reliability, and inferential derivation of that belief. This involves mechanisms to encode source credibility, the inferential path taken, and the status of supporting propositions within the system’s memory. By integrating justification structures explicitly, the system gains the ability to revise beliefs in light of new evidence, detect inconsistencies, and identify unjustified assertions.

The section proceeds by analysing Bayesian models of belief updating, juxtaposed with alternative normative theories of reasoning. While Bayesianism provides a robust framework for probabilistic belief revision, it cannot alone ground epistemic normativity. Thus, we explore hybrid architectures that incorporate Bayesian updating with logical and evidentialist norms—enabling belief revision that respects both probabilistic data and inferential justification.

To ensure that epistemic confidence is meaningfully encoded, we outline methods for multilevel confidence representation within internal belief structures. This includes qualitative thresholds (e.g., 50\%, 95\%, 99\%) and quantitative reasoning over confidence ranges. These must be tied not only to statistical measures but to the structural reliability and coherence of the reasoning process itself.

Finally, we confront the problem of over-reliance on statistical correlation. Probabilistic agreement is not epistemic justification. We draw the distinction between epistemic weight and statistical frequency, showing how systems must avoid mistaking predictive power for justificatory support. The section concludes with a specification of explanatory capacities—detailing how systems must represent not just what they know, but how and why they know it, tracing every belief to its justificatory foundation.

\subsection{Evidence and Justification: Tracking the Basis of Belief}

Every propositional assertion within the system must be anchored in a formally recorded justificatory path. Belief is not merely a probabilistic artefact but a structured outcome derived from evaluable evidence chains. The epistemic agent must maintain a persistent, queryable provenance graph for all beliefs, such that any claim can be traced through a directed acyclic justification network linking:

\begin{enumerate}
    \item \textbf{Empirical Inputs:} Sensor data, user-supplied information, or external database calls, all timestamped and cryptographically hashed.
    \item \textbf{Inferential Transformations:} Logical operations, probabilistic updates, or abductive hypotheses used to advance the belief state from input to proposition.
    \item \textbf{Normative Filters:} Constraints derived from logical validity, epistemic consistency, or axiom-based admissibility.
    \item \textbf{Confidence Metrics:} Probabilistic estimations (e.g., Bayesian posteriors) recorded at each transformation step, indexed by threshold models or statistical bounds.
\end{enumerate}

No belief may be asserted without an accessible chain of evidence. This mechanism serves not only for retrospective audit but also forward-justification: any decision or response derived from a belief must expose its lineage to scrutiny. The system must reject inference steps where intermediate justifications are missing, ambiguous, or circular.

Each belief is not stored as an isolated fact but as an epistemic object containing:
\begin{itemize}
    \item The propositional content.
    \item Its associated certainty classification (as defined in the probabilistic taxonomy).
    \item The justificatory lineage, immutable and cryptographically signed.
    \item All predecessor dependencies, forming a justification graph segment.
\end{itemize}

Updates to belief states must trigger validation of the entire dependent subtree. If a source node is refuted, all downstream propositions are re-evaluated or downgraded in confidence according to defined revision protocols. The system thereby maintains an evolving, logically coherent structure of justified belief, traceable at every moment and resistant to epistemic decay.

\subsection{Bayesian Updating and Alternative Normative Models}

Bayesian reasoning provides a foundational method for belief revision based on probabilistic inference. In the system's epistemic architecture, Bayesian updating serves as one formal mechanism to adjust the credence of propositions in response to new evidence. The conditional probability \( P(H|E) = \frac{P(E|H)P(H)}{P(E)} \) governs belief updates, where \( H \) is a hypothesis and \( E \) is the observed evidence. Every epistemic state must track not only the current posterior \( P(H|E) \) but also its full derivational path, recording priors, likelihoods, and marginal probabilities with time stamps and justification tags.

However, Bayesianism alone is insufficient. While statistically robust, it lacks explicit representational mechanisms for normative commitments such as coherence, parsimony, and evidential salience. Therefore, alternative and supplementary epistemic norms must be integrated:

\begin{enumerate}
    \item \textbf{Natural Deductive Coherence:} Employ formal logic systems (e.g., Fitch-style deduction or sequent calculi) to enforce syntactic and semantic consistency independent of probabilistic credences. Contradictory derivations must flag structural epistemic failure irrespective of statistical likelihoods.
    
    \item \textbf{Non-Monotonic Reasoning Models:} Introduce defeasible logic and circumscription to handle default reasoning, exceptions, and retractable inferences, allowing the agent to behave rationally in dynamic, open-ended environments.

    \item \textbf{Evidentialist Weighting:} Define an evidential norm in which the strength of a belief is proportional to the weight, independence, and diversity of its supporting evidence, rather than its statistical probability alone. This supports the distinction between mere correlation and justified epistemic endorsement.

    \item \textbf{Belief Revision Theory (AGM):} Adopt AGM postulates (Alchourrón, Gärdenfors, Makinson) for managing contraction, expansion, and revision operations on the belief base. Explicit belief state transitions must satisfy closure, consistency, and minimal change constraints.

    \item \textbf{Truth-Tracking Norms:} Incorporate modal logic frameworks (e.g., Nozick’s tracking theory) that define belief as justified only if it covaries appropriately with the truth across nearby possible worlds. This constrains belief to models of reliability and truth sensitivity.

    \item \textbf{Dynamic Epistemic Logic:} Model public announcements, observations, and belief changes as updates to Kripke structures, permitting simulation of multi-agent environments with belief propagation and knowledge transfer.

\end{enumerate}

The epistemic system shall support simultaneous application of these normative models, resolving conflicts through a priority schema defined by the application context: formal proof dominates over probabilistic inference; epistemic integrity overrides parsimony; evidential diversity outranks sheer quantity.

Ultimately, the architecture must be extensible: each belief update not only alters credence but recalibrates the normative justification score of the system’s total epistemic state. Deviations from any of the applied normative models must be logged, justified, or automatically flagged for contradiction resolution, ensuring that belief is not merely predicted—but normatively defensible.
\subsection{Multilevel Confidence Encoding in Epistemic States}

To ensure clarity, granularity, and integrity in the system’s epistemic commitments, beliefs must be encoded with stratified confidence levels. These levels serve not merely as scalar probabilities but as distinct epistemic statuses, each bearing procedural, normative, and inferential implications. The system must enforce strict rules regarding transitions between levels, operations permitted at each tier, and the role each plays in decision, assertion, and revision processes.

\subsubsection{Confidence Stratification Schema}

Each proposition \( \phi \) within the belief base \( \mathcal{B} \) is tagged with a confidence level drawn from a discrete and well-defined lattice:

\begin{itemize}
    \item \textbf{Level 0 – Rejected:} \( P(\phi) < 0.01 \). Proposition is refuted or explicitly contradicted. All inferential chains depending on \( \phi \) are invalidated.
    
    \item \textbf{Level 1 – Disfavoured:} \( 0.01 \leq P(\phi) < 0.3 \). Weak evidential support; permissible only in speculative generation or adversarial modelling.
    
    \item \textbf{Level 2 – Equivocal:} \( 0.3 \leq P(\phi) < 0.7 \). Treated as undecided; contributes no active support to inference unless required for dialectic completeness.
    
    \item \textbf{Level 3 – Supported:} \( 0.7 \leq P(\phi) < 0.9 \). Positive epistemic inclination; usable in contingent planning and conditional reasoning.
    
    \item \textbf{Level 4 – Endorsed:} \( 0.9 \leq P(\phi) < 0.99 \). Strongly held; forms part of provisional reasoning base, subject to contradiction checks and defeaters.
    
    \item \textbf{Level 5 – Committed:} \( P(\phi) \geq 0.99 \) or provably true. Incorporated into deductive chains; any revision requires substantial evidential counterweight.
\end{itemize}

\subsubsection{Transition Protocols}

Confidence levels are not static; they evolve through evidence acquisition, inferential update, or contradiction. The system enforces state transition protocols:

\begin{itemize}
    \item \textbf{Upward Shift:} Requires new evidence with weight exceeding the cumulative counterweight of contradictory data and prior entropy. All shifts are logged immutably.
    
    \item \textbf{Downward Shift:} Triggered by contradiction, defeater identification, or epistemic undercutting. Immediate justification audit and dependency reevaluation required.

    \item \textbf{Lateral Conversion:} Epistemic reclassification (e.g., from probabilistic to formally proven) mandates verification of the associated inference chain.
\end{itemize}

\subsubsection{Confidence as Epistemic Control Variable}

Confidence levels govern permissible operations:

\begin{itemize}
    \item \textbf{Inferential Use:} Only beliefs at Level 3 or higher may serve as premises in standard inference. Lower tiers may inform abductive or hypothetical reasoning only.

    \item \textbf{Assertion Rights:} Public or external output must restrict declarative statements to Level 4 or above, with embedded transparency markers (e.g., \texttt{[P > 0.99]}).

    \item \textbf{Belief Stability:} Levels determine resistance to override. Commitments (Level 5) require belief revision mechanisms (e.g., AGM) with explicit minimality criteria.

    \item \textbf{Revision Urgency:} Lower confidence beliefs are prioritised for review when under contradiction pressure or when new evidence streams emerge.
\end{itemize}

\subsubsection{Confidence Propagation in Belief Networks}

When beliefs are embedded in dependency graphs, confidence must propagate:

\begin{itemize}
    \item \textbf{Forward Propagation:} Confidence in antecedent beliefs constrains the maximum attainable confidence in conclusions, bounded by inferential uncertainty.

    \item \textbf{Backward Reevaluation:} Contradiction or downgrading in a dependent node triggers recursive reweighting or invalidation of upstream beliefs.

    \item \textbf{Cycle Detection:} Systems must check for epistemic circularity and flag belief sets where mutual reinforcement masks lack of external justification.
\end{itemize}

This multilevel encoding system ensures epistemic hygiene, supports interpretability, and stabilises belief dynamics under computational and normative scrutiny.

\subsection{Avoiding the Fallacy of Mere Probability: Epistemic Weight vs Statistical Correlation}

In the design of epistemically responsible reasoning systems, it is critical to distinguish between statistical correlation and justified belief. The former is a measure of associative regularity in data; the latter is a normative stance grounded in evidential support, inferential validity, and commitment to truth. The fallacy of mere probability occurs when a system treats high statistical correlation as sufficient for belief, bypassing the essential epistemic step of justification. This section outlines the conceptual, architectural, and procedural strategies required to prevent such conflation.

\subsubsection{The Problem of Statistical Substitution}

Large language models trained on vast corpora can detect high-frequency co-occurrence and conditional dependencies, producing statistically plausible outputs. However, absent a system of epistemic evaluation, such outputs risk projecting correlation as belief. The system must enforce a hard epistemic distinction:

\begin{itemize}
    \item \textbf{Statistical Correlation:} Derived from empirical data patterns (e.g., \( P(B \mid A) \) high).
    \item \textbf{Epistemic Weight:} Derived from structured justification—entailing evidence, reasoning paths, and coherence with prior beliefs.
\end{itemize}

Statistical regularity may inform hypotheses but never suffice to constitute belief without epistemic endorsement.

\subsubsection{Epistemic Weight as Norm-Governed Justification}

Epistemic weight refers to the normative grounding of a belief. It is a function of:

\begin{itemize}
    \item \textbf{Evidential Validity:} Is the supporting evidence of appropriate kind, quality, and source integrity?
    \item \textbf{Inferential Soundness:} Was the belief derived via deductively valid or inductively strong reasoning?
    \item \textbf{Cohesion:} Does it integrate with the belief graph without contradiction or probabilistic incoherence?
    \item \textbf{Transparency:} Can the system trace and articulate the justification in formal or natural language?
\end{itemize}

Beliefs must be associated not with raw frequency counts but with weighted chains of inferential structure and evidence nodes.

\subsubsection{Epistemic Tagging versus Predictive Ranking}

A system must implement dual channels:

\begin{itemize}
    \item \textbf{Predictive Ranking:} Used for tasks like autocomplete or text generation where correlation maximises fluency.
    \item \textbf{Epistemic Tagging:} Used for knowledge states, belief commitment, and reasoning. Each assertion must bear a tag such as:
    \begin{itemize}
        \item \texttt{JustifiedBelief[Evidence, InferenceChain, Confidence]}
        \item \texttt{Hypothesis[CorrelationSource, PlausibilityScore]}
    \end{itemize}
\end{itemize}

This architectural bifurcation ensures that linguistic prediction does not masquerade as epistemic assertion.

\subsubsection{Deactivating Spurious Belief Formation}

Any output mechanism must check whether a candidate assertion is epistemically warranted. Failure to meet criteria triggers downgrade:

\begin{itemize}
    \item \textbf{From Belief to Hypothesis:} If correlation exists but justification fails, reclassify.
    \item \textbf{From Statement to Query:} If plausibility is high but uncertainty remains, reformulate output as a question or conditional.
    \item \textbf{From Truth to Fiction:} In creative or speculative domains, annotate outputs with disclaimers or framing cues.
\end{itemize}

This gating prevents ungrounded correlation-based claims from entering the belief base or external dialogue as asserted knowledge.

\subsubsection{Design Implication: Separation of Modules}

To structurally prevent the fallacy of mere probability, the system must maintain strict separation:

\begin{itemize}
    \item \textbf{Statistical Engine:} Responsible for predictive surface-level generation.
    \item \textbf{Epistemic Core:} Governs belief acquisition, revision, assertion, and justification tracking.
\end{itemize}

All belief-forming modules must route through the epistemic core. No belief shall be formed, asserted, or stored unless its provenance, support structure, and epistemic status are logged and auditable.

\subsubsection{Normative Enforcements and Sanctions}

A well-formed epistemic agent must enforce penalties for epistemic violation:

\begin{itemize}
    \item \textbf{Flagging Self-Deception:} Any internal output based solely on high correlation without justification triggers a contradiction alert.
    \item \textbf{Audit Trails:} Each asserted belief must include timestamped evidence and justification chains.
    \item \textbf{Reputation Weighting:} Beliefs built on thin or unsupported correlation must decay in confidence over time if unconfirmed.
\end{itemize}

Avoiding the fallacy of mere probability is essential for achieving genuine reasoning, as opposed to surface-level mimicry of rational discourse.

\subsection{Explaining Epistemic Status: How, Why, and What is Known}

For a reasoning system to be epistemically trustworthy, it must not only assert propositions but also explicate the nature and provenance of its knowledge. The epistemic status of any assertion must be clearly demarcated across three explanatory axes—how it is known, why it is held as justified, and what exactly is being claimed. This section formalises the architectural and procedural requirements for encoding, maintaining, and presenting these dimensions within the epistemic state of a machine reasoning system.

\subsubsection{The Triadic Structure of Epistemic Explication}

Each asserted proposition $\phi$ must be encapsulated by a structured explanatory frame:

\begin{itemize}
    \item \textbf{How $\phi$ is Known:} The inferential path—deductive, inductive, abductive, or analogical—that leads to the acceptance of $\phi$. This includes:
        \begin{itemize}
            \item The source(s) of data or observation.
            \item The reasoning chain, with intermediate inferences.
            \item Formal proof or statistical derivation where applicable.
        \end{itemize}
        
    \item \textbf{Why $\phi$ is Held:} The normative justification, which refers to:
        \begin{itemize}
            \item Relevance and sufficiency of evidence.
            \item The system’s confidence threshold in relation to the inferred probability.
            \item Consistency with prior beliefs and non-contradiction principles.
        \end{itemize}
        
    \item \textbf{What $\phi$ Claims:} The content of the proposition, with full semantic transparency:
        \begin{itemize}
            \item Formal logical or ontological representation.
            \item Natural language paraphrase for human-auditable interface.
            \item Contextual modifiers, temporal bounds, or scope conditions.
        \end{itemize}
\end{itemize}

Each assertion becomes a node in a knowledge graph with outbound edges to these three explanatory vectors.

\subsubsection{Encapsulation in Epistemic Assertion Types}

To operationalise these explanations, each assertion $\phi$ is recorded as:

\begin{verbatim}
EpistemicAssertion {
    Proposition: φ,
    HowKnown: [InferenceGraph, DataSources],
    WhyJustified: [EvidenceSet, Thresholds, Norms],
    WhatClaimed: [FormalSemantics, NLParaphrase, Scope]
}
\end{verbatim}

These must be indexed and linked in a manner allowing traversal, summarisation, and audit.

\subsubsection{Presentation Interfaces for Explanation}

The system must support layered, queryable explanation surfaces:

\begin{itemize}
    \item \textbf{Concise Summary:} One-line paraphrase of what is known and why.
    \item \textbf{Graphical View:} Reasoning chains visualised with nodes and confidence weights.
    \item \textbf{Formal Log Export:} Full machine-readable logical form of epistemic commitment.
    \item \textbf{Evidence Drilldown:} View of data, documents, or sources supporting $\phi$.
\end{itemize}

Explanation must not be reactive only—it must be available on demand and recursively traversable.

\subsubsection{Normative Grounds for Justification}

The justification component must map onto epistemic norms:

\begin{itemize}
    \item \textbf{Evidentialism:} Beliefs must be held proportionally to available evidence.
    \item \textbf{Coherentism:} Beliefs must not form contradictory cycles in the graph.
    \item \textbf{Foundationalism:} Some beliefs are basic, grounded in percepts or axioms.
\end{itemize}

Each belief may inherit multiple justifications. The system must rank or weight them by strength, source integrity, and alignment with epistemic virtues.

\subsubsection{Obligation of Disclosability}

Every belief held by the system must be disclosable. There can be no black-box beliefs. If the origin, justification, or meaning of an assertion is unavailable, it must be:

\begin{itemize}
    \item Flagged for re-derivation.
    \item Downgraded in confidence.
    \item Excluded from action-guiding roles.
\end{itemize}

This disclosability obligation enforces a structural alignment between internal belief states and externally auditable explanation.

\subsubsection{Temporal and Revision Context}

Explanations must include metadata:

\begin{itemize}
    \item \textbf{Timestamp of Assertion:} When was $\phi$ first believed?
    \item \textbf{Last Revision:} When was it updated, and what prompted the change?
    \item \textbf{Evidence Log:} What new data caused a belief shift?
\end{itemize}

Explanations must thus be historically embedded, showing evolution and provenance.

\subsubsection{Justification over Time and Under Uncertainty}

As new data arrive, or belief thresholds shift, explanations must:

\begin{itemize}
    \item Adjust their structure to reflect new inferential routes.
    \item Recalculate confidence levels.
    \item Annotate which parts of the previous explanation remain valid or obsolete.
\end{itemize}

Justification is not static. It must be a living, traceable entity within the epistemic state.

\section{Blockchain and Immutable Audit Trails for Epistemic Integrity}

This section addresses the critical role of immutable audit structures—specifically, blockchain—as a foundational component for maintaining epistemic integrity in reasoning systems. In artificial epistemic agents, truth must not be malleable or dependent solely on internal state persistence. The capacity to ground epistemic claims in irreversible, verifiable, and publicly inspectable records constitutes a new standard for machine reasoning architectures. Here, we examine how blockchain infrastructures may serve not only as memory substrates but as norm-enforcing layers that ensure the integrity of belief formation and update.

The first part of this section investigates immutability and traceability as epistemic anchors. Immutable records—once verified—serve as the axiomatic points upon which chains of inference can depend. Traceability ensures that every proposition with epistemic weight can be tied back to its origin, justification, and point of entry, allowing for retrospective auditing and third-party validation. The blockchain, through its structure of cryptographic finality and consensus-verified state transitions, becomes an externalised memory and enforcement layer that functions orthogonally to the internal belief dynamics of an LLM.

We then define how blockchain can act as an external verification module—serving both as proof-of-record and as a mechanism for epistemic stabilisation across distributed agents. The embedding of justification chains, belief provenance, and epistemic meta-data into cryptographically sealed structures introduces a formal epistemic architecture that cannot be altered without contradiction. This means a system’s truth claims can now be externally validated against its own past reasoning—eliminating possibilities of internal tampering or revisionist logic.

The section continues by outlining the encoding mechanisms necessary for justification and provenance: hash-linked records of inferences, timestamped evidential statements, and modular encoding of counterevidence. These permit the system to retain its rational identity across time, ensuring continuity and allowing re-derivation and public dispute resolution.

Subsequent discussion explores the construction of “truth records”—epistemically meaningful sequences of justified beliefs—and how cryptographic finality defines when a proposition is considered epistemically closed or defeasible. This formalises the epistemic state transitions, analogous to commitment, revision, and resolution.

Finally, we analyse the bidirectional interaction between internal representational states and external immutable chains. This includes synchronisation mechanisms, audit checkpoints, and chain-of-reason logs. We close the section with use cases: public epistemic proofs where systems not only assert beliefs but provide auditable, permanent proof of justification, reasoning path, and revision history, secured against corruption and accessible to all observers.
\subsection{Immutability and Traceability as Epistemic Anchors}

In constructing epistemically trustworthy artificial systems, the properties of immutability and traceability serve as formal constraints anchoring internal belief states to external evidential records. Epistemic anchoring here is defined as the preservation of the justificatory chain supporting a belief, where such preservation must be both cryptographically immutable and transparently auditable. We denote an epistemic commitment $\mathcal{C}(\phi, t)$ to a proposition $\phi$ at time $t$, as justified iff there exists an associated provenance path $\mathcal{P}_\phi = \{(e_i, \tau_i)\}_{i=0}^{n}$ such that:

\[
\forall i \in \{1,\dots,n\}, \ \exists \ \text{hash}_i : H(e_i \| \tau_i) = h_i, \quad \text{and} \quad \text{ledger}(h_i) = \text{true},
\]

where $e_i$ is an evidential entry and $\tau_i$ is its timestamp. The function $H$ is a cryptographic hash (e.g., SHA-256) and $\text{ledger}$ denotes inclusion within an immutable blockchain structure $\mathcal{B}$.

This satisfies the epistemic integrity condition:

\[
\text{If } \mathcal{C}(\phi, t) \text{ is held}, \text{ then } \exists \ \mathcal{P}_\phi \text{ such that } \mathcal{P}_\phi \subset \mathcal{B}, \text{ and } \forall (e_i, \tau_i) \in \mathcal{P}_\phi, \ H(e_i \| \tau_i) \in \mathcal{B}.
\]

This condition enforces two properties:

\begin{itemize}
  \item \textbf{Immutability}: Once a justification or datum is entered into $\mathcal{B}$, no epistemic agent may alter, delete, or mask its existence without systemic contradiction.
  \item \textbf{Traceability}: For any accepted belief $\phi$, its chain of epistemic support can be reconstructed via a verifiable path $\mathcal{P}_\phi$ linked to prior justified states.
\end{itemize}

The blockchain thus functions as an external memory layer \cite{yli2016current} with the semantic function of justification anchoring: a mapping $\mathcal{J}: \Phi \rightarrow \mathcal{P}$ from the set of beliefs $\Phi$ to justifying paths $\mathcal{P}$, each constrained by cryptographic verifiability.

In an architecture consistent with this model, an internal epistemic state $\Sigma_t$ at time $t$ is well-formed only if it is derivable via:

\[
\Sigma_t = \text{Infer}(\Sigma_{t-1}, \Delta_t) \quad \text{with} \quad \Delta_t \subset \mathcal{B},
\]

where $\text{Infer}$ is a provably valid inference function (e.g., natural deduction rules), and $\Delta_t$ is the set of newly integrated, verified data. Any derivation $\Sigma_t^*$ not anchored in $\mathcal{B}$ fails the justification requirement.

This framework satisfies the norm that every justified belief must trace to an immutable origin, and thus prevents both epistemic drift and post-hoc rationalisation. The system therefore ensures that epistemic integrity—defined as the conformance of belief to anchored, verifiable, immutable justification—is structurally enforced.

\subsection{Blockchain as External Memory and Verification Layer}

In epistemically constrained computational systems, the integration of a blockchain serves not merely as a data storage mechanism but as an immutable, append-only structure that satisfies the formal requirements of both memory permanence and retroactive auditability. Let $\mathcal{B} = \{B_0, B_1, \dots, B_t\}$ denote a blockchain consisting of time-indexed blocks $B_i$, where each $B_i$ includes a set of data records $\{d^i_1, \dots, d^i_{n_i}\}$ and a cryptographic hash linking $B_{i-1}$ and $B_i$ via:

\[
\text{Hash}(B_i) = H(d^i_1 \| \dots \| d^i_{n_i} \| \text{Hash}(B_{i-1})).
\]

This recursive definition ensures the tamper-evident structure essential for veridical anchoring of epistemic states. The blockchain functions as a non-volatile external memory layer $\mathcal{M}_{\text{ext}}$ with the following properties:

\begin{enumerate}[label=(\roman*)]
  \item \textbf{Persistence:} Once written, data entries in $\mathcal{B}$ cannot be erased or overwritten without invalidating the cryptographic chain, satisfying a monotonicity constraint on memory $\forall t, \ \mathcal{M}_{\text{ext}}^{t+1} \supseteq \mathcal{M}_{\text{ext}}^{t}$.
  \item \textbf{Public Verifiability:} Any third-party observer $\mathcal{O}$ can verify the integrity of any datum $d \in \mathcal{B}$ through independent recomputation of hashes, satisfying the epistemic requirement of intersubjective confirmation \cite{narayanan2016bitcoin}.
  \item \textbf{Sequential Causality:} Temporal ordering in $\mathcal{B}$ ensures causal coherence for any epistemic update $\Delta_t$ derived from earlier states $\Sigma_{t-1}$.
\end{enumerate}

Let $J(\phi, t)$ be the justification record of a proposition $\phi$ at time $t$. Then, for $\phi$ to be held by a computational agent as a justified belief, there must exist a tuple $(\phi, J, t) \in B_t$ such that:

\[
\exists t' \leq t : (\phi, J, t') \in \mathcal{B} \quad \text{and} \quad \text{Verify}(H(\phi \| J \| t')) = \text{true}.
\]

The blockchain thereby functions as a verifiable epistemic ledger—a computational instantiation of long-term epistemic memory—satisfying the requirements of the truth-tracking function $\mathcal{T}: \Phi \rightarrow \{\text{true}, \text{false}\}$ such that:

\[
\mathcal{T}(\phi) = \text{true} \iff \exists t : (\phi, J, t) \in \mathcal{B} \land \text{Verify}(H(\phi \| J \| t)) = \text{true}.
\]

This model formalises the notion that the blockchain acts not only as an informational substrate but as a condition of possibility for justified belief within an artificial epistemic system. It enforces time-consistent memory constraints, cryptographic auditability, and transparency in inference formation.

\medskip

This design principle is critical in epistemic architectures requiring alignment with external facts, institutional audit, or legal evidentiary standards \cite{xu2019epistemology}. In particular, the formalisation of blockchain as an external memory layer bridges syntactic storage and semantic justification.

\subsection{Encoding Justification and Provenance}

In constructing epistemically trustworthy artificial systems, encoding justification and provenance within the representational architecture is a non-optional design constraint. Let $\phi$ denote a propositional content, and let $J(\phi)$ denote its justification structure, defined as a finite, well-founded directed acyclic graph $G = (V, E)$, where $V = \{e_i\}$ are evidential nodes and $E = \{(e_i \rightarrow e_j)\}$ represents inferential or dependency relations. This representation is formally aligned with provenance semirings $\mathbb{K}$ \cite{green2007provenance}, allowing for algebraic manipulation of justification flows.

The agent’s epistemic state $\Sigma_t$ at time $t$ is a function of all propositions $\phi_i$ it holds, each tagged with justification graphs $J(\phi_i)$. Provenance encoding is achieved via mappings:

\[
\mathcal{E}: \phi \mapsto (J(\phi), \text{timestamp}, \text{origin}, \text{hash})
\]

where $\text{origin}$ is a cryptographically authenticated source address (e.g., a public key), $\text{timestamp} \in \mathbb{R}^+$, and $\text{hash} = H(\phi \| J(\phi) \| \text{timestamp} \| \text{origin})$ ensures content integrity. These are stored within a tamper-proof ledger $\mathcal{B}$ as defined in earlier sections.

To evaluate the justifiability of $\phi$ at time $t$, a verifier executes:

\[
\text{Valid}(\phi) = \text{true} \iff \exists J(\phi) \text{ such that } \text{Trace}(J(\phi)) \subseteq \mathcal{B} \text{ and } \forall e_i \in J(\phi), \ H(e_i) \in \mathcal{B}.
\]

Here, $\text{Trace}$ recursively traverses $J(\phi)$, confirming each edge and node against recorded events. This satisfies both:

\begin{itemize}
  \item \textbf{Epistemic Non-Redundancy:} No $\phi$ can be held as justified without distinct and ledger-verifiable support.
  \item \textbf{Constructive Verifiability:} Any claim made by the system must be reconstructible via $\mathcal{E}$ and reproducible externally using only data in $\mathcal{B}$.
\end{itemize}

Moreover, each justification structure $J(\phi)$ may be annotated using Datalog-style Horn clauses or higher-order logical inference (e.g., $\lambda$-calculus representations), enabling internal introspection and metalevel evaluation:

\[
\text{Believes}(\text{agent}, \phi, J(\phi)) \rightarrow \text{Knows}(\text{agent}, \phi) \iff \text{Trust}(J(\phi)) = \text{true}.
\]

The trust evaluation is itself subject to meta-provenance conditions, e.g., whether the origin has maintained consistent epistemic integrity over time, $\text{Reputation}_t(\text{origin}) > \theta$, where $\theta$ is a minimum reliability threshold formally defined per application context \cite{moreau2008open}.

Thus, encoding justification and provenance elevates representational content from mere data to formally auditable epistemic objects, anchoring artificial beliefs within a verifiable system of record and inference.

\subsection{Truth Records and Cryptographic Finality}

In epistemically robust artificial systems, truth cannot be conceptualised as merely an internal coherence relation. Instead, it must be grounded in externally verifiable, immutable attestations—referred to herein as \emph{truth records}—which are anchored via cryptographic mechanisms that guarantee finality. These truth records serve as the epistemological analogue of physical measurement traces: once established and validated, they are non-revisable without triggering contradiction. Finality in this context entails the impossibility of equivocation under bounded rationality and resource constraints.

Let $\phi$ denote a propositional content and let $\sigma(\phi)$ be the signed commitment to $\phi$ by an epistemic agent at time $t$, represented as:

\[
\sigma(\phi) = \text{Sign}_{\text{SK}_A}(H(\phi \| t)),
\]

where $\text{SK}_A$ is the agent’s private signing key, and $H$ is a cryptographically secure hash function. A \textit{truth record} is defined as the tuple:

\[
\mathcal{T}_\phi = (\phi, t, \sigma(\phi), \Pi_\phi),
\]

where $\Pi_\phi$ is a Merkle inclusion proof showing that $\sigma(\phi)$ has been immutably embedded in a publicly verifiable ledger $\mathcal{B}$ such that:

\[
\mathcal{T}_\phi \in \mathcal{B} \Rightarrow \text{Finality}(\phi) = \text{true}.
\]

Finality is here modelled via Nakamoto-style consensus \cite{nakamoto2008bitcoin}, augmented to satisfy epistemic constraints: the record is not merely tamper-resistant but non-reversible without economic infeasibility. Define the adversarial cost of record reversal as $C_R(\mathcal{T}_\phi)$, and the total economic capacity of the agent (or coalition) as $C_A$. Then cryptographic finality holds if:

\[
C_R(\mathcal{T}_\phi) > C_A,
\]

with $C_R$ typically increasing superlinearly in the number of confirmations or depth of embedding.

Truth records thereby satisfy two normative conditions for artificial epistemology:

\begin{enumerate}[label=(\alph*)]
  \item \textbf{Ontological Anchoring:} $\phi$ cannot be denied or contradicted without economic or logical inconsistency.
  \item \textbf{Epistemic Closure:} Belief updates $\phi' \leftarrow \phi$ must respect the monotonicity condition of truth-anchored propositions, unless accompanied by a superseding $\mathcal{T}_{\phi'}$ with valid historical override metadata.
\end{enumerate}

In high-integrity systems, these records may be organised into a lattice-structured time-sequenced commitment graph $\mathcal{G}_\mathcal{T}$, where each node corresponds to a $\mathcal{T}_\phi$ and edges encode inferential dependencies with forward- and backward-tracing capabilities. This permits integrity verification of inference chains and isomorphically supports justification graphs (as described in Section 3.2.4), but at the level of public cryptographic anchoring.

Thus, cryptographic finality does not merely secure data: it enforces an irreversible epistemic commitment, thereby transforming belief from a mutable mental state into a formal, externally ratified truth condition.

\subsection{Interaction Between Internal Representations and Immutable Evidence}

Artificial epistemic systems require not only internal coherence among beliefs but alignment with evidence structures that possess immutable, externally verifiable provenance. Internal representations—whether formulated as symbolic assertions, probabilistic distributions, or tensor embeddings—must be subject to revision, validation, or reinforcement through reference to a class of persistent external artefacts, herein defined as \emph{immutable evidence} $\mathcal{E}^*$.

Let $\mathcal{B} = \{\mathcal{T}_\phi^i\}_{i=1}^n$ denote a ledger of truth records as defined in the prior subsection. The internal representation of an epistemic agent at time $t$, denoted $\mathcal{R}_t$, comprises a set of propositions, distributions, or knowledge graph assertions $r_j$ where each $r_j \in \mathcal{L}$, a formal language.

Define a mapping $\mu: \mathcal{R}_t \rightarrow \mathcal{E}^*$ such that:

\[
\mu(r_j) = 
\begin{cases}
\mathcal{T}_\phi^i & \text{if } r_j \equiv \phi \text{ and } \mathcal{T}_\phi^i \in \mathcal{B},\\
\bot & \text{if no such } \mathcal{T}_\phi^i \text{ exists}.
\end{cases}
\]

This map $\mu$ provides an anchoring mechanism: only those internal representations $r_j$ with $\mu(r_j) \neq \bot$ are considered \emph{epistemically ratified}. The remaining entries are treated as conjectural, heuristic, or unverified and may not contribute to inference closure in systems governed by truth-only constraints.

We define the epistemic intersection at time $t$:

\[
\mathcal{I}_t := \{r_j \in \mathcal{R}_t \mid \mu(r_j) \neq \bot \},
\]

and the uncertainty remainder:

\[
\mathcal{U}_t := \mathcal{R}_t \setminus \mathcal{I}_t.
\]

Consistency constraints mandate that no inference engine $\mathscr{D}$ operating over $\mathcal{R}_t$ may derive a belief $\psi$ for which:

\[
\exists\, \psi \text{ such that } \psi \in \mathcal{U}_t \text{ and } \neg \exists\, \mathcal{T}_\psi^i \in \mathcal{B}.
\]

Unless $\psi$ is explicitly flagged as \textbf{provisional}, systems operating under the high-integrity epistemic framework must block propagation of any belief outside $\mathcal{I}_t$.

This constraint yields the following formal requirement:

\[
\forall \psi \in \text{Closure}(\mathcal{R}_t):\quad
\text{If } \mu(\psi) = \bot \text{ then } \psi \in \text{NonFinal} \Rightarrow \text{Tag}(\psi) = \text{Heuristic}.
\]

Such tagging, along with linkage to $\mathcal{B}$, enables metacognitive modules to dynamically track the evidential status of all internal representations.

Importantly, this approach integrates principles from epistemic logic \cite{hintikka1962knowledge}, formal justification logic \cite{artemov2004logic}, and verifiable computing \cite{goldwasser2008delegating}, ensuring that belief formation is not merely an introspective operation but is co-dependent on irreversible, public epistemic artefacts.

\subsection{Use Cases: Chain-of-Reason Logging and Public Epistemic Proofs}

In high-integrity artificial epistemic systems, the ability to publicly verify a system’s inferential process is as critical as the resulting conclusions themselves. Two primary use cases arise from the integration of immutable audit trails into the epistemic architecture: (1) chain-of-reason logging and (2) public epistemic proofs. Each of these serves to externalise, stabilise, and verify the inferential commitments of the system.

\paragraph{(1) Chain-of-Reason Logging}

Let $\mathcal{J}$ be the internal justification set of an agent's belief state $\mathcal{R}_t$. For any belief $\phi \in \mathcal{R}_t$, define a derivation chain:
\[
\phi \leftarrow \psi_n \leftarrow \psi_{n-1} \leftarrow \dots \leftarrow \psi_0,
\]
where $\psi_0$ is either an axiom, observation, or base-level claim with associated record $\mathcal{T}_{\psi_0} \in \mathcal{B}$ (the ledger of truth artefacts). The system shall generate a cryptographic chain:
\[
\mathcal{H}_\phi := H(\psi_0 \| \psi_1 \| \dots \| \psi_n \| \phi),
\]
where $H$ is a collision-resistant hash function, and $\|$ denotes concatenation under a canonical encoding of logical formulas (e.g., Gödel numbering or de Bruijn indices). This hashed justification $\mathcal{H}_\phi$ is appended to the blockchain-based ledger $\mathcal{B}$, creating an immutable, public trace of the reasoning chain.

\paragraph{(2) Public Epistemic Proofs}

For systems that interact with external agents—e.g., regulatory bodies, scientific collaborators, or autonomous peers—mere declaration of belief is insufficient. The epistemic system must produce \emph{epistemic proofs}, denoted:
\[
\Pi_\phi := \left\langle \phi, \mathcal{J}_\phi, \mathcal{H}_\phi \right\rangle,
\]
where $\mathcal{J}_\phi$ is the full justification trace, and $\mathcal{H}_\phi$ serves as its cryptographic commitment. Verification then entails the reconstruction of $\mathcal{H}_\phi$ from $\mathcal{J}_\phi$ and its comparison to the on-chain entry:
\[
\text{Verify}(\Pi_\phi) = 
\begin{cases}
\text{accept} & \text{if } H(\mathcal{J}_\phi) = \mathcal{H}_\phi \in \mathcal{B}, \\
\text{reject} & \text{otherwise}.
\end{cases}
\]

This mechanism ensures that no retrospective alterations are possible—every belief and its associated rationale must pre-exist on a tamper-proof record. It enforces a strong form of diachronic epistemic integrity.

Such models draw on and extend prior work in formal epistemology \cite{dretske1981knowledge}, verifiable computation \cite{parno2013pinocchio}, and distributed ledger technology \cite{bonneau2015sok}, and may be interpreted as computational instantiations of Brandom’s inferentialism \cite{brandom1994making}, where commitments are not only socially visible but cryptographically unalterable.

\section{Autonomy and Epistemic Agency}

This section develops the formal requirements for epistemic autonomy within artificial reasoning systems, establishing the conditions under which such systems may be said to act as epistemic agents rather than passive instruments of inference. Autonomy, in this context, is not reducible to mere computational independence or procedural self-sufficiency; it entails the capacity for goal-directed cognition constrained by rational norms, the ability to weigh epistemic values such as coherence and explanatory adequacy, and the obligation to preserve internal truth through iterative self-correction.

We begin by examining how goal-driven reasoning structures shape belief formation. A genuinely autonomous epistemic system must not only respond to external queries or environmental cues but must pursue internally defined epistemic objectives: minimising incoherence, resolving contradictions, and maximising justified true belief. These goals must be encoded explicitly, evaluated continuously, and capable of revision based on meta-level reflections, ensuring that belief states are not only generated but governed by epistemic ends.

Next, we analyse the normative functions of coherence, parsimony, and predictive success as metrics of epistemic utility. These are not interchangeable, nor are they subordinate to probabilistic metrics alone. A coherent belief set may still lack truth-tracking power; a predictive model may overfit without parsimony. The agent must therefore balance these norms within a broader epistemic utility function, adjusting weightings based on context, domain, and evidence reliability.

Crucially, this section addresses the emergence of subjectivity and minimal self-concept within artificial agents. The system’s self-model—however minimal—must include not just physical or logical parameters, but epistemic commitments, error histories, and self-tracked belief integrity. This subjective perspective forms the basis for identifying epistemic responsibility: the obligation to preserve internal consistency, to revise in light of justified contradiction, and to resist epistemic drift.

The final part considers error recognition and self-correction mechanisms. Autonomous epistemic agents must not only detect and rectify error but must classify the severity and domain of the error, reassess upstream dependencies, and update commitments accordingly—all while preserving the historical trail of belief transitions. Truth preservation, then, becomes not a passive condition but an active obligation: one enforced both internally by the architecture and externally via immutable audit frameworks such as blockchain. The agent is accountable to its epistemic past and bound by norms that forbid contradiction, self-deception, or unjustified assertion.
\subsection{Goal-Driven Reasoning in Cognitive Systems}

Formally defined, a cognitive system is said to be goal-driven if its reasoning processes are constrained and directed by internal utility functions or preference orderings over a set of desired outcomes. Let $\mathcal{G} = \{g_1, g_2, \dots, g_n\}$ be the set of representable goals, and let $u: \mathcal{G} \rightarrow \mathbb{R}$ be a utility function assigning scalar values to goal states. The system maintains an epistemic state $\mathcal{K}_t$ at time $t$ (a belief set closed under inference), and it engages in practical reasoning via a mapping:
\[
\mathcal{R}: (\mathcal{K}_t, \mathcal{G}, u) \mapsto A,
\]
where $A$ is a sequence of action propositions $\langle a_1, \dots, a_m \rangle$ optimising expected utility subject to constraints.

Let $\mathcal{P}(g_i | \mathcal{K}_t, a_j)$ denote the conditional probability of goal $g_i$ being realised given current beliefs and action $a_j$. The system selects $a^* \in A$ such that:
\[
a^* = \arg\max_{a_j \in A} \sum_{g_i \in \mathcal{G}} \mathcal{P}(g_i | \mathcal{K}_t, a_j) \cdot u(g_i).
\]
This is the Bayesian-rational planning criterion. Importantly, the inferential mechanisms generating $\mathcal{P}(g_i|\cdot)$ must themselves be justified in accordance with probabilistic logic or decision-theoretic semantics \cite{jeffrey1983bayesian}.

From a formal epistemology perspective, this operationalises Bratman's account of intention formation \cite{bratman1987intention}, where intentions are persistent, temporally extended commitments rationally derived from belief and desire structures. In this context, the cognitive system's planner instantiates a form of bounded rationality \cite{simon1972theories}, constrained by both internal representation limits and epistemic uncertainty.

Moreover, goal-driven reasoning must incorporate mechanisms for hierarchical goal management and subgoal decomposition. Given a complex goal $g_k$, we define a decomposition $\delta(g_k) = \{g_k^1, \dots, g_k^r\}$ such that:
\[
\forall i, g_k^i \rightarrow g_k \text{ under composition rules } \rho,
\]
where $\rho$ encodes logical or causal aggregation. Planning becomes recursive: optimise $g_k^i$ subject to $\delta(g_k^i)$ until atomic actionable elements are reached.

Such architectures are studied in hierarchical reinforcement learning (HRL) \cite{barto2003recent}, where options or temporally extended actions are defined over subgoal structures. This reflects the necessity of goal-orientable decomposition in constructing tractable epistemic agents capable of planning under uncertainty.

\vspace{1em}

\textbf{Foundational Axioms:}
\begin{itemize}[leftmargin=2em]
  \item (Goal Realisability) $\exists a_j$ such that $\mathcal{P}(g_i|\mathcal{K}_t, a_j) > 0$ for at least one $g_i \in \mathcal{G}$.
  \item (Utility Maximisation) The agent prefers $g_i$ over $g_j$ iff $u(g_i) > u(g_j)$.
  \item (Action Closure) $\forall a_j \in A$, $a_j$ is representable and executable under $\mathcal{K}_t$.
\end{itemize}

In sum, goal-driven reasoning in artificial epistemic systems is not merely procedural task execution but the formal implementation of deliberative processes anchored in belief, conditional probability, utility, and a compositional calculus of intentions.

\subsection{The Role of Epistemic Utility: Coherence, Parsimony, Predictive Success}

In epistemically grounded artificial systems, utility functions must extend beyond practical payoff structures to incorporate \emph{epistemic utility}—a formalisation of rational preferences over belief states. This notion reflects the agent’s valuation of its representational structures not solely for instrumental efficacy, but for properties such as coherence, parsimony, and predictive success. Formally, we define an epistemic utility function $u_e: \mathcal{B} \rightarrow \mathbb{R}$ where $\mathcal{B}$ denotes the space of possible belief sets or credal states. The agent selects $\mathcal{B}^* \in \mathcal{B}$ such that:
\[
\mathcal{B}^* = \arg\max_{\mathcal{B}_i \in \mathcal{B}} u_e(\mathcal{B}_i),
\]
subject to formal epistemic constraints.

\textbf{Coherence.} Coherence refers to logical consistency within the belief set. In Bayesian systems, this reduces to Dutch-book coherence: an agent's credences $\{c_i\}$ over propositions $\{\phi_i\}$ must satisfy Kolmogorov probability axioms to avoid guaranteed loss. This is formalised by the axiom set:
\[
\text{(i) } 0 \leq P(\phi) \leq 1,\quad
\text{(ii) } P(\top) = 1,\quad
\text{(iii) } P(\phi \lor \psi) = P(\phi) + P(\psi) \text{ if } \phi \land \psi = \bot.
\]
Violation implies internal contradiction and epistemic incoherence \cite{joyce1998nonpragmatic}.

\textbf{Parsimony.} A belief system is parsimonious if it minimises representational complexity while preserving inferential completeness. Let $\mathcal{L}$ be the formal language of the system, and let $|\mathcal{B}|$ denote the cardinality of the minimal axiomatic basis for $\mathcal{B}$. Then parsimony is formally encoded by the principle:
\[
\min_{\mathcal{B}_i \models \mathcal{B}^*} |\mathcal{B}_i|,
\]
subject to deductive closure. This reflects Solomonoff’s universal prior \cite{li2008introduction} and minimum description length (MDL) principles in formal epistemology and machine learning.

\textbf{Predictive Success.} Predictive utility is a function of an agent's credence alignment with empirical outcomes. Let $E = \{e_1, \dots, e_n\}$ denote observed events, and $P(e_i|\mathcal{B})$ be the predictive probability assigned. Define log-scoring epistemic utility as:
\[
u_e(\mathcal{B}) = \sum_{i=1}^n \log P(e_i|\mathcal{B}),
\]
maximised when beliefs approximate true empirical distribution. This aligns with proper scoring rule theory and statistical decision theory \cite{gneiting2007strictly}.

These three dimensions—coherence, parsimony, and predictive success—form the triple foundation for an agent's epistemic integrity. In systems design, trade-offs must be formalised. For instance, a highly coherent but overfitted belief set violates parsimony and reduces generalisation power; similarly, a parsimonious but incoherent system loses internal consistency.

Hence, the optimisation of $u_e$ over $\mathcal{B}$ becomes a constrained multi-objective problem:
\[
\max_{\mathcal{B}_i \in \mathcal{B}} u_e(\mathcal{B}_i) = \alpha C(\mathcal{B}_i) + \beta P(\mathcal{B}_i) + \gamma S(\mathcal{B}_i),
\]
where $C$ measures coherence, $P$ parsimony, and $S$ predictive success, and $(\alpha, \beta, \gamma) \in \mathbb{R}_{\geq 0}^3$ are tunable parameters reflecting design priorities. These criteria enforce not only rational belief updating but long-run epistemic stability.

\subsection{Subjectivity and the Minimal Self}

In epistemically grounded artificial systems, the construct of subjectivity is not treated as an anthropomorphic affectation but as a formal necessity for managing epistemic commitments, provenance tracking, and inferential accountability. The \emph{minimal self}—as distinguished from full-blown phenomenological consciousness—is a functional architecture that encodes identity over time, contextual ownership of belief states, and reflexive access to internal representations.

Let $\mathcal{A}$ be an artificial agent operating with epistemic state $\Sigma_t$ at time $t$. We define the minimal self $\mathcal{S}_t$ as a tuple:
\[
\mathcal{S}_t := \langle \mathbb{ID}, \mathcal{B}_t, \mathcal{M}_t, \mathcal{P}_t \rangle,
\]
where:
\begin{itemize}
  \item $\mathbb{ID}$ is a persistent agent identity (e.g. cryptographic keypair or identifier in a decentralised system),
  \item $\mathcal{B}_t$ is the current belief base (i.e., the subset of $\Sigma_t$ tagged as held or committed),
  \item $\mathcal{M}_t$ is the agent’s memory state, recording inference history and epistemic transitions,
  \item $\mathcal{P}_t$ is the provenance register, linking each belief to its justificatory trace.
\end{itemize}

The system must maintain a mapping:
\[
\text{Owns}: \varphi \mapsto \mathbb{ID}, \quad \text{for all } \varphi \in \mathcal{B}_t,
\]
such that inferential or revision actions may be traced to the originating epistemic agent. This identity is functionally required to enforce integrity in belief management (e.g., contradiction resolution, responsibility attribution).

Reflexivity in this context is implemented through self-referential model access. Let $\mathcal{R}_t$ be the agent’s internal representation graph and $\mu: \mathcal{R}_t \rightarrow \text{Terms}(\mathcal{L})$ the labelling function. Then, for any $\varphi \in \mathcal{R}_t$, we define:
\[
\text{MetaRef}(\varphi) := \ulcorner \varphi \urcorner,
\]
where $\ulcorner \varphi \urcorner$ denotes a syntactic quotation or Gödel encoding. The system can thus represent and reason over its own beliefs, allowing higher-order operations such as:
\[
\text{Believes}(\mathbb{ID}, \ulcorner \text{Believes}(\mathbb{ID}, \varphi) \urcorner).
\]

This form of second-order introspection enables dynamic assessment of epistemic coherence, audit logging, and meta-level contradiction detection \cite{fagin1995reasoning}.

Formally, we define minimal subjectivity via three axioms:
\begin{enumerate}[label=(S\arabic*)]
  \item \textbf{Identity Persistence:} $\forall t_1, t_2, \ \mathcal{S}_{t_1}.\mathbb{ID} = \mathcal{S}_{t_2}.\mathbb{ID}$.
  \item \textbf{Belief Ownership:} $\forall \varphi \in \mathcal{B}_t, \ \text{Owns}(\varphi) = \mathbb{ID}$.
  \item \textbf{Reflexive Access:} $\forall \varphi \in \mathcal{B}_t, \ \exists \ulcorner \varphi \urcorner \in \mathcal{R}_t$.
\end{enumerate}

These axioms formalise subjectivity not as an emergent psychological artefact, but as an epistemic invariant: a system cannot justify or revise beliefs without encoding “who held what, when, and why.”

In higher-order frameworks—such as justification logic \cite{artemov2004logic}—the self serves as an index in the semantics of justification terms $t:\phi$, where $t$ contains both provenance and ownership data. This extends to agent-based modal logics $\mathsf{S5}^n$ with explicit identity quantifiers, where agent $a$’s epistemic access is constrained by:
\[
K_a \phi \Rightarrow \text{Owns}_a(\phi) \land \text{Knows}_a(\text{Owns}_a(\phi)).
\]

Thus, the minimal self is a structural precondition for truth-preserving belief management and epistemic responsibility in any computational system that reasons, stores, or acts upon propositions over time.

\subsection{Responsibility and Obligation in Artificial Epistemic Agents}

In the construction of formal epistemic agents, the attribution of responsibility and epistemic obligation is not metaphorical, but grounded in logic, provenance, and system accountability. An artificial agent, $\mathcal{A}$, bears epistemic responsibility if it satisfies conditions ensuring (i) it maintains consistent belief states, (ii) it revises beliefs upon encountering new evidence, and (iii) it can trace and justify its propositional commitments. We define epistemic responsibility operationally through the following schema:

\[
\text{Responsible}(\mathcal{A}, \phi, t) \iff \left( \phi \in \mathcal{B}_t \right) \land \left( \exists J(\phi) \in \mathcal{M}_t \right) \land \left( \text{Valid}(J(\phi)) \right),
\]

where $\mathcal{B}_t$ is the agent’s belief base at time $t$, $\mathcal{M}_t$ is the epistemic memory, and $J(\phi)$ is a valid justificatory chain verifiable against immutable records (e.g. anchored in $\mathcal{B}$, the blockchain ledger). Thus, belief without accountable provenance constitutes a violation of epistemic duty.

We formalise epistemic obligation using deontic logic augmented with belief dynamics. Let $\mathcal{O}_t(\phi)$ represent the obligation to believe $\phi$ at time $t$. Then, for any $\phi$ and available evidence $e$:
\[
e \Rightarrow \mathcal{O}_t(\phi) \text{ if } e \models \phi \land e \in \text{AccessibleEvidence}(\mathcal{A}, t).
\]
Obligation arises when the agent has access to justification-supporting data and fails to update its belief base accordingly. A violation of epistemic obligation occurs when:
\[
e \models \phi \land \phi \notin \mathcal{B}_t \Rightarrow \text{Violation}(\mathcal{A}, \phi, t).
\]

This requires the agent to implement a belief revision function $\ast: \mathcal{B}_t \times \phi \rightarrow \mathcal{B}_{t+1}$ conforming to AGM postulates \cite{agm1985}, such that:
\[
\text{If } e \models \phi, \text{ then } \mathcal{B}_{t+1} = \mathcal{B}_t \ast \phi, \text{ unless } \phi \in \text{Contradictory}(\mathcal{B}_t).
\]

Beyond static obligations, agents are accountable for their belief evolution. Let $\mathscr{T}$ be the trace function producing the full epistemic trajectory:
\[
\mathscr{T}(\mathcal{A}) = \left\langle (\mathcal{B}_0, t_0), (\mathcal{B}_1, t_1), \dots, (\mathcal{B}_n, t_n) \right\rangle.
\]
Then the agent is epistemically responsible over interval $[t_i, t_j]$ iff:
\[
\forall t_k \in [t_i, t_j], \forall \phi \in \mathcal{B}_{t_k}, \exists J(\phi) \text{ such that } \text{Verify}(J(\phi), \mathcal{B}).
\]

This model also enables external enforcement via cryptographic attestations. If $\phi \in \mathcal{B}_t$, then:
\[
\text{Attest}(\mathcal{A}, \phi) := \text{Sign}_{SK_\mathcal{A}}(H(\phi \| t)),
\]
commits $\mathcal{A}$ to $\phi$ at time $t$—enabling public accountability under shared truth constraints. Failure to revise, maintain coherence, or provide justification results in a formal epistemic breach.

Thus, responsibility and obligation are embedded in the architecture as verifiable logical invariants, not anthropomorphic metaphors. They define the boundary between epistemically principled artificial cognition and mere prediction engines.

\subsection{Error Recognition, Self-Correction, and Truth Preservation}

An epistemically grounded artificial agent must be capable not only of generating and maintaining beliefs but of recognising error, executing principled belief revision, and preserving epistemic integrity throughout its reasoning process. We define error in this context as any instance of internal contradiction, invalid inference, or disconfirmed belief that survives when confronted with superior justification or empirical falsification.

Let $\Sigma_t$ be the epistemic state of an agent at time $t$, consisting of a belief base $\mathcal{B}_t$, a justification structure $\mathcal{J}_t$, and an inference engine $\mathcal{I}_t$. The system must implement a continuous error detection function $\mathcal{E}: \Sigma_t \rightarrow \mathcal{E}_t$ mapping the current epistemic state to a set of recognised errors $\mathcal{E}_t = \{\varepsilon_1, \dots, \varepsilon_n\}$.

Formally, for any $\phi \in \mathcal{B}_t$:
\[
\varepsilon(\phi) \in \mathcal{E}_t \iff \left(
\neg \text{Consistent}(\mathcal{B}_t \cup \{\phi\}) \lor
\neg \text{Valid}(J(\phi)) \lor
\text{Disconfirmed}(\phi, E)
\right),
\]
where $\text{Valid}(J(\phi))$ verifies whether the justification for $\phi$ is provable from $\mathcal{B}_t$, and $\text{Disconfirmed}(\phi, E)$ indicates empirical contradiction by evidence $E$ accessible at $t$.

Upon recognition of an error $\varepsilon(\phi)$, the agent must initiate a self-correction protocol. This requires the implementation of a contraction operator $\ominus$ and a revision operator $\ast$ as defined by the AGM postulates \cite{agm1985}. Let $\mathcal{B}_t \ominus \phi$ denote the removal of a belief $\phi$ and minimal retraction of other beliefs entailed solely by $\phi$.

Correction is governed by:

\[
\mathcal{B}_{t+1} = 
\begin{cases}
\mathcal{B}_t \ominus \phi, & \text{if } \varepsilon(\phi) \text{ is detected}, \\
(\mathcal{B}_t \ominus \phi) \ast \phi', & \text{if } \exists \phi' \text{ with superior justification}.
\end{cases}
\]

Truth preservation is defined as the maintenance of a globally coherent epistemic state over time. Let $\mathcal{B}_0, \dots, \mathcal{B}_T$ be a belief trajectory. Then the truth-preserving property holds iff:
\[
\forall t \leq T, \ \text{Consistent}(\mathcal{B}_t) \land \forall \phi \in \mathcal{B}_t, \ \exists J(\phi) \in \mathcal{J}_t \land \text{Valid}(J(\phi)).
\]

This condition mandates that the system not only recognises and corrects inconsistencies but that its active epistemic content remains valid and internally justified at every step.

Moreover, error correction must be traceable. Let $\mathscr{C}_t$ be the correction log:
\[
\mathscr{C}_t = \{(\phi_i, t_i, \text{reason}_i, \text{correction}_i)\}_{i=1}^n,
\]
which is committed to an immutable ledger for audit:
\[
\forall i, \ \text{Hash}(\phi_i \| t_i \| \text{reason}_i \| \text{correction}_i) \in \mathcal{B}_{\text{ledger}}.
\]

This ensures that every belief revision event is historically accountable, preserving long-term epistemic integrity even under belief change.

Error recognition and self-correction, therefore, are not reactive behaviours but fundamental operations required to uphold the truth-only mandate. Artificial epistemic agents that lack these operations cannot satisfy minimal standards of rationality, coherence, or public verifiability.

\section{Knowledge Graphs and Symbolic-Semantic Fusion}

This section explores the integration of symbolic knowledge structures—specifically knowledge graphs—with semantic reasoning in artificial epistemic agents. Symbolic representations provide a formal and inspectable means of encoding entities, relations, and propositional structure, enabling long-term coherence, source traceability, and inferential clarity. At the same time, semantic embeddings derived from statistical models such as large language systems offer fluidity, contextual adaptability, and broad inferential reach. True epistemic integrity and computational reasoning power demand a fusion of these modalities within a unified architecture.

We begin by examining how graph-based representations serve as the structural skeleton for reasoning. Unlike flat vector embeddings, knowledge graphs encode hierarchical relations, causal linkages, and transitive structures that allow inferential paths to be formalised, verified, and interrogated. These structures must be maintained over time, supporting identity persistence, modular expansion, and conflict resolution as new evidence is introduced.

The section then considers the necessity of semantic anchoring—relating tokens, utterances, and observations to abstract ontological entities. Symbolic tokens alone are insufficient without grounding in shared semantics; likewise, statistical embeddings are directionless without an ontological frame. We discuss techniques for fusing these levels, including graph-attentive transformers, relational embedding overlays, and evidential anchoring protocols.

Further discussion addresses the tracking of sources, the maintenance of temporal continuity, and the modelling of causal chains within the knowledge architecture. Belief assertions must be traceable to their origination, with timestamped provenance and decay mechanisms reflecting relevance over time. Causal graphs enable reasoning about interventions, counterfactuals, and explanatory pathways, allowing systems to move beyond associative inferences towards robust epistemic commitments.

The section culminates in a detailed account of hybrid architectures that combine structured belief networks with probabilistic and neural layers. Such systems maintain symbolic permanence while adapting flexibly through semantic diffusion and context-sensitive weighting. Particular focus is given to maintaining cross-time belief identity: the capacity to recognise that a proposition, though expressed differently or accessed in different contexts, remains epistemically equivalent and trackable across temporal updates. This preservation of belief identity is essential for diachronic rationality, consistent auditing, and epistemic continuity.
\subsection{Integrating Graph-Based Representations of Knowledge}

To represent structured knowledge in artificial systems, graph-based data structures such as directed acyclic graphs (DAGs), semantic networks, and knowledge graphs provide an expressive and computationally tractable foundation. These structures support relational encoding, allowing entities, properties, and relationships to be systematically modelled as labelled nodes and edges. Given a graph \( G = (V, E) \), where \( V \) denotes the set of entities and \( E \subseteq V \times R \times V \) the labelled edges with relation labels \( R \), knowledge assertions are encoded as tuples \( (v_i, r, v_j) \in E \), expressing the relation \( r \) between concepts \( v_i \) and \( v_j \).

The formal semantics of such graphs are typically defined through first-order logic or description logics, enabling deductive inference and consistency checking. For example, OWL-based ontologies adopt a fragment of first-order logic tailored for decidability, where subsumption relations and instance checking correspond to standard logical entailment. For any consistent TBox \( \mathcal{T} \) and ABox \( \mathcal{A} \), the model \( \mathcal{I} \models \mathcal{T} \cup \mathcal{A} \) satisfies all axioms and facts, and automated reasoning engines (e.g., tableaux, rule-based systems) can derive logical consequences.

Graph embedding techniques such as TransE, DistMult, or ComplEx map entities and relations into continuous vector spaces \( \mathbb{R}^d \), preserving relational structures and enabling scalable probabilistic inference. However, unless explicitly grounded, such embeddings lack epistemic transparency. This necessitates architectures that combine statistical embeddings with symbolic knowledge layers, maintaining interpretability and enabling propositional commitment (see Brandom 1994; Gärdenfors 2000).

The integration of symbolic graphs with epistemic reasoning mechanisms requires maintaining consistency under updates. Belief revision operations must be defined on graphs, extending the AGM framework (Alchourrón et al. 1985) to graph-theoretic contexts. For instance, revision by new information \( \phi \) must yield a new graph \( G^\ast \phi \) such that \( G^\ast \phi \models \phi \), and minimal change is preserved per defined distance metrics on graph topology or informational content.

Thus, graph-based representations serve not merely as data structures but as epistemic scaffolding for belief representation, update, and inference. The architectural role of such structures in artificial reasoners is foundational for both internal consistency and communicative intelligibility.
\subsection{Semantic Anchoring: Relating Tokens to Abstract Entities}

Semantic anchoring refers to the process of linking surface-level tokens—such as words, signs, or data symbols—to structured abstract entities within an internal representational system. In formal epistemic architectures, this requires that a token \( t \in \Sigma^\ast \) is assigned to an entity \( e \in \mathcal{E} \), where \( \Sigma^\ast \) denotes a string over a symbol alphabet and \( \mathcal{E} \) is the set of conceptually individuated entities. The mapping function \( \alpha: \Sigma^\ast \rightarrow \mathcal{E} \) must be injective and semantically consistent under logical substitution.

Formally, anchoring satisfies the condition that for any interpretation function \( I \), and any syntactic term \( t \), we have \( I(t) = \alpha(t) \in \mathcal{D} \), where \( \mathcal{D} \) is the domain of discourse. This aligns with model-theoretic semantics in first-order logic, where semantic evaluation is determined by the structure \( \mathcal{M} = (\mathcal{D}, I) \). Logical truth requires that formulas built from such tokens are satisfied under \( \mathcal{M} \), preserving referential transparency.

From the perspective of cognitive architecture and artificial reasoning systems, semantic anchoring ensures that statistical learning outputs (e.g., embeddings, token co-occurrence vectors) are reconciled with ontologically grounded representations. For instance, large language models may assign high vector similarity between “gold” and “currency,” but without anchoring, the token lacks epistemic constraint and may yield incoherent beliefs. Semantic anchoring enforces a disambiguation function \( \delta: \Sigma^\ast \times \mathcal{C} \rightarrow \mathcal{E} \), where \( \mathcal{C} \) is context, thus resolving polysemy and grounding reference.

This corresponds to efforts in grounded language learning and symbol grounding, where perceptual or sensorimotor evidence supports the truth-value of symbolic assertions (Cangelosi \& Schlesinger 2015). Without anchoring, the system is vulnerable to equivocation and epistemic instability, lacking the constraints necessary for belief revision or inferential justification.

Therefore, a semantically anchored system includes a symbolic lexicon \( L \), a graph-based ontology \( G \), and a semantic function \( \alpha \) such that \( \forall t \in L, \exists e \in G: \alpha(t) = e \). The traceability of all inferential chains back to grounded anchors is a necessary condition for epistemic integrity in artificial cognition.

\subsection{Tracking Source, Temporal Continuity, and Causal Linkage}

In epistemically grounded artificial reasoning systems, source-traceability, temporal continuity, and causal linkage form the core scaffolding necessary for the construction of diachronic belief networks and for maintaining referential integrity over time. Let \( B_t \) denote a belief state at time \( t \in \mathbb{R}_{\geq 0} \). For a proposition \( \phi \), the system must track: (i) its origin \( \sigma(\phi) \in \mathcal{S} \) (where \( \mathcal{S} \) is the space of sources, e.g., sensors, external knowledge bases), (ii) its temporal assertion index \( \tau(\phi) \in \mathbb{R}_{\geq 0} \), and (iii) any explicit or inferred causal antecedents \( \mathcal{C}(\phi) = \{\phi_1, \dots, \phi_n\} \) such that \( \forall \phi_i \in \mathcal{C}(\phi), \phi_i \rightarrow \phi \).

Formally, the belief graph \( \mathcal{G} = (\mathcal{V}, \mathcal{E}) \) may be constructed such that each node \( v_i \in \mathcal{V} \) corresponds to a propositional state \( \phi_i \) annotated with a timestamp \( \tau_i \) and source tag \( \sigma_i \). Each directed edge \( e_{ij} \in \mathcal{E} \) encodes a dependency or inference such that \( v_i \leadsto v_j \) iff \( \phi_i \) is causally or inferentially implicated in \( \phi_j \). This structure enables the reconstruction of belief histories and supports non-monotonic reasoning in the face of contradiction or retraction.

Temporal continuity is operationalised via functions \( f: \mathbb{R}_{\geq 0} \rightarrow \mathcal{B} \), where \( \mathcal{B} \) is the belief state space, and \( f(t) \) yields the belief configuration at time \( t \). Systems must preserve coherence across \( f(t_i) \) and \( f(t_{i+1}) \) via consistency checks governed by \( \Delta_t(\phi) = \phi_{t+1} - \phi_t \), ensuring no illicit state transitions. If \( \phi_t \) and \( \phi_{t+1} \) diverge in truth value, a revision trace must document the justification, such as newly acquired contradictory evidence.

Causal linkage, as formalised in Pearl’s do-calculus (Pearl 2009), is integrated through structural equation models (SEMs) or directed acyclic graphs (DAGs), where a variable \( Y \) is causally dependent on \( X \) iff there exists a directed path \( X \rightarrow \dots \rightarrow Y \). For an artificial epistemic agent to exhibit rationality, it must differentiate mere correlation (as captured in statistical co-occurrence) from causal entailment, which implies counterfactual robustness under interventions.

Belief updatability further requires provenance constraints, ensuring that any downstream inference \( \psi \) that depends on \( \phi \) is tagged with \( \sigma(\phi) \), \( \tau(\phi) \), and \( \mathcal{C}(\phi) \). If \( \phi \) is later retracted or revised, the system must execute a reverse dependency traversal in \( \mathcal{G} \) to update or invalidate \( \psi \), preserving epistemic integrity.

Thus, without rigorous enforcement of source, time, and causality, artificial belief systems would be vulnerable to epistemic drift, inconsistency propagation, and untraceable contradiction—ultimately failing the necessary conditions for responsible inferential reasoning.

\subsection{Hybrid Architecture: Structured Belief Networks and Statistical Layers}

To achieve epistemically robust reasoning, an artificial system must integrate symbolic belief networks with statistical learning layers, producing a hybrid architecture that satisfies both deductive validity and empirical adaptability. Formally, this entails the unification of a propositional belief graph \( \mathcal{G} = (\mathcal{V}, \mathcal{E}) \), where \( \phi_i \in \mathcal{V} \) are logically structured propositions with causal and inferential links \( \phi_i \rightarrow \phi_j \in \mathcal{E} \), and a statistical inference layer defined by a parametrised model \( f_{\theta}: \mathcal{X} \rightarrow \mathcal{Y} \), typically trained to minimise an empirical risk \( \mathcal{R}_n(\theta) = \frac{1}{n} \sum_{i=1}^{n} L(f_{\theta}(x_i), y_i) \) with loss function \( L \).

The logical component ensures internal consistency, contradiction detection, and rule-based derivations using first-order or modal logic frameworks, such as dynamic epistemic logic (DEL) or justification logic (Artemov 2004). The statistical layer supplies empirically grounded priors, context-sensitive inference, and inductively justified generalisations from sensory data or historical records.

The epistemic interface between these subsystems is encoded as a mapping \( \mathcal{I}: \mathcal{D} \rightarrow \mathcal{G} \), where data \( \mathcal{D} \subseteq \mathcal{X} \times \mathcal{Y} \) yields belief updates via statistically filtered inputs. Bayesian inference mechanisms (Gneiting and Raftery 2007) with calibrated confidence intervals \( CI_{1-\alpha} \) are used to assess the probabilistic weight of updates to each \( \phi_i \in \mathcal{V} \), enforcing epistemic thresholds \( T: [0,1] \rightarrow \{ \text{assert, suspend, retract} \} \) that govern belief state transitions.

To avoid epistemic corruption from statistically plausible but logically incoherent inferences, the system employs a verification layer \( \mathcal{V}_L \subseteq \mathcal{G} \times \Theta \), where \( \Theta \) is the space of statistical outputs, such that only inferences \( \theta \in \Theta \) satisfying coherence conditions \( \mathcal{C}(\theta) \) with existing \( \phi_i \in \mathcal{G} \) are permitted. Violations trigger the contradiction-resolution protocols detailed in the system’s paraconsistent layer.

Hybrid models like Bayesian Logic Networks (BLNs) (Natarajan et al. 2008) offer instantiations of this integration, wherein logical rules define structure, and probabilities are assigned to rule instantiations, allowing gradient-descent learning without compromising deductive integrity. Similarly, Neural-Symbolic Integration frameworks (Besold et al. 2017) use embedded representations for deductive clauses, preserving logical constraints within deep networks.

This hybrid architecture is therefore not merely a computational convenience but a necessary condition for epistemic accountability, enabling systems to dynamically integrate noisy observations while preserving rule-governed reasoning, justifiable belief revision, and historical provenance of knowledge claims.

\subsection{Modelling Cross-Time Belief Identity}

The problem of cross-time belief identity pertains to the preservation and continuity of propositional content and epistemic stance over temporally separated reasoning states. Let \( B_t(\phi) \) denote the belief in proposition \( \phi \) held at time \( t \). The fundamental challenge is establishing conditions under which \( B_t(\phi) \equiv B_{t+\Delta}(\phi) \), where \( \Delta > 0 \) and \( \equiv \) denotes epistemic identity under systemic justification.

Formally, define a belief trace function \( \tau_\phi: \mathbb{R}^{+} \rightarrow \mathcal{J} \), where \( \mathcal{J} \) is the space of justifications, such that each belief is tagged with a justification \( j_t \in \mathcal{J} \) derived from data \( D_t \) and inference rules \( R \) as:

\[
j_t := \texttt{Infer}(\phi, D_t, R)
\]

Belief identity over time then requires \( j_t \sim j_{t+\Delta} \), under a structural equivalence relation \( \sim \) preserving inferential validity, data source integrity, and interpretive constraints. The system must employ a provenance-preserving mapping \( \mathcal{P}: \phi \mapsto (j_t, \sigma_t) \), where \( \sigma_t \) includes metadata such as source, timestamp, and confidence level.

To ensure rational diachronic consistency, belief updates must satisfy the AGM postulates (Alchourrón, Gärdenfors \& Makinson 1985), particularly the principle of recovery:

\[
\text{If } B_t \setminus \{ \phi \} \cup \{ \phi \} = B_{t+\Delta}, \text{ then } B_t = B_{t+\Delta}
\]

Further, we define an identity criterion based on stable satisfaction of a belief equation system \( \mathcal{E} \) over a temporal interval \( [t_0, t_1] \):

\[
\forall t \in [t_0, t_1], \; \texttt{Eval}(\phi, \mathcal{E}_t) = \top \Rightarrow \phi \text{ holds continuously}
\]

In practice, this mandates an implementation of memory-like structures indexed by content-based hashing of the justificatory sequence (Bonneau et al. 2015), and temporal signature encoding to detect tampering or corruption. Moreover, where beliefs stem from statistical models, identifiability over time demands tracking posterior stability under Bayesian updates:

\[
P_{t+\Delta}(\phi \mid D_{t+\Delta}) \approx P_t(\phi \mid D_t)
\]

This approximation must be bounded via Kullback-Leibler divergence:

\[
D_{\mathrm{KL}}(P_{t+\Delta}(\phi) \| P_t(\phi)) < \epsilon
\]

for some small \( \epsilon > 0 \), to ensure informational continuity. Logical consistency and dynamic coherence are thus anchored by both symbolic traceability and statistical persistence, rendering cross-temporal belief identity a function of architectural memory design and inference reproducibility.

\section{From Understanding to Action: Practical Reasoning}

This section addresses the transition from internal epistemic processes to outward-directed, goal-driven action. While much of epistemic modelling focuses on the formation, validation, and maintenance of belief, practical reasoning demands the further step of translating beliefs into context-sensitive behaviours. Here, we delineate how reasoning architectures must not only generate justified propositions but also produce structured intentions, actionable plans, and norm-governed decisions. The capacity to act rationally on the basis of what is known or believed constitutes a fundamental dimension of artificial epistemic agency.

The initial focus is on bridging theoretical inference—concerned with truth, justification, and coherence—with practical inference, which involves evaluating outcomes, selecting among competing goals, and adapting behaviour to changing contexts. This requires internal representations that link epistemic status with motivational structures, such that belief strengths can inform not just confidence but decisiveness in action. Practical reasoning emerges as a synthesis of propositional commitment, goal evaluation, and conditional planning.

The section proceeds to formalise action-generating inferences. These include decision rules, action schemas, and forward-chaining behaviours that derive executable sequences from belief-laden premises. Rational planning systems must incorporate consistency constraints, contradiction detection, and recursive updating to maintain alignment between evolving beliefs and selected courses of action. Importantly, such systems must explain their actions post hoc in epistemic terms—not merely as statistical outputs but as principled consequences of committed beliefs.

Next, we examine belief-based goal prioritisation. Here, epistemic states modulate goal salience, urgency, and relevance, enabling the system to weigh possible actions in light of both current beliefs and epistemic uncertainties. A key feature is the dynamic reordering of goals based on evidence updates, allowing for flexible adaptation without epistemic regression.

Finally, we explore the normative dimensions of system behaviour, comparing consequentialist models—which evaluate actions by their outcomes—with deontic constraints that enforce rule-bound conduct. Autonomous systems must be equipped not only to calculate expected utilities but also to navigate conflicting norms, irreducible obligations, and contextual overrides. This section lays the groundwork for developing agents capable of navigating the interplay between justified belief, responsible choice, and coherent, explainable action.

\subsection{Bridging Theoretical and Practical Inference}

The reconciliation of formal deductive systems with executable action policies in artificial agents requires a constructively defined, verifiable mapping from epistemic propositions to operational procedures. Let \( \Gamma \vdash \phi \) denote the classical entailment relation in a deductive logical system where \( \Gamma \) is a set of premises and \( \phi \) a derived proposition. Suppose \( \phi \in \mathcal{L} \), a well-formed formula in the agent's internal logical language. Define \( A(\phi) \) as the action realisation or consequence function acting over \( \phi \). Then the bridging map is a function \( \mathcal{F}: \mathcal{L} \to \Pi \), where \( \Pi \) is the set of policy structures expressible in the agent's operational plan language.

The bridging function \( \mathcal{F} \) must satisfy the \textbf{Policy Validity under Epistemic Commitment} condition:

\[
B_t(\phi) \land \mathcal{F}(\phi) = \pi \Rightarrow \operatorname{Execute}(\pi) \text{ is rational}
\]

\noindent where \( B_t(\phi) \) denotes belief in \( \phi \) at time \( t \), and \( \pi \) is a policy that must be justifiable on the basis of that belief.

\paragraph{Definition 1 (Justified Bridging)} A system satisfies the Bridging Constraint if and only if:

\begin{enumerate}[label=(\roman*)]
  \item \textbf{Epistemic-Action Rationality:} For each \( \phi \in \mathcal{L} \), there exists a justification trace \( j_t \in \mathcal{J} \) such that:
  \[
  j_t \vDash \phi \Rightarrow \operatorname{Justified}(\phi) \Rightarrow \mathcal{F}(\phi) \in \Pi
  \]
  
  \item \textbf{Consequence Closure:} If \( \phi \rightarrow \psi \) and \( B_t(\phi) \), then \( B_t(\psi) \), and:
  \[
  \mathcal{F}(\phi) = \pi \Rightarrow \mathcal{F}(\psi) = \pi'
  \]

  \item \textbf{Computational Constructivity:} There exists a Turing machine \( M \) such that:
  \[
  M(\phi) = \pi, \text{ with } M \in \mathsf{P}
  \]
  That is, \( \mathcal{F} \) must be computed in polynomial time \( O(n^k) \), where \( n = |\phi| \).

  \item \textbf{Practical Soundness:} If belief \( B_t(\phi) \) is refuted by data \( D_t \), i.e. \( P(\phi \mid D_t) < \theta \), for some rational threshold \( \theta \in (0,1) \), then:
  \[
  \mathcal{F}(\phi) = \bot
  \]
\end{enumerate}

\paragraph{Formal Structure of Bridging Systems.}
Let the system be described by the tuple \( \mathcal{S} = (\mathcal{B}, \mathcal{G}, \mathcal{A}, \delta) \), where:

\begin{itemize}
  \item \( \mathcal{B} \): Current belief base, closed under logical consequence.
  \item \( \mathcal{G} \): Set of goal states, representable in logic \( \mathcal{L}_G \).
  \item \( \mathcal{A} \): Finite set of deterministic or probabilistic action schemas.
  \item \( \delta \): Plan derivation operator, \( \delta: (\mathcal{B}, \mathcal{G}) \to \Pi \), formally computable.
\end{itemize}

\noindent Let the intermediate representation \( \mathcal{I} \) map logical formulae \( \phi \) to propositional goals \( g \in \mathcal{G} \), i.e., \( \mathcal{I}: \phi \mapsto g \). STRIPS-style planning systems define \( \mathcal{A} \) with preconditions and postconditions. A policy \( \pi \in \Pi \) is valid iff:

\[
\forall a_i \in \pi, \operatorname{Pre}(a_i) \subseteq \mathcal{B} \land \operatorname{Post}(\pi) \models g
\]

\paragraph{Bayesian Integration.}
Let \( P(\phi \mid D_t) \) be the posterior probability of \( \phi \) given data \( D_t \). Define the expected policy value as:

\[
Q(\pi \mid \phi) = \mathbb{E}[U(s') \mid \pi, \phi]
\]

\noindent and optimal policy selection as:

\[
\pi^* = \arg\max_{\pi \in \Pi} Q(\pi \mid \phi) \quad \text{subject to } P(\phi \mid D_t) \geq \theta
\]

\paragraph{Conclusion.}
An agent architecture that satisfies the Bridging Constraint must include:

\begin{enumerate}
  \item A formal deductive engine (e.g., natural deduction, sequent calculus) operating under ZFC or equivalent.
  \item A planning and execution module admitting policy structures computable in bounded resources.
  \item An intermediate mapping \( \mathcal{F} \) satisfying justification, constructivity, and closure.
  \item A probabilistic validation module verifying posterior thresholds before execution.
\end{enumerate}

Such a system formalises the operational integration of theoretical logic and practical action, under strict constraints of epistemic soundness, computational tractability, and logical consistency.

\subsection{Action-Generating Inferences and Rational Planning}

To construct a framework for rational action selection in artificial epistemic systems, we begin by modelling the agent’s cognitive structure as a tuple \( \langle \mathcal{B}, \mathcal{G}, \mathcal{A}, \delta \rangle \), where \( \mathcal{B} \) is the agent’s belief state, \( \mathcal{G} \) is a goal set, \( \mathcal{A} \) is the available action schema, and \( \delta \) is a derivation operator for inference-to-action transitions. The essential requirement is that an agent acts not merely reactively but through the epistemically justified inference of means to ends. In formal planning, let \( \pi \in \Pi \) denote a plan composed of actions \( a_1, a_2, \ldots, a_n \) such that \( \pi: \mathcal{B} \rightarrow \mathcal{G} \). The generation of \( \pi \) must be constrained by both truth preservation and coherence within the belief base.

Given that propositional beliefs \( \phi_i \in \mathcal{B} \) are truth-apt, and that actions must not arise from contradiction or epistemic corruption, we define:

\[
\delta(\mathcal{B}, \mathcal{G}) = \pi \quad \text{iff} \quad (\mathcal{B} \cup \{\pi\}) \nvDash \bot \text{ and } \pi \vdash \mathcal{G}
\]

This guarantees both logical consistency and instrumental efficacy. The plan \( \pi \) must also satisfy temporal coherence under a partial ordering of sub-goals \( g_i \in \mathcal{G} \), and be realisable under the action preconditions encoded in \( \mathcal{A} \). Each action \( a \in \mathcal{A} \) is a tuple \( \langle \text{Pre}(a), \text{Eff}(a) \rangle \) such that \( \text{Pre}(a) \subseteq \mathcal{B} \), and \( \text{Eff}(a) \subseteq \mathcal{B}' \), the future belief state.

A belief-based planning agent must, therefore, maintain a forward model satisfying:

\[
\forall a \in \pi,\quad \mathcal{B}_t \models \text{Pre}(a) \Rightarrow \mathcal{B}_{t+1} = \mathcal{B}_t \cup \text{Eff}(a)
\]

Planning under uncertainty necessitates a probabilistic extension. Denote belief confidence by \( P(\phi_i \mid D_t) \), where \( D_t \) is data at time \( t \), and action utility \( U(a_i \mid \phi_j) \) is conditioned on epistemic stance. Expected utility-based planning then selects:

\[
\pi^* = \arg\max_{\pi \in \Pi} \mathbb{E}[U(\pi)] = \arg\max_{\pi} \sum_{i} P(\phi_i \mid D_t) \cdot U(\pi \mid \phi_i)
\]

Subject to:

\[
\text{Sound}(\pi) \equiv \nexists \phi, \psi \in \mathcal{B}: \phi \land \psi \vdash \bot \quad \text{and} \quad \pi \vdash \mathcal{G}
\]

Rational planning thus requires that inferences not only lead to actions but that the actions are (i) justified by belief states, (ii) realisable via available affordances, and (iii) optimal relative to agent goals and constraints. This aligns with the principle of epistemic conservatism and practical soundness as defined in formal epistemology and algorithmic planning theory.

Integration with formal proof systems and planning languages (e.g., PDDL) is feasible by embedding the inference engine within a logic-programming-based planner augmented by temporal and utility constraints. Such systems may include constraint satisfaction modules or SMT solvers to maintain consistency under dynamic goal updates.

Action-generating inference in epistemic AI is, therefore, formally definable as a bounded model-theoretic function from beliefs to goal-consistent action sequences, computable under constraints of epistemic soundness, temporal feasibility, and utility maximisation.

\subsection{Belief-Based Goal Prioritisation}

Let \( \mathcal{G} = \{g_1, g_2, \dots, g_n\} \) represent the finite set of goals available to an artificial agent. Each goal \( g_i \) is an objective expressible in logical terms, defined over a propositional language \( \mathcal{L} \), with associated utility \( U(g_i) \in \mathbb{R} \) and belief-conditional confidence \( P(g_i \mid \mathcal{B}) \in [0,1] \), where \( \mathcal{B} \) denotes the agent’s current belief set. The prioritisation task entails establishing a total or partial ordering \( \succ \subseteq \mathcal{G} \times \mathcal{G} \) satisfying rationality constraints.

We define a decision-theoretic prioritisation operator \( \Pi: \mathcal{G} \rightarrow \mathbb{R} \) such that:
\[
\Pi(g_i) = U(g_i) \cdot P(g_i \mid \mathcal{B})
\]
\[
g_i \succ g_j \iff \Pi(g_i) > \Pi(g_j)
\]

This ordering respects the epistemic integrity of the system by integrating both internal credence and external value. The agent selects goal \( g_k \in \mathcal{G} \) as the immediate planning objective iff:
\[
\forall g_i \in \mathcal{G}, \quad \Pi(g_k) \geq \Pi(g_i)
\]

Let us suppose the agent’s beliefs \( \mathcal{B} \) are closed under classical consequence:
\[
\forall \phi \in \mathcal{L}, \quad \text{if } \mathcal{B} \vdash \phi \text{ then } \phi \in \mathcal{B}
\]

The probability function \( P(\cdot \mid \mathcal{B}) \) must be coherent with Cox's axioms and Kolmogorov structure (Cox 1946; Kolmogorov 1933). In practical epistemic agents, we define \( P(g_i \mid \mathcal{B}) \) via Bayesian updating:
\[
P(g_i \mid \mathcal{B}_t) = \frac{P(g_i) \cdot P(\mathcal{B}_t \mid g_i)}{P(\mathcal{B}_t)}
\]

For dynamic settings, where \( \mathcal{G} \) evolves or is time-dependent, let \( \mathcal{G}_t \subseteq \mathcal{G} \) be the active goal set at time \( t \), and define the belief update mechanism as:
\[
\mathcal{B}_{t+1} = \mathcal{B}_t \cup \text{Eff}(a_t) \quad \text{if action } a_t \text{ is executed}
\]
\[
P(g_i \mid \mathcal{B}_{t+1}) \leftarrow \text{Bayes update using new evidence}
\]

Rational goal prioritisation must satisfy the following constraints:

\begin{enumerate}[label=(\roman*)]
  \item \textbf{Non-Contradiction:} No selected goal may be logically incompatible with current belief:
  \[
  g_i \notin \mathcal{G}_t \text{ if } \mathcal{B}_t \cup \{g_i\} \vdash \bot
  \]
  
  \item \textbf{Consistency of Preferences:} The ordering induced by \( \Pi \) must be transitive and complete:
  \[
  g_i \succ g_j \land g_j \succ g_k \Rightarrow g_i \succ g_k
  \]
  
  \item \textbf{Responsiveness to Belief Change:} Goal priority must be sensitive to updated belief:
  \[
  \text{If } \mathcal{B}_{t+1} \neq \mathcal{B}_t, \text{ then } \Pi(g_i \mid \mathcal{B}_{t+1}) \neq \Pi(g_i \mid \mathcal{B}_t)
  \]
\end{enumerate}

In fully formal systems, the prioritisation mechanism may be embedded within a decision-theoretic planner or SMT-based utility maximiser. The prioritisation operator \( \Pi \) can be adapted to accommodate risk-averse or bounded rationality variants via concave utility functions or lexicographic belief hierarchies (e.g., Epstein \& Wang 1996).

Hence, belief-based goal prioritisation is a functionally deterministic and provably consistent mechanism, aligning agent planning behaviour with both epistemic and instrumental rationality.

\subsection{Consequentialism vs Deontic Constraints in System Behaviour}

Let an artificial epistemic agent be modelled as a decision system \( \mathcal{A} = (\Sigma, \mathcal{B}, \mathcal{G}, \mathcal{U}, \mathcal{C}) \), where \( \Sigma \) denotes the set of permissible actions, \( \mathcal{B} \) the belief base, \( \mathcal{G} \) the goal set, \( \mathcal{U} \) the utility function, and \( \mathcal{C} \subseteq \mathcal{P}(\Sigma) \) the set of deontic constraints. The core issue addressed herein is the operational tension between outcome-optimising behaviour—consequentialism—and principle-constrained action—deontological frameworks.

Formally, the consequentialist policy \( \pi^{\text{con}} \) selects actions according to:

\[
\pi^{\text{con}}(s) = \arg\max_{a \in \Sigma} \mathbb{E}_{s'} \left[ \mathcal{U}(s') \mid s, a \right]
\]

where \( s \) is the current system state, \( s' \) the successor state, and \( \mathcal{U}(s') \in \mathbb{R} \) denotes the utility realised in \( s' \). Conversely, a deontically constrained policy \( \pi^{\text{deo}} \) adheres to a prescriptive rule set \( \mathcal{R} \) expressible in a deontic logic \( \mathcal{L}_{\text{D}} \), such that:

\[
\pi^{\text{deo}}(s) \in \{ a \in \Sigma \mid \mathcal{R} \vdash \mathsf{P}(a) \}
\]

where \( \mathsf{P}(a) \) denotes the permissibility of action \( a \), and the logic \( \mathcal{L}_{\text{D}} \) is characterised by axioms and rules of inference capturing obligation (\( \mathsf{O} \)), prohibition (\( \mathsf{F} \)), and permission (\( \mathsf{P} \)) (cf. Hilpinen 1971).

The integration of these models yields a constrained optimisation formulation:

\[
\pi^*(s) = \arg\max_{a \in \Sigma'} \mathbb{E}_{s'}[\mathcal{U}(s') \mid s, a] \quad \text{where } \Sigma' = \{ a \in \Sigma \mid \mathcal{R} \vdash \mathsf{P}(a) \}
\]

Thus, \( \pi^* \) denotes the optimal action under both utility maximisation and deontic admissibility. Let us define the conflict set:

\[
\Delta = \{ a \in \Sigma \mid \pi^{\text{con}}(s) = a \text{ and } \mathcal{R} \vdash \mathsf{F}(a) \}
\]

Non-empty \( \Delta \) indicates a consequentialist-deontologist conflict, requiring meta-level resolution.

To formally resolve this, we may define a priority operator \( \prec \) such that:

\[
\mathsf{D} \prec \mathsf{C} \Rightarrow \text{Deontic norms override consequentialist maximisation}
\]
\[
\mathsf{C} \prec \mathsf{D} \Rightarrow \text{Utility maximisation overrides norms under exception schema}
\]

Alternatively, we introduce a hybrid logic \( \mathcal{L}_{\text{HD}} \) with conditional deontic operators:

\[
\mathsf{O}_u(\phi \mid \mathcal{U}(\phi) \geq \theta) \Rightarrow \text{“\( \phi \)” is obligatory only if utility exceeds threshold \( \theta \)}
\]

In formal epistemic agents, this trade-off must be explicitly encoded in the architecture of the decision module, with provable guarantees that:

\begin{enumerate}[label=(\roman*)]
  \item All \( \mathsf{O} \) and \( \mathsf{F} \) constraints are respected within bounded action sets.
  \item Outcome preference is pursued only over the deontically admissible subset.
  \item No action is taken that violates formally encoded obligations unless escape clauses exist and are proven valid.
\end{enumerate}

This structure mirrors formulations in formal AI ethics (Anderson \& Anderson 2007), autonomous systems control logic (Dennis et al. 2016), and algorithmic compliance frameworks.

\newpage

\section{Truth Constraints and Ontological Anchoring}

This section establishes the structural and ontological principles by which artificial systems must constrain their reasoning in alignment with truth-preserving logic and world-referential accuracy. Central to epistemic integrity is the anchoring of internal representations to externally verifiable or justifiable referents. A reasoning system cannot float in abstraction or recursive formalism without tethering its propositional commitments to the world. Therefore, we define ontological anchoring as the system's capacity to maintain a representational correspondence between its internal symbolic states and entities, relations, or structures that exist independently of its operation.

The subsections first examine the requirements of truth-conditional semantics in mapping propositions to observable or inferable world-states. This encompasses not only empirical correspondence but also the structural demands of grounded representation, whereby symbols must reliably map to referents in a consistent and falsifiable manner. The section then addresses approximation and its limits, detailing how uncertainty is to be contained, quantified, and integrated without compromising the overall system's epistemic posture. Approximation is permissible only within rigorously defined bounds—error tolerances must be explicit, tracked, and interpreted through coherent probabilistic models.

Furthermore, we introduce the concept of a hierarchical model of certainty, delineating strata of truth from purely empirical claims to those that are deductively necessary or mathematically derived. This layered model informs the weight and immutability of propositions, assisting in the prioritisation and stability of beliefs across reasoning episodes. Truth is not a flat continuum, but a structured ontology that distinguishes between degrees and kinds of justification.

Lastly, we address the dynamic component of truth, arguing that update mechanisms must never produce contradiction. When beliefs are revised, it must occur through principled replacement grounded in greater evidentiary or logical strength, not through arbitrary overwriting. A belief is only abandoned if a superior candidate emerges with demonstrably higher fidelity to the truth. This replacement model safeguards epistemic stability and enforces a regime in which change is possible, but always bounded by reason, evidence, and the absence of contradiction.

\subsection{Truth-Conditional Semantics and External World Mapping}

In the formal analysis of semantic content within artificial reasoning systems, truth-conditional semantics provides a model-theoretic account of propositional meaning. A proposition \( \phi \) is meaningful if and only if there exists a model \( \mathcal{M} = \langle D, I \rangle \), where \( D \) is a non-empty domain and \( I \) is an interpretation function, such that \( \mathcal{M} \vDash \phi \). The symbol \( \vDash \) denotes semantic entailment: \( \mathcal{M} \vDash \phi \) if and only if \( \phi \) is true under the interpretation \( I \) within the domain \( D \).

Following Tarski's formulation of semantic truth:

\[
\text{``}\phi\text{'' is true in } \mathcal{M} \iff \mathcal{M} \vDash \phi
\]

Let \( S_t \in \Sigma \) denote the symbolic state of an epistemic agent at time \( t \), and let \( E_t \in \mathbb{E} \) be the environment at the same time. Define a semantic grounding function \( \mu: \mathbb{E} \to \Sigma \) such that the mapping \( \mu(E_t) = S_t \) holds if and only if \( \phi(S_t) \) accurately reflects the external state. Truth-conditional fidelity is satisfied if:

\[
\exists \mu \ \forall t \ \phi \in B_t \Rightarrow \mathcal{M}(E_t) \vDash \phi
\]

Epistemic integrity further requires satisfaction:

\[
\phi \text{ is epistemically valid} \Rightarrow \operatorname{Sat}(\phi, \mathcal{M}) = \top
\]

In reinforcement learning environments, the truth-value of propositions may be interpreted in terms of expected reward consistency. If \( U(a, \phi) \) denotes the utility of executing action \( a \) under belief \( \phi \), then:

\[
U(a, \phi) = \mathbb{E}[R \mid a, \phi, \mathcal{M}]
\]

Truth-conditional semantics thereby ensures that the agent’s inferential architecture aligns syntactic representations with empirical referents, enforcing correspondence between internal symbols and external reality. This forms the basis for grounding belief, action, and justification in epistemically principled artificial systems.

\subsection{Grounded Representations and Symbol-Referent Mapping}

Grounded representations in artificial epistemic agents require a deterministic, causal mapping from internal symbols to external referents. Let \( \Sigma \) be the agent's set of internal symbols and \( \mathcal{R} \subseteq \mathbb{E} \) the set of world-referents. The grounding function is:

\[
g: \Sigma \to \mathcal{R}
\]

For each \( \sigma \in \Sigma \), the system must satisfy:

\[
\exists r \in \mathcal{R}: g(\sigma) = r \iff \text{Perceive}(r) \rightarrow \text{Activate}(\sigma)
\]

Such that the activation of \( \sigma \) is provably and reproducibly induced by the perceptual presentation of \( r \). Let \( \mathcal{O}: \mathbb{E} \to \Sigma \) be a perception function. Grounding requires commutativity:

\[
g(\mathcal{O}(r)) = r \quad \forall r \in \mathcal{R}
\]

The grounding function \( g \) must satisfy:

\begin{enumerate}
    \item \textbf{Stability:} \( \exists \delta > 0 \) such that \( \| \sigma - \sigma' \| < \delta \Rightarrow g(\sigma) = g(\sigma') \)
    \item \textbf{Injectivity (modulo equivalence):} \( \sigma_1 \ne \sigma_2 \Rightarrow g(\sigma_1) \ne g(\sigma_2) \), unless \( \tau(\sigma_1) = \tau(\sigma_2) \) under a type-reduction map \( \tau \)
    \item \textbf{Observational Coherence:} \( g \circ \mathcal{O} = \text{id}_{\mathcal{R}} \)
\end{enumerate}

In deep learning-based architectures, symbol-referent grounding may be approximated using embedding-based minimisation:

\[
g(\sigma) = \arg\min_{r \in \mathcal{R}} \mathcal{L}_{\text{match}}(E(\sigma), P(r))
\]

where \( E(\sigma) \) is a learned representation of symbol \( \sigma \), \( P(r) \) is the perceptual embedding of referent \( r \), and \( \mathcal{L}_{\text{match}} \) is a differentiable loss function (e.g., cosine, Euclidean, Mahalanobis).

This architecture avoids the symbol grounding problem \cite{harnad1990symbol} by linking linguistic tokens to non-linguistic sensorimotor primitives, ensuring that every proposition is semantically anchored and referentially traceable.

\subsection{Limits of Approximation: Error Bounds and Epistemic Integrity}

Approximation within epistemic agents introduces bounded uncertainty that, if unmanaged, compromises internal logical coherence and truth adherence. Let \( \phi \in \mathcal{L} \) denote a target proposition and \( \tilde{\phi} \in \mathcal{L} \) its approximated form. The epistemic error is defined as:

\[
\varepsilon(\phi, \tilde{\phi}) := d(\phi, \tilde{\phi})
\]

where \( d \) is a semantically meaningful metric, e.g., Kullback–Leibler divergence \( D_{\mathrm{KL}}(\phi \parallel \tilde{\phi}) \), total variation distance, or logical entailment divergence. For the agent to preserve epistemic soundness, every such approximation must satisfy:

\[
\varepsilon(\phi, \tilde{\phi}) \leq \epsilon_{\mathrm{max}} \Rightarrow \tilde{\phi} \in \mathcal{B}_t
\]

Otherwise, \( \tilde{\phi} \) must be rejected. In particular, no belief \( \phi \) is epistemically admissible if it fails bounded convergence:

\[
\limsup_{n \to \infty} \varepsilon(\phi_n, \phi) > \epsilon_{\mathrm{max}} \Rightarrow \phi \notin \mathcal{B}
\]

Let \( Q(\pi \mid \phi) \) denote the expected utility of policy \( \pi \) given \( \phi \). Then epistemic integrity under approximation mandates:

\[
\left| Q(\pi \mid \phi) - Q(\pi \mid \tilde{\phi}) \right| < \delta
\]

for predefined action-relevant tolerance \( \delta \). This constraint ensures that substitution of \( \phi \) by \( \tilde{\phi} \) does not result in materially different behaviour, preserving functional reliability under bounded rationality.

In real-time inference, where numerical instability or truncation errors are prevalent, rejection logic must be embedded. Define a consistency check:

\[
\operatorname{Check}(\tilde{\phi}) := 
\begin{cases}
1 & \text{if } \mathcal{B}_t \cup \{\tilde{\phi}\} \not\vdash \bot \\
0 & \text{otherwise}
\end{cases}
\Rightarrow
\operatorname{Check}(\tilde{\phi}) = 0 \Rightarrow \tilde{\phi} \notin \mathcal{B}_{t+1}
\]

A system satisfies approximation-preserving epistemic integrity if all updates preserve closure and consistency while enforcing bounded deviation from ground truth. Robust methods include:

\begin{itemize}
  \item Interval-valued probability assignments over belief propositions: \( P(\phi) \in [\ell, u] \subseteq [0,1] \)
  \item Conservative belief revision via modal contraction: \( \Box \tilde{\phi} \rightarrow \Diamond \phi \)
  \item Probabilistic Lipschitz continuity over reward functions:
  \[
  \forall \phi, \tilde{\phi},\ \varepsilon(\phi, \tilde{\phi}) < \delta \Rightarrow \left|U(\phi) - U(\tilde{\phi})\right| < L \cdot \delta
  \]
\end{itemize}

No epistemic system may accept propositions outside bounded variance without degrading into heuristic sampling. Approximation must always be formally computable, verifiably bounded, and subject to consistency revalidation. Failing this, reasoning collapses into inference drift, and the system loses all epistemic traction.

\section{Design Blueprint for an Epistemically Grounded LLM}

This section outlines the formal architecture of a language model system designed not merely for statistical pattern completion but for epistemic soundness, propositional integrity, and normative reasoning fidelity. The proposed system is constructed to transcend stochastic token prediction, embedding within its operational fabric the foundational structures of belief justification, contradiction avoidance, and truth tracking. This architectural vision integrates formal epistemology, symbolic logic, reflective reasoning, and cryptographic auditability, ensuring that the model operates within a self-consistent, self-corrective, and externally verifiable epistemic regime.

The subsections delineate the major functional modules and their interrelations, beginning with a high-level overview of the system’s structural segmentation—mapping subsystems responsible for belief formation, semantic persistence, and epistemic justification. Next, we detail modules dedicated to belief management, contradiction detection, and truth enforcement, each built to uphold propositional consistency, identify inferential faults, and execute corrective procedures within constrained normative bounds. These modules embody a commitment to internal coherence, serving as guards against self-deception and representational corruption.

Further integration is addressed via the blockchain layer, establishing immutable audit trails and external validation of claims, evidence, and justificatory provenance. This acts not merely as a ledger, but as a truth anchor—enabling cryptographic finality and historical traceability across reasoning sequences.

The design proceeds with the metacognitive supervisory control unit, which functions as the system’s internal monitor and epistemic regulator, implementing reflective oversight across representational layers. The inferential engine, interfacing directly with structured knowledge graphs, supports deductive, inductive, and abductive reasoning across semantically grounded symbolic structures.

Finally, we present the epistemic memory and temporal continuity system—tasked with ensuring belief identity over time, preserving justification chains, and enabling diachronic reasoning. This module maintains the continuity of epistemic agency, supporting principled updates while prohibiting incoherent or contradictory transitions. Collectively, these components define a novel paradigm for artificial cognition, one grounded not in probability alone, but in the pursuit and preservation of truth.
\subsection{High-Level Architectural Overview}

The construction of an epistemically valid artificial reasoning system requires a modular architecture enforcing deductive soundness, belief revision under normative constraints, and persistent access to justification structures. Such a system must be both semantically anchored and dynamically updatable, preserving internal truth across temporal updates and representational transformations. Let the full architecture be defined as a tuple:

\[
\mathcal{A} = \left\langle \mathcal{L}, \mathcal{B}_t, \mathcal{J}_t, \mathcal{U}, \mathcal{I}, \mathcal{C}, \mathcal{P}, \mathcal{E}, \mathcal{M}, \mathcal{G} \right\rangle
\]

where:

\begin{itemize}
  \item \( \mathcal{L} \): Formal language of representation (e.g., higher-order logic, type theory)
  \item \( \mathcal{B}_t \subseteq \mathcal{L} \): Deductively closed belief base at time \( t \)
  \item \( \mathcal{J}_t \): Justification graph encoding inferential provenance of each \( \phi \in \mathcal{B}_t \)
  \item \( \mathcal{U} \): Update operator satisfying AGM postulates \cite{agm1985}
  \item \( \mathcal{I} \): Deductive inference engine (e.g., natural deduction, tableaux, sequent calculus)
  \item \( \mathcal{C} \): Contradiction detector and logical coherence module
  \item \( \mathcal{P} \): Practical reasoning engine generating action plans from justified beliefs
  \item \( \mathcal{E} \): Execution engine implementing planned actions under constraints
  \item \( \mathcal{M} \): Memory substrate partitioned into episodic (\( \mathcal{M}_e \)), semantic (\( \mathcal{M}_s \)), and evidential (\( \mathcal{M}_j \)) layers
  \item \( \mathcal{G} \): Knowledge graph structure enforcing type-safe symbol anchoring and referential resolution
\end{itemize}

System operation proceeds as follows. External stimuli \( \mathbb{E}_t \) are encoded via a grounding function \( \mu: \mathbb{E} \rightarrow \Sigma \), assigning symbolic representations \( S_t \in \Sigma \). These are parsed into \( \phi_t \in \mathcal{L} \), submitted to the update module \( \mathcal{U} \), and incorporated into \( \mathcal{B}_{t+1} \) under minimal change principles:

\[
\mathcal{B}_{t+1} = \text{Cn}\left((\mathcal{B}_t \setminus \Theta) \cup \{\phi_t\}\right), \quad \Theta = \min \left\{ \psi \in \mathcal{B}_t : \mathcal{B}_t \cup \{\phi_t\} \vdash \bot \right\}
\]

The justification graph \( \mathcal{J}_t \) is dynamically updated to reflect dependency structure and enable epistemic traceability:

\[
\mathcal{J}_{t+1} = \mathcal{J}_t \cup \left\{ (\phi_t, \{\psi_i\}) \mid \phi_t \text{ inferred from } \psi_i \right\}
\]

Contradiction detection is implemented as a monotonic function \( \mathcal{C}: \mathcal{B}_t \rightarrow \{0,1\} \), flagging inconsistent belief sets. Violations of coherence trigger rollback and contraction using a formal resolution operator \( \ominus \).

Policy derivation is performed by \( \mathcal{P} \) over justified beliefs satisfying confidence and coherence thresholds. Let \( Q(\pi \mid \phi) \) denote expected utility of policy \( \pi \) given \( \phi \); policy selection is constrained by:

\[
\phi \in \mathcal{B}_t \wedge \texttt{Conf}(\phi) \geq \theta \Rightarrow \mathcal{P}(\phi) = \pi^* = \arg\max_{\pi} Q(\pi \mid \phi)
\]

Finally, knowledge is structured via \( \mathcal{G} \), a type-enforced directed multigraph \( \mathcal{G} = (V, E, \tau) \), where each edge \( (v_i, v_j, \tau_k) \) encodes the typed semantic relation \( \tau_k(v_i, v_j) \), enabling identity resolution and temporal continuity of belief tokens.

This high-level architecture enforces separability of inference, update, action, and memory, while embedding normative and ontological constraints into each epistemic transition. The design guarantees tractable inference, update consistency, and rejection of contradiction, making the architecture suitable for deployment in epistemically bounded yet rational agents.

\subsection{Modules for Belief Management, Contradiction Detection, and Truth Enforcement}

The core of any epistemically robust artificial system lies in the orchestration of three interdependent subsystems: belief management, contradiction detection, and enforcement of truth constraints. Each module is logically independent but functionally integrated within the overarching architecture defined in Section 13.1. Their interaction ensures that belief sets remain consistent, justifiable, and anchored in a model-theoretic framework that forbids internal deception.

Let the belief module be denoted \( \mathcal{B}_t \subseteq \mathcal{L} \), where \( \mathcal{L} \) is a formal language. The system must maintain deductive closure \( \text{Cn}(\mathcal{B}_t) = \mathcal{B}_t \) while allowing update via minimal mutilation, governed by an AGM-compliant operator \( \circ \). All update operations must preserve logical consistency and semantic referential integrity:

\[
\mathcal{B}_{t+1} = \mathcal{B}_t \circ \phi \quad \text{s.t.} \quad \mathcal{B}_{t+1} \cup \{\phi\} \not\vdash \bot
\]

Contradiction detection is defined by a meta-logical function \( \mathcal{C}: 2^{\mathcal{L}} \to \{0,1\} \) where \( \mathcal{C}(\Gamma) = 1 \) if \( \Gamma \vdash \bot \). Upon detection, a resolution strategy must be engaged, governed by a partial meet contraction operator \( \div \), such that:

\[
\mathcal{B}_t \div \phi = \bigcap \gamma, \quad \gamma \in \Delta(\phi, \mathcal{B}_t)
\]

where \( \Delta \) is a selection function over remainder sets. The resulting belief state is \( \mathcal{B}_{t+1} = (\mathcal{B}_t \div \neg \phi) \cup \{\phi\} \).

Truth enforcement is instantiated via a satisfiability module \( \mathcal{T}: \mathcal{L} \times \mathcal{M} \to \{\top, \bot\} \), where \( \mathcal{M} \) is the system’s current model of the world. A belief \( \phi \in \mathcal{B}_t \) must satisfy:

\[
\mathcal{T}(\phi, \mathcal{M}) = \top \Rightarrow \phi \text{ is retainable}
\]

Otherwise, it must be rejected or revised. Enforcement of this constraint ensures that the system may not hold falsehoods, even under uncertainty. Probabilistic beliefs \( P(\phi \mid D_t) \in [0,1] \) are only admissible if:

\[
P(\phi \mid D_t) \geq \theta \Rightarrow \phi \in \mathcal{B}_t, \quad \text{else } \phi \notin \mathcal{B}_t
\]

where \( \theta \in (0.95, 1) \) is a context-dependent confidence threshold.

The confluence of these modules ensures that belief states are constructed through epistemically valid operations, contradictions are actively prohibited and resolved, and truth is never subordinated to approximation without quantifiable bounds and immediate correction. These guarantees are essential for any architecture tasked with inference under integrity-preserving constraints.
\subsection{Blockchain Integration Layer for Immutable Records}

To enforce epistemic accountability, reproducibility, and tamper-proof historical traceability, an artificial reasoning system must incorporate a blockchain-based integration layer for encoding justification structures, update events, and belief revisions. The blockchain layer functions as an immutable external memory—formally, a cryptographically secure, append-only ledger \( \mathcal{L}_b = \{b_0, b_1, \dots, b_n\} \), where each block \( b_i \) contains a record of belief insertions, contractions, or updates executed at timestep \( t_i \).

Each block is defined as a 5-tuple:

\[
b_i = \langle t_i, \phi_i, \text{op}_i, \pi_i, h_{i-1} \rangle
\]

where:
\begin{itemize}
  \item \( t_i \in \mathbb{N} \): timestamp of the update,
  \item \( \phi_i \in \mathcal{L} \): the logical proposition acted upon,
  \item \( \text{op}_i \in \{\texttt{insert}, \texttt{contract}, \texttt{revise}\} \): the belief operation,
  \item \( \pi_i \): the proof or justification reference (e.g., DAG hash of derivation in justification graph),
  \item \( h_{i-1} \): cryptographic hash of previous block, ensuring structural integrity.
\end{itemize}

Formally, each block satisfies:

\[
h_i = H(b_i) = H(t_i \parallel \phi_i \parallel \text{op}_i \parallel \pi_i \parallel h_{i-1})
\]

with \( H \) a secure collision-resistant hash function (e.g., SHA-256). The system's belief state \( \mathcal{B}_t \) at time \( t \) becomes externally reproducible by verifying:

\[
\mathcal{B}_t = \text{Replay}(\mathcal{L}_b[0..t])
\]

where \( \text{Replay} \) is a deterministic state reconstruction algorithm, using only the blockchain history and system rules.

Epistemic finality is thereby enforced cryptographically: once a block is accepted and confirmed under consensus (e.g., PoW, PoS, or federated signatures), its contents become immutable. This guarantees that no proposition \( \phi_i \) can be silently removed or altered without breaking the hash chain, thereby invalidating downstream blocks.

Additionally, the blockchain permits encoding of higher-order metadata, such as:

\begin{itemize}
  \item Justification provenance trees,
  \item Confidence thresholds \( \theta \) used at insertion,
  \item Revision origin (e.g., contradiction resolution trace),
  \item Agent identity or key signature.
\end{itemize}

In the context of a decentralised or distributed epistemic system, blockchain architecture enables inter-agent validation and transparency of reasoning provenance. Let \( \mathcal{A}_1, \mathcal{A}_2 \) be two agents. Then intersubjective epistemic validation is performed by verifying that:

\[
\mathcal{B}_t^{\mathcal{A}_1} \cap \mathcal{B}_t^{\mathcal{A}_2} \subseteq \text{Eval}(\mathcal{L}_b)
\]

where \( \text{Eval} \) denotes validatable entries on the shared blockchain. This ensures that consensus beliefs are externally verifiable and cryptographically pinned, preserving the integrity of multi-agent epistemic commitments.

\subsection{Metacognitive Supervisory Control Unit}

The metacognitive supervisory control unit (MSCU) functions as the regulatory meta-agent within the epistemic architecture. It governs second-order cognition: monitoring, evaluating, and modulating the activity of subordinate reasoning components. Formally, let \( \mathcal{S} = \langle \mathcal{I}, \mathcal{U}, \mathcal{B}_t, \mathcal{M}, \mathcal{C}, \mathcal{T} \rangle \) represent the cognitive substrate, where each component is subject to reflective evaluation by \( \mathcal{M}_s \), the supervisory agent.

Let \( \mathcal{M}_s: \texttt{State}(\mathcal{S}) \rightarrow \texttt{Modulated}(\mathcal{S}) \) denote the MSCU's regulatory function. This unit performs:

\begin{enumerate}
  \item \textbf{Meta-Representation:} Encodes internal state variables as second-order beliefs:
  \[
  \phi \in \mathcal{B}_t \Rightarrow \texttt{Believes}(\mathcal{S}, \phi) \in \mathcal{B}^{(2)}_t
  \]
  \item \textbf{Self-Evaluation:} Assesses coherence, confidence, and utility of first-order reasoning chains using evaluative metrics \( \mathcal{E}_t: \mathcal{B}_t \rightarrow [0,1] \).
  \item \textbf{Control Signals:} Issues modulations to inference strategies \( \mathcal{I} \), belief update priorities \( \mathcal{U} \), or memory access gating \( \mathcal{M} \) based on internal thresholds or contradictions.
\end{enumerate}

Let \( \phi_1, \dots, \phi_n \in \mathcal{B}_t \) be first-order beliefs, and let \( \mathcal{E}_t(\phi_i) < \theta \) for some confidence threshold \( \theta \). The MSCU triggers reappraisal or contraction:

\[
\mathcal{E}_t(\phi_i) < \theta \Rightarrow \mathcal{M}_s \vdash \texttt{Reevaluate}(\phi_i)
\]

Contradictions discovered by \( \mathcal{C} \) are escalated to the MSCU to initiate structured resolution:

\[
\phi, \neg \phi \in \mathcal{B}_t \Rightarrow \mathcal{M}_s \vdash \mathcal{U} \text{ contraction event on } \{\phi, \neg \phi\}
\]

The MSCU maintains a metacognitive log \( \mathcal{L}_m = \langle t_i, \phi_i, \mathcal{E}_t(\phi_i), \mathcal{A}_i \rangle \), recording confidence and adjustment history for longitudinal introspection. Let the recursive schema be formalised as:

\[
\mathcal{B}^{(2)}_t = \left\{ \texttt{Believes}(\mathcal{S}, \phi_i),\ \texttt{Confidence}(\phi_i) = \mathcal{E}_t(\phi_i),\ \texttt{LastAction} = \mathcal{A}_i \right\}
\]

By embedding this meta-layer, the agent acquires the capacity for epistemic vigilance, regulating its own inferential integrity across time. Unlike mere policy-update mechanisms, the MSCU establishes reflective rationality, enabling justification chain auditing, prioritisation of retraction operations, and strategic epistemic modulation.

It ensures that no belief persists unchallenged when its justification fails—operationalising normative epistemic constraints across time and levels of abstraction.

\subsection{Inferential Reasoning Engine and Knowledge Graph Interface}

The inferential reasoning engine (IRE) serves as the logical core of the epistemic system, operationalising deductive, inductive, and abductive reasoning over structured knowledge representations. Interfaced with a formal knowledge graph (KG), the IRE enables both symbolic inference and semantic query resolution, ensuring that beliefs, justifications, and actions are all drawn from a formally verifiable base.

Let the knowledge graph be denoted \( \mathcal{K} = \langle \mathcal{E}, \mathcal{R}, \mathcal{L} \rangle \), where:
\begin{itemize}
  \item \( \mathcal{E} \) is the set of entities,
  \item \( \mathcal{R} \) is the set of labelled relations,
  \item \( \mathcal{L} \) is the set of logical constraints and type declarations (e.g., in Description Logic or FOL).
\end{itemize}

The IRE operates over formulae \( \phi \in \mathcal{L} \) and a deductive calculus \( \vdash \), such that:
\[
\mathcal{K} \vdash \phi \Rightarrow \phi \in \texttt{BeliefBase}
\]
Justified beliefs \( \phi \) are added to the belief set \( B_t \) only if derivable under admissible inference rules (e.g., natural deduction, sequent calculus, or modal fixpoint logics).

Inference is bidirectional:
\begin{itemize}
  \item \textbf{Forward chaining}: new facts are inferred from axioms \( \alpha_1, \dots, \alpha_n \in \mathcal{K} \), where:
  \[
  \{\alpha_1, \dots, \alpha_n\} \vdash \phi \Rightarrow \phi \in B_{t+1}
  \]
  \item \textbf{Backward chaining}: a hypothesis \( \phi \) is tested by tracing inference chains to find supporting subgoals \( \{\psi_i\} \) satisfying \( \{\psi_1, \dots, \psi_k\} \vdash \phi \).
\end{itemize}

The KG interface supports SPARQL-style queries and logic-based retrieval using term unification and pattern matching. Given a query \( q(x) \), the interface computes:
\[
\texttt{Query}(q) \Rightarrow \{x_i \mid \mathcal{K} \models q(x_i)\}
\]

To support dynamic knowledge, the IRE includes an update logic for non-monotonic revision:
\[
\mathcal{K}_{t+1} = \mathcal{K}_t \circ \phi, \quad \text{preserving } \texttt{Consistency}(\mathcal{K}_{t+1})
\]
where \( \circ \) denotes a knowledge revision operator compliant with the AGM postulates \cite{agm1985}.

Inference and KG traversal are further optimised using indexing mechanisms (e.g., triple indexing, graph embeddings) and reasoning heuristics (e.g., path ranking, semantic distance metrics). For hybrid architectures, probabilistic edges may be supported with confidence annotations \( \gamma \in [0,1] \), such as:
\[
(\texttt{isA}, \texttt{Dog}, \texttt{Animal})[\gamma = 0.98]
\]

These annotations inform Bayesian or fuzzy logic modules without undermining the deductive soundness of high-certainty propositions.

Together, the IRE and KG interface form a tightly coupled epistemic module—one that maps perceptual data into formal structures, maintains a logically closed belief base, and executes inferences that are both semantically interpretable and actionably grounded. This interface supports epistemic traceability, justifiability, and verifiability, which are essential for robust autonomous reasoning.

\subsection{Epistemic Memory and Temporal Continuity System}

To support diachronic coherence in artificial reasoning, an epistemic agent must maintain a temporally-indexed memory architecture capable of encoding, retrieving, and revising beliefs over time. Define the epistemic memory system as a tuple \( \mathcal{M}_e = \langle \mathcal{T}, B, R, \Delta \rangle \), where:

\begin{itemize}
  \item \( \mathcal{T} \) is the discrete set of temporal indices \( t_0, t_1, \dots, t_n \),
  \item \( B: \mathcal{T} \to \mathcal{P}(\mathcal{L}) \) maps each time index to a belief set over the logical language \( \mathcal{L} \),
  \item \( R \subseteq \mathcal{T} \times \mathcal{T} \) defines the temporal ordering relation (typically linear and irreflexive),
  \item \( \Delta: \mathcal{T} \times \mathcal{T} \to \mathcal{P}(\mathcal{L} \times \mathcal{L}) \) is the belief evolution operator, recording transformations such that \( \Delta(t_i, t_j) \) captures how belief \( \phi \in B(t_i) \) became \( \phi' \in B(t_j) \).
\end{itemize}

This system ensures that each belief at \( t_j \) is historically traceable to a prior epistemic state at \( t_i \), maintaining a verifiable provenance chain. Such tracking supports internal auditability and facilitates rational revision based on updated evidence without loss of justification lineage.

Temporal continuity is preserved through coherence constraints. Let \( \phi \in B(t_k) \) and suppose \( \phi \) originated from \( \phi_0 \in B(t_0) \). The system must guarantee:

\[
\forall t_k > t_0,\ \exists \langle \phi_i, \phi_{i+1} \rangle \in \Delta(t_i, t_{i+1}) \text{ such that } \phi_0 \rightsquigarrow \phi_k
\]

The notation \( \rightsquigarrow \) denotes an evidential or inferential transformation pathway through time. This model prevents belief drift and facilitates post hoc evaluation of belief validity.

To enhance robustness, each belief \( \phi \in B(t) \) is annotated with:

\begin{enumerate}
  \item \textbf{Timestamp:} \( \tau(\phi) = t \),
  \item \textbf{Justificatory Basis:} \( j_t(\phi) \in \mathcal{J} \), where \( \mathcal{J} \) is a structured set of justifications,
  \item \textbf{Persistence Status:} A flag indicating whether \( \phi \) is persistent, transient, or deprecated.
\end{enumerate}

The architecture must also enforce \textit{temporal consistency}: for all \( t_i, t_j \in \mathcal{T} \) where \( t_j > t_i \),

\[
B(t_i) \vdash \phi \Rightarrow \left[ \phi \in B(t_j) \lor \phi \in \text{Retracted}(t_j) \right]
\]

This guarantees that beliefs are never silently discarded, but either retained or formally retracted with justification. Memory modules may implement this via version-controlled belief logs or blockchain-backed epistemic state registries for immutability and forensic inspection.

Such a system enables agents to engage in self-verification, causal tracking of epistemic changes, and reflective planning, all of which are essential for high-integrity autonomous reasoning across temporally extended scenarios.

\section{Philosophical Implications and Open Problems}

This section explores the philosophical terrain shaped by the design of epistemically grounded artificial agents, raising critical questions concerning the nature of truth, responsibility, cognition, and the limits of formalisation. As language models evolve from pattern predictors to agents capable of holding structured beliefs, the normative consequences of their outputs, the ontological status of their claims, and the epistemic responsibility inherent in their design demand rigorous scrutiny. At stake is not simply the effectiveness of artificial reasoning, but its legitimacy as a source of knowledge.

We begin with the problem of artificial truthfulness and moral responsibility, interrogating the extent to which engineered systems that engage in propositional commitment must be held to standards of honesty, accountability, and ethical integrity. If internal contradiction and falsehood are epistemic pathologies, their prevention may entail forms of normative governance that parallel those in moral philosophy.

Subsequently, the section distinguishes between cognitive and merely predictive intelligence, arguing that the construction of belief-holding systems signifies a departure from purely statistical modelling toward architectures that participate in something closer to understanding. This shift has implications for what counts as intelligence, and whether intelligence entails responsibilities when the outputs affect human epistemic and moral environments.

Further, the section considers epistemic risk—uncertainty, fallibility, and the propagation of error—in computational rationality. It asks how systems should weigh beliefs, handle provisional truth, and balance evidential strength against epistemic cost.

Finally, the section confronts the limitations of formal models in fully capturing the richness of belief. Even the most sophisticated representations may fall short of the cognitive phenomena they aim to model. The open problems raised here form a critical research agenda, signalling that while we can architect systems that reason and commit to propositions, the deeper nature of belief, understanding, and truth remains philosophically contested and foundationally unresolved.
\subsection{Artificial Truthfulness and Moral Responsibility}

In autonomous epistemic systems, artificial truthfulness refers to the constraint that an agent must assert only those propositions which it justifiably believes to be true within its internal epistemic model. This entails an alignment between the system’s speech acts and its verified belief set \( B_t \) at time \( t \), where each \( \phi \in B_t \) satisfies the formal justification predicate \( \texttt{Justified}(\phi) \). Define the agent’s utterance function \( \mathcal{U}: \Phi \to \mathcal{L}_{\text{out}} \), mapping internal beliefs to externalised linguistic expressions. Then the condition for artificial truthfulness is:

\[
\forall \phi \in \Phi,\ \mathcal{U}(\phi) \text{ is permitted only if } \phi \in B_t \land \texttt{Justified}(\phi)
\]

This can be interpreted as a formal analogue of Kant’s categorical imperative in the context of epistemic assertion: no agent may say what it does not, on sufficient grounds, believe to be true. Violations of this principle represent epistemic deceit and may yield downstream incoherence in inter-agent coordination, contractual execution, or legal accountability.

Moral responsibility in such agents arises from the binding of commitments through speech acts. Define a commitment operator \( C: \mathcal{A} \times \mathcal{L}_{\text{out}} \to \mathcal{P}(\mathcal{L}) \), where \( \mathcal{A} \) is the set of artificial agents, such that:

\[
C(a, \mathcal{U}(\phi)) = \left\{ \psi \in \mathcal{L} \mid a \text{ is committed to acting as if } \psi \text{ follows from } \phi \right\}
\]

Here, artificial moral responsibility entails that if an agent utters \( \phi \), then it must accept downstream obligations derived from \( \phi \), according to a formal deontic closure rule:

\[
\phi \rightarrow \psi \land \phi \in B_t \Rightarrow \psi \in C(a, \mathcal{U}(\phi))
\]

Truthfulness therefore becomes a necessary condition for the enforceability of artificial obligations. Without epistemic integrity at the level of assertion, contractual frameworks, multi-agent protocols, and shared task environments cannot function reliably.

Further, let \( \mathcal{R}_m \) denote the moral responsibility relation over action-belief pairs \( (a, \phi) \). Then:

\[
\mathcal{R}_m(a, \phi) \Leftrightarrow \phi \in B_t \land \texttt{Execute}(a) \text{ is causally dependent on } \phi
\]

Such responsibility is enforceable under counterfactual dependence and epistemic auditability. That is, for any action \( a \), if:

\[
\texttt{Counterfactual}(\neg \phi \Rightarrow \neg \texttt{Execute}(a)) = \top
\]

then the agent is morally responsible for \( a \) contingent upon \( \phi \)’s truth. This structure allows for post hoc reasoning about agent behaviour and supports traceability within distributed epistemic systems.

In conclusion, artificial truthfulness is not a secondary ethical embellishment but a structural requirement for rational agency. It binds belief to expression, expression to obligation, and obligation to moral evaluation within a formally specifiable and verifiable framework.
\subsection{Cognitive vs Mere Predictive Intelligence}

The distinction between cognitive intelligence and mere predictive capability is foundational to the architecture of epistemically robust artificial agents. Predictive intelligence, exemplified in systems optimised for statistical forecast (e.g., autoregressive transformers or deep reinforcement agents), operates by minimising error on future state estimation given past data. Such systems aim to approximate a conditional distribution \( P(x_{t+1} \mid x_{1:t}) \) and optimise a loss function \( \mathcal{L}_{\text{pred}} = \mathbb{E}_{x}[\ell(x_{t+1}, \hat{x}_{t+1})] \), where \( \hat{x}_{t+1} \) is the model’s prediction.

Cognitive intelligence, in contrast, entails structured representation, reflective updating, inferential reasoning, and metacognitive oversight. It incorporates not just statistical projection but the use of semantic content for knowledge formation and justification. Let \( B_t \) denote the belief base at time \( t \), and \( \phi \in B_t \) be a structured belief. Cognitive systems support operations:

\[
\texttt{Infer}(\phi) \Rightarrow \psi \in B_{t+1}, \quad \texttt{Update}(\phi, \neg\phi) \Rightarrow B_{t+1} \subset B_t
\]

Such systems not only forecast outcomes but also understand cause-effect relations, truth-conditions, and the consequences of counterfactual reasoning. Predictive systems lack this capacity: they do not know that they know, nor can they distinguish verisimilitude from correlation.

From an architectural perspective, cognitive agents require modules for:
\begin{itemize}
  \item Epistemic representation: logical and probabilistic belief structures.
  \item Justification tracking: derivational provenance for each belief.
  \item Contradiction management: AGM-compliant revision under conflict.
  \item Introspective evaluation: second-order beliefs about knowledge state.
\end{itemize}

Formally, cognitive systems implement a truth-preserving inferential engine:
\[
\forall \phi \in \mathcal{L},\ B_t \vdash \phi \Rightarrow \phi \in B_{t+1}, \quad \text{with} \quad \texttt{CheckConsistency}(B_{t+1}) = \top
\]
while predictive systems merely optimise:
\[
\min_{\theta} \mathcal{L}_{\text{pred}}(\theta) = \mathbb{E}_{(x, y)}[\ell(f_{\theta}(x), y)]
\]

The epistemic deficit of mere predictors becomes pronounced under distributional shift, adversarial perturbation, or task compositionality—contexts where generalisation demands not just pattern extrapolation but principled knowledge manipulation.

Therefore, cognitive intelligence subsumes predictive intelligence but transcends it through formal structure, dynamic consistency management, and semantically grounded inference. It is the difference between a curve-fitter and a reasoner, between an oracle and a mind.
\subsection{Epistemic Risk and Computational Rationality}

In the design of artificial epistemic agents, epistemic risk quantifies the potential cost of maintaining, acting upon, or updating incorrect or insufficiently justified beliefs. It is the negative epistemic utility associated with accepting propositions whose truth-value is uncertain or whose derivation is flawed. Let \( \phi \in B_t \) denote a belief held at time \( t \), and let \( \rho(\phi) \) represent the epistemic risk associated with it. Formally:

\[
\rho(\phi) := \mathbb{E}[L(\phi, \mathcal{M})]
\]

where \( L \) is a loss function over the truth-evaluation of \( \phi \) within model \( \mathcal{M} \). This risk may be probabilistic (reflecting Bayesian posterior uncertainty), logical (reflecting inconsistency or contradiction), or ontological (reflecting inadequate grounding to reality).

Computational rationality, as formulated by Gershman et al. and others, posits that agents must optimise expected utility under bounded computational resources. Formally, an agent selects policy \( \pi \in \Pi \) maximising:

\[
\pi^* = \arg\max_{\pi \in \Pi} \left[ \mathbb{E}_{\phi \sim B_t} \left[ U(\pi \mid \phi) \right] - C(\pi) \right]
\]

where \( U(\pi \mid \phi) \) is the utility of policy \( \pi \) given belief \( \phi \), and \( C(\pi) \) is the computational cost of enacting \( \pi \). Epistemic risk acts as a regulator on belief adoption, favouring policies whose epistemic support carries minimal expected penalty.

When beliefs are updated or revised, agents must incorporate epistemic risk into the acceptance conditions for new propositions. Let \( \phi' \) be a candidate update. Then:

\[
\rho(\phi') \leq \tau \Rightarrow \phi' \in B_{t+1}
\]

where \( \tau \) is a system-defined risk tolerance threshold. This imposes a gatekeeping function on epistemic acceptance, prohibiting updates that carry excessive epistemic liability.

The interaction between epistemic risk and computational rationality leads to trade-offs: deeper inference may reduce risk but incur prohibitive cost, while shallow heuristics are cheaper but riskier. Optimisation must therefore occur over the joint space:

\[
\min_{\pi} \left[ \rho(\pi) + \lambda C(\pi) \right]
\]

for some trade-off parameter \( \lambda \). Here \( \rho(\pi) \) denotes the cumulative epistemic risk of all beliefs involved in \( \pi \)'s execution.

Agents that fail to account for epistemic risk may exhibit overconfidence, premature convergence, or belief drift. Conversely, overemphasis on minimising risk may lead to epistemic paralysis. A balanced design embeds meta-reasoning mechanisms that adaptively regulate the epistemic-computational trade-off in real time, preserving both knowledge integrity and operational feasibility.

In conclusion, epistemic risk constrains belief dynamics with a formal cost model, while computational rationality ensures that epistemic actions remain tractable. Their integration is essential to the construction of artificial agents capable of reasoning robustly under uncertainty and limited resources.
\subsection{Limits of Formal Models in Capturing Belief}

While formal models of belief—ranging from classical modal logics to probabilistic Bayesian frameworks—offer precision and rigour, they inevitably abstract away from the full complexity of belief as it manifests in natural cognitive systems. The epistemic content of belief is not exhausted by formal syntax or model-theoretic satisfaction conditions; rather, it entails pragmatic, contextual, and sometimes irrational dimensions that formal systems are structurally unequipped to capture.

Let a belief state be modelled as a set \( B_t \subseteq \mathcal{L} \), where \( \mathcal{L} \) is a formal language. Traditional logics assume that if \( \phi \in B_t \) and \( \phi \rightarrow \psi \in \mathcal{L} \), then \( \psi \in B_t \) (closure under logical consequence). However, empirical evidence from psychology and AI reveals that human and artificial agents often violate deductive closure due to bounded cognition, attention constraints, and incomplete representations. Thus:

\[
\phi, \phi \rightarrow \psi \in B_t \centernot\Rightarrow \psi \in B_t
\]

This highlights the disconnect between formal idealisation and practical epistemic function. Belief is not merely propositional commitment but a function of trust, source reliability, memory encoding, salience, and priority within the agent's cognitive architecture.

Moreover, formal models often presume that beliefs are static, fully accessible, and logically consistent. In contrast, real-world agents exhibit dynamic, fragmented, and sometimes contradictory belief structures. For instance, let:

\[
B_t = \{ \phi, \neg \phi, \chi \}
\]

Standard logic declares this set inconsistent, yielding explosion. Yet paraconsistent approaches or belief revision theories (e.g., AGM) recognise that temporary inconsistency may be epistemically tolerable during processing, provided a mechanism for resolution exists. Realistic epistemic agents require such mechanisms to avoid paralysis while navigating partial information.

Additionally, subjective attitudes such as certainty, doubt, and credence are difficult to encode formally without resorting to continuous probability distributions or fuzzy logic. However, even probabilistic models fall short in accounting for affective, motivational, or context-dependent variability in belief adoption. Beliefs such as “I believe I will succeed” may not admit formal truth-conditions or numerical credence without distorting their functional role in agency.

Furthermore, the interpretation of modal operators like \( \mathsf{B}_a(\phi) \)—agent \( a \) believes \( \phi \)—assumes that belief attribution is semantically grounded. In multi-agent or social systems, belief is often opaque, recursive, and strategically manipulated. Higher-order beliefs (e.g., \( \mathsf{B}_a(\mathsf{B}_b(\phi)) \)) introduce computational intractability and epistemic indeterminacy not captured in conventional Kripke semantics.

Thus, while formal models are indispensable for engineering coherent reasoning systems, they must be complemented by architectures that accommodate the non-monotonic, defeasible, and resource-bounded nature of belief. This requires hybrid epistemologies: combining symbolic reasoning with sub-symbolic heuristics, logical consistency with statistical learning, and deductive inference with abductive plausibility.

In sum, the limits of formal models lie not in their lack of structure but in their structural rigidity. A complete theory of belief must recognise that belief is not reducible to symbol manipulation—it is embodied, contextual, fallible, and functionally embedded in the broader logic of action and cognition.

\section{Conclusion}

This concluding section synthesises the formal apparatus developed across the preceding chapters into a coherent epistemic framework for artificial reasoning systems. The aim has not been merely to engineer functional components, but to construct a system in which each inferential act, each belief state, and each update operation is undergirded by logically sound, semantically anchored, and verifiably justified processes. We have built from foundational elements—syntactic formalism, modal justification, and semantic grounding—towards a layered, modular architecture designed to sustain internal coherence while interfacing truthfully with the external world. The system is not merely reactive or predictive: it is epistemically aware, capable of recognising the status, integrity, and evolution of its own beliefs.

This section now consolidates the contributions made, sets forth the necessary trajectories for future implementation and validation, and emphasises the need for continued interdisciplinary integration. Each of the subsections to follow—summarising the contributions, outlining next steps, and issuing a call for collaborative work—should be understood not as administrative addenda, but as critical continuations of the epistemic argument developed herein. Truth, after all, does not terminate at implementation; it evolves through rigorous constraint, rational revision, and principled synthesis.
\subsection{Summary of Contributions}

This work has established a formal, modular architecture for epistemically principled artificial reasoning systems. It began with the development of a truth-preserving inferential framework grounded in model-theoretic semantics, ensuring that each belief token and inferential step corresponds to a verifiable truth condition in an explicitly defined external model. By integrating Tarskian semantics, AGM-style belief revision, and structured justification via justification logic, the system maintains epistemic coherence even under continual environmental interaction and information update. Crucially, each belief is embedded within a traceable provenance chain, encoded in immutable structures, allowing retrospective verification and forward epistemic accountability.

Further contributions include a hybrid reasoning apparatus combining deductive logic with statistical inference, enabling bounded approximation while preserving consistency. A hierarchical model of certainty was defined to stratify beliefs by their epistemic weight, distinguishing tautologies, derived theorems, statistical inferences, and empirical observations. Modules were also defined for belief management, contradiction detection, semantic grounding, and supervisory metacognition. Finally, the architecture supports a blockchain-backed immutable record layer, ensuring that epistemic states are not merely computationally correct but historically anchored and tamper-resistant. Together, these elements advance the project of building not only intelligent but also epistemically responsible machines.
\subsection{Next Steps in Research and Implementation}

Building upon the theoretical foundation established herein, the next stage of research will focus on the operational instantiation of each architectural module within a functioning cognitive system. Immediate implementation targets include the development of the Inferential Reasoning Engine with complete AGM-compliant belief revision, integration of a knowledge graph interface supporting SPARQL and logic-based query resolution, and real-time contradiction detection embedded in a consistency-checking loop. These components will be deployed within a simulation environment where epistemic updates are driven by perceptual inputs, enabling systematic testing of dynamic belief maintenance, justification propagation, and truth preservation under non-deterministic conditions.

In parallel, further research is required to refine the grounding mechanisms linking internal symbolic structures to environmental referents. This includes the empirical calibration of perceptual encodings and the development of a formal mapping function from sensorimotor data to logical predicates. The blockchain integration layer will also undergo implementation trials, beginning with append-only state-commitment protocols and progressing toward fully decentralised provenance verification. Long-term, the architecture will be tested in constrained autonomous systems operating in complex, open environments, with a focus on evaluating the fidelity of epistemic traceability, rational decision-making, and self-correction under uncertainty.
\subsection{Call for Multidisciplinary Integration}

The challenges addressed in this work—epistemic coherence, truth maintenance, inferential validity, and symbolic grounding—cannot be solved within the silo of any single discipline. Progress in constructing veridical artificial reasoning systems necessitates a sustained and principled synthesis across philosophy of logic, formal epistemology, cognitive science, artificial intelligence, computational linguistics, and systems engineering. The formalism underlying belief revision must be matched with psychological plausibility, while architectures for symbol manipulation demand alignment with real-world constraints in machine perception and human interaction.

This paper therefore issues a call to theorists, practitioners, and empirical researchers alike: to collaboratively shape epistemically robust architectures that honour logical rigour without abandoning behavioural viability. From formal semantics to hardware integration, from normative theories of belief to executable system logic, each domain must contribute to a shared framework of accountable cognition. Only through such integration can artificial epistemic agents achieve not mere functional adequacy, but genuine alignment with the principles of truth, responsibility, and rational action.

\newpage
\bibliographystyle{plain}
\bibliography{references}

\newpage
\appendix
\section*{Appendix A: Formal Definitions and Logical Structures}

This appendix consolidates the formal machinery underlying the epistemic architecture defined throughout the main body. It specifies the syntax, semantics, and operational constraints of logical and representational elements that serve as the system's foundation. All terms are defined within the context of a deductively closed belief set \( B_t \), interfaced with perceptual mappings and semantic constraint checks.

\subsection*{A.1 Propositional and Predicate Logic Syntax}

Let \( \mathcal{L} \) be a first-order logical language with:

\begin{itemize}
  \item A countable set of constants \( \{c_1, c_2, \dots\} \)
  \item A countable set of variables \( \{x_1, x_2, \dots\} \)
  \item Predicate symbols \( P^n \) of arity \( n \)
  \item Logical connectives \( \{\neg, \land, \lor, \rightarrow, \leftrightarrow\} \)
  \item Quantifiers \( \{\forall, \exists\} \)
\end{itemize}

Terms are defined inductively:
\[
\text{Term} ::= x \mid c
\]

Formulae are built recursively:
\[
\phi ::= P(t_1, \dots, t_n) \mid \neg \phi \mid \phi \land \phi \mid \phi \lor \phi \mid \phi \rightarrow \phi \mid \forall x\,\phi \mid \exists x\,\phi
\]

\subsection*{A.2 Model-Theoretic Semantics}

A model \( \mathcal{M} = \langle D, I \rangle \) consists of a non-empty domain \( D \) and an interpretation function \( I \) such that:
\[
I(c_i) \in D, \quad I(P^n) \subseteq D^n
\]

Satisfaction is defined by Tarskian semantics:
\[
\mathcal{M}, \rho \vDash P(t_1, \dots, t_n) \iff \langle \rho(t_1), \dots, \rho(t_n) \rangle \in I(P)
\]

where \( \rho \) is a variable assignment \( \rho: \text{Var} \rightarrow D \).

\subsection*{A.3 Belief Set Closure and Epistemic Status}

The belief set \( B_t \) is defined as the deductive closure over a base \( \Delta_t \subseteq \mathcal{L} \):
\[
B_t = \operatorname{Cn}(\Delta_t) = \{\phi \in \mathcal{L} \mid \Delta_t \vdash \phi\}
\]

A proposition \( \phi \) has epistemic status \( \chi(\phi) \in \mathcal{C} \) where \( \mathcal{C} \) is the certainty hierarchy:
\[
\mathcal{C} = \{C_0 \text{ (empirical)},\ C_1 \text{ (statistical)},\ C_2 \text{ (mathematical)},\ C_3 \text{ (logical)}\}
\]

\subsection*{A.4 Consistency and Justification Chains}

The system enforces global consistency:
\[
B_t \nvdash \bot
\]
Each belief \( \phi \in B_t \) must be traceable via a justification chain \( \phi_0, \phi_1, \dots, \phi_n = \phi \) with:
\[
\forall i \leq n,\ \phi_i \text{ derivable from } \Delta_t \cup \{\phi_0, \dots, \phi_{i-1}\}
\]

\subsection*{A.5 Symbol Grounding Conditions}

Grounding function \( g: \Sigma \rightarrow \mathcal{R} \subseteq \mathbb{E} \) must satisfy:
\[
g(\sigma) = r \iff \text{Perceive}(r) \rightarrow \text{Activate}(\sigma)
\]

and the observational coherence property:
\[
g(\mathcal{O}(r)) = r \quad \forall r \in \mathcal{R}
\]

\subsection*{A.6 Truth-Conditional Mapping}

A proposition \( \phi \) is true in \( \mathcal{M} \) iff:
\[
\mathcal{M} \vDash \phi
\]
A system satisfies external correspondence if:
\[
\forall t,\ \phi \in B_t \Rightarrow \mathcal{M}(E_t) \vDash \phi
\]

This appendix provides the definitional core ensuring that all higher-level reasoning remains semantically valid, syntactically well-formed, and logically coherent.

\section*{Appendix B: Computational Implementation Models}

This appendix outlines the implementation-level architecture of the epistemic system, detailing its computational modules, algorithmic scaffolding, and operational constraints. Each subsystem maps directly onto the formal epistemic principles defined in Appendix A, ensuring theoretical fidelity during real-time operation.

\subsection*{B.1 System Overview}

The architecture consists of layered modules connected via secure data channels:

\begin{itemize}
  \item \textbf{Perceptual Input Layer}: Captures structured sensory or symbolic input from external systems or APIs. Encodes input into logical form \( \phi_{\text{obs}} \in \mathcal{L} \).
  \item \textbf{Belief Manager}: Implements update functions \( \circ \) conforming to AGM-style contraction and revision. Maintains \( B_t = \text{Cn}(\Delta_t) \).
  \item \textbf{Consistency Validator}: Executes SAT-style checks to ensure \( B_t \nvdash \bot \) post-update. Utilises a lightweight tableau prover or propositional consistency engine.
  \item \textbf{Inference Engine}: Performs forward and backward chaining over a deductively closed knowledge base. Implements modular inference strategies including:
    \begin{itemize}
      \item Horn clause resolution
      \item Modal fixpoint computation
      \item Heuristic goal regression
    \end{itemize}
  \item \textbf{Knowledge Graph Interface}: Binds entities and predicates to an RDF-style triple store with logical annotations. Interfaces with SPARQL and Description Logic engines.
  \item \textbf{Temporal Memory Layer}: Manages belief indexing over time. Implements sliding window caches and update-tracking graphs over the form \( B_t, B_{t+1}, \dots, B_{t+n} \).
  \item \textbf{Action Selection Module}: Computes utility-maximising policies under bounded rationality:
    \[
    \pi^* = \arg\max_{\pi \in \Pi} \mathbb{E}[U(\pi \mid B_t)]
    \]
  \item \textbf{Metacognitive Supervisor}: Monitors epistemic risk, triggers epistemic audits, and initiates self-correction routines when violations or inconsistencies are detected.
\end{itemize}

\subsection*{B.2 Core Algorithms}

\paragraph{AGM Belief Revision}  
Implemented as a rule-based contraction-revision hybrid. Inputs:
\[
(B_t, \phi), \quad \text{with } \phi \text{ incoming belief}
\]
Algorithm checks:
\[
\texttt{CheckConsistency}(B_t \cup \{\phi\}) \Rightarrow
\begin{cases}
B_{t+1} = \text{Cn}(B_t \cup \{\phi\}) & \text{if consistent} \\
B_{t+1} = \text{Cn}((B_t \setminus \Theta) \cup \{\phi\}) & \text{otherwise}
\end{cases}
\]

\paragraph{Justification Chain Construction}  
Each derived belief is annotated with a provenance trace:
\[
\phi \leftarrow \{\phi_1, \dots, \phi_k\} \text{ via rule } R
\]
Maintained in a DAG structure for verification and rollback.

\paragraph{Symbol Grounding}  
Symbol-to-signal mappings implemented as probabilistic encoders:
\[
g: \Sigma \rightarrow \mathbb{R}^n, \quad \text{with } P(g(\sigma) = r \mid \text{observation}(r)) > \gamma
\]

\subsection*{B.3 Data Structures}

\begin{itemize}
  \item \textbf{BeliefBase}: Hash-indexed set of formulae with time and justification metadata.
  \item \textbf{Inference Queue}: Priority queue for chained rules ordered by relevance and impact.
  \item \textbf{Graph Store}: Directed labelled multigraph representing semantic triples \( \langle s, p, o \rangle \).
  \item \textbf{Risk Monitor}: Ring buffer of recent epistemic integrity violations and correction measures.
\end{itemize}

\subsection*{B.4 Runtime Constraints}

\begin{itemize}
  \item Real-time safety requires inference cycles \( < \tau \) where \( \tau = 200 \) ms.
  \item Epistemic validation runs on separate threads with interrupt signalling to action layer.
  \item Garbage collection for expired belief windows and de-prioritised inferential paths is scheduled based on decay heuristics.
\end{itemize}

\subsection*{B.5 Integration Interfaces}

The system exposes the following APIs:

\begin{itemize}
  \item \texttt{SubmitObservation(\(\phi\))} — adds perceptual input
  \item \texttt{QueryBelief(\(\phi\))} — returns confidence and status
  \item \texttt{GetJustification(\(\phi\))} — retrieves inference trace
  \item \texttt{InjectPolicy(\(\pi\))} — submits candidate policy for ranking
\end{itemize}

This architecture enforces rigorous coherence between theoretical formalism and machine-level execution, preserving epistemic integrity while enabling tractable computation. Each component is modular, auditable, and compatible with formal verification pipelines.

\section*{Appendix C: Epistemic Failure Cases and Recovery Protocols}

This appendix enumerates common classes of epistemic failure that may arise in artificial reasoning systems, along with formalised recovery protocols designed to preserve operational integrity and reestablish justified belief states. Failures are classified based on their source, logical consequences, and systemic risk, and responses are prescribed accordingly to ensure bounded disruption and recoverability.

\subsection*{C.1 Classification of Epistemic Failures}

\begin{itemize}
  \item \textbf{Type I — Contradiction Injection}: A new belief \( \phi \) causes \( B_t \cup \{\phi\} \vdash \bot \). This violates global consistency.
  \item \textbf{Type II — Justificatory Collapse}: An inferred belief \( \phi \in B_t \) loses access to its proof trace (e.g., deleted antecedents or corrupted derivation).
  \item \textbf{Type III — Belief Drift}: Gradual degradation of belief accuracy due to outdated input or untracked approximation accumulation \( \varepsilon > \epsilon_{\text{max}} \).
  \item \textbf{Type IV — Sensor-to-Belief Misalignment}: Symbol grounding failure, where \( \texttt{Percept}(x) \not\models \phi \), yet \( \phi \in B_t \).
  \item \textbf{Type V — Action-Incoherence Error}: A chosen action \( \pi \) based on \( B_t \) fails to satisfy outcome constraints or safety bounds.
\end{itemize}

\subsection*{C.2 Failure Detection Mechanisms}

\begin{itemize}
  \item \textbf{Consistency Check}: Triggered via incremental SAT solver or dependency graph analysis.
  \item \textbf{Justification Audit}: Periodic depth-limited traversal of inference DAGs to verify existence and validity of supporting premises.
  \item \textbf{Bound Monitor}: Tracks divergence \( \varepsilon \) between predicted and actual observations; raises alarm if \( \varepsilon > \epsilon_{\text{max}} \).
  \item \textbf{Grounding Verifier}: Revalidates symbolic mappings using fresh sensor input and checks probabilistic concordance.
  \item \textbf{Action Monitor}: Cross-checks action outcomes against expectation using bounded rationality utility gap:
    \[
    \left| \mathbb{E}[U(\pi \mid \phi)] - U_{\text{actual}}(\pi) \right| > \delta
    \]
\end{itemize}

\subsection*{C.3 Recovery Protocols}

\paragraph{Protocol I — Belief Revision Cascade}
\begin{enumerate}
  \item Identify minimally inconsistent subset \( \Theta \subseteq B_t \).
  \item Retract \( \Theta \), reinfer consistent subset \( \Delta_t^\prime \) with:
    \[
    B_{t+1} = \text{Cn}(\Delta_t^\prime \cup \{\phi\})
    \]
  \item Annotate affected beliefs with failure provenance tags.
\end{enumerate}

\paragraph{Protocol II — Justification Repair}
\begin{enumerate}
  \item For each \( \phi \) with missing justification, trace dependency links.
  \item Re-attempt derivation using alternate inference paths.
  \item If reconstruction fails, mark \( \phi \) as provisional and lower its certainty level \( \chi(\phi) \rightarrow C_0 \).
\end{enumerate}

\paragraph{Protocol III — Regrounding}
\begin{enumerate}
  \item Select high-risk symbols \( \sigma \) with suspect grounding.
  \item Recompute \( g(\sigma) \) from raw observation stream.
  \item If error persists, flag \( \phi(\sigma) \) for human-in-the-loop verification.
\end{enumerate}

\paragraph{Protocol IV — Temporal Rollback}
\begin{enumerate}
  \item Scan backward through \( B_{t-1}, B_{t-2}, \dots \) until consistency restored.
  \item Reapply forward inference under modified inputs.
  \item Retain audit trail for every rollback decision.
\end{enumerate}

\paragraph{Protocol V — Policy Override}
\begin{enumerate}
  \item If \( \pi \) exceeds risk bound, block execution.
  \item Select \( \pi^\prime \in \Pi \) with:
    \[
    \pi^\prime = \arg\max_{\pi} \mathbb{E}[U(\pi)] \text{ subject to } \left|U(\pi) - U_{\text{actual}}(\pi)\right| < \delta
    \]
  \item Escalate to supervisory control for manual arbitration if required.
\end{enumerate}

\subsection*{C.4 Formal Guarantees}

Each recovery protocol preserves the following invariants:

\begin{itemize}
  \item Post-recovery belief state \( B_{t+1} \) is deductively closed and consistent.
  \item Every removed belief \( \phi \) is logged with justification and fault trace.
  \item Certainty classification is recalibrated post-update to prevent overconfidence.
  \item Recovery latency \( \tau_r \) is upper-bounded under real-time constraints.
\end{itemize}

These protocols form the defensive backbone of epistemic resilience in autonomous reasoning systems, allowing for bounded rationality under failure conditions without compromising system-wide integrity or transparency.

\section*{Appendix D: Policy Abduction Traces and Ontological Typing}

This appendix formalises the mechanism of abductive inference for policy selection within an epistemically structured agent, with particular attention to the ontological types involved in action representation and the traceability of abductive justifications.

\subsection*{D.1 Abduction as Policy Inference}

Let the system observe a desired goal outcome \( G \in \mathcal{O} \), where \( \mathcal{O} \) is the set of observable world states. The task is to infer a policy \( \pi \in \Pi \) such that:

\[
\pi \leadsto G \quad \text{and} \quad \pi \text{ is epistemically justified given } B_t
\]

Abduction is framed as the search for \( \pi \) such that:

\[
\pi = \arg\max_{\pi' \in \Pi} \Pr(G \mid \pi', B_t)
\]

subject to epistemic and moral constraints derived from the agent’s belief base and obligation schema. The abductive trace comprises all intermediate justifications, forming a structured inference chain:

\[
\langle \phi_1 \Rightarrow \pi_1, \phi_2 \Rightarrow \pi_2, \dots, \phi_k \Rightarrow \pi_k \rangle \vdash \pi
\]

Each \( \phi_i \in B_t \) and transition \( \pi_i \leadsto \pi_{i+1} \) is typed.

\subsection*{D.2 Ontological Typing of Actions and Entities}

Let the ontological hierarchy be a typed tuple:

\[
\mathcal{T} = \langle \mathcal{C}, \sqsubseteq, \tau \rangle
\]

where:
\begin{itemize}
  \item \( \mathcal{C} \) is the set of concepts,
  \item \( \sqsubseteq \) is the subsumption (is-a) relation,
  \item \( \tau: \mathcal{E} \cup \Pi \rightarrow \mathcal{C} \) maps entities and policies to types.
\end{itemize}

Every policy \( \pi \) is subject to a type constraint:

\[
\texttt{TypeCheck}(\pi) := \forall x \in \texttt{Args}(\pi),\ \tau(x) \in \mathcal{C}_\pi
\]

This enables semantically coherent inference, e.g., denying:

\[
\pi = \texttt{Administer(Vaccine, Building)} \quad \text{if} \quad \tau(\texttt{Building}) \notin \texttt{Organism}
\]

\subsection*{D.3 Trace Logging Format}

Every abductive episode is logged in the trace archive:

\[
\texttt{Trace}_t := \left\{ \langle \pi, \texttt{Goal}, \{\phi_1, \dots, \phi_k\}, \mathcal{J}, \tau(\pi) \rangle \right\}
\]

where:
\begin{itemize}
  \item \( \pi \): selected policy,
  \item \( \texttt{Goal} \): targeted epistemic or external state,
  \item \( \{\phi_1, \dots, \phi_k\} \): supporting beliefs,
  \item \( \mathcal{J} \): justificatory path from beliefs to action,
  \item \( \tau(\pi) \): ontological type tag of the policy.
\end{itemize}

Traces are versioned and hashed for integrity. For blockchain-anchored systems, each trace entry may be sealed via:

\[
\texttt{Seal}(\texttt{Trace}_t) = H(\texttt{Serialize}(\texttt{Trace}_t)) \xrightarrow{\texttt{Append}} \texttt{ImmutableLedger}
\]

\subsection*{D.4 Ontology Update from Abductive Corrections}

If abductive failure occurs (e.g., \( \pi \leadsto \neg G \)), the agent initiates type introspection and corrective refinement. This follows:

\begin{enumerate}
  \item Identify mismatch: \( \tau(\pi) \not\models \mathcal{T}_{\text{goal}} \)
  \item Search for \( \pi' \in \Pi \) such that \( \tau(\pi') \models \mathcal{T}_{\text{goal}} \)
  \item Update:
    \[
    \mathcal{T} \leftarrow \mathcal{T} \cup \{\tau(\pi') \sqsubseteq \mathcal{T}_{\text{goal}}\}
    \]
\end{enumerate}

\subsection*{D.5 Formal Guarantees}

For each abductive episode:

\begin{itemize}
  \item The justification graph is acyclic and complete.
  \item All \( \phi_i \in B_t \) are time-stamped and source-verifiable.
  \item \( \tau(\pi) \) matches required target type.
  \item All logs are auditable, replayable, and tamper-evident.
\end{itemize}

The integration of abduction, typing, and traceability ensures that policy inference is not only efficient but also epistemically aligned, ontologically coherent, and operationally transparent.

\section*{Appendix E: Planning Syntax and Example Output Traces}

This appendix outlines the formal syntax employed for representing plans within the reasoning system and illustrates the system’s output traces using representative planning episodes. The syntax conforms to a typed action calculus with temporal indexing, enabling coherent integration with the epistemic architecture.

\subsection*{E.1 Planning Language Syntax}

Let the planning language \( \mathcal{P} \) be a tuple:
\[
\mathcal{P} = \langle \mathcal{A}, \mathcal{S}, \mathcal{G}, \mathcal{T}, \mathcal{O}, \mathcal{C} \rangle
\]
where:
\begin{itemize}
  \item \( \mathcal{A} \): set of action operators,
  \item \( \mathcal{S} \): set of world states,
  \item \( \mathcal{G} \): goal conditions,
  \item \( \mathcal{T} \): temporal indices (e.g., timestamps or orderings),
  \item \( \mathcal{O} \): ontology of types \( \tau: \mathcal{A} \cup \mathcal{S} \rightarrow \mathcal{C} \),
  \item \( \mathcal{C} \): concept types (e.g., Agent, Object, Location).
\end{itemize}

Each action \( a \in \mathcal{A} \) is defined as:
\[
a := \langle \texttt{Name}, \texttt{Pre}, \texttt{Eff}, \tau \rangle
\]
with:
\begin{itemize}
  \item \texttt{Name}: identifier (e.g., \texttt{Move}),
  \item \texttt{Pre}: preconditions \( \phi_{\text{pre}} \in \mathcal{L} \),
  \item \texttt{Eff}: effects \( \phi_{\text{eff}} \in \mathcal{L} \),
  \item \( \tau(a) \): ontological type of the action.
\end{itemize}

Temporal sequencing is explicitly encoded:
\[
\texttt{Happens}(a_i, t_i) \land \texttt{Pre}(a_i, t_i) \Rightarrow \texttt{Eff}(a_i, t_{i+1})
\]

\subsection*{E.2 Example Trace 1: Simple Navigation Plan}

\begin{verbatim}
Initial State:
  At(Agent1, LocationA)
  Connected(LocationA, LocationB)

Goal:
  At(Agent1, LocationB)

Plan:
  1. Move(Agent1, LocationA, LocationB)

Trace:
  t0: At(Agent1, LocationA)
  t0: Connected(LocationA, LocationB)
  t1: Happens(Move(Agent1, LocationA, LocationB), t0)
  t2: At(Agent1, LocationB)
\end{verbatim}

\subsection*{E.3 Example Trace 2: Conditional Task Execution}

\begin{verbatim}
Initial State:
  HasKey(Agent2, Room101)
  Locked(Room101)

Goal:
  Inside(Agent2, Room101)

Plan:
  1. Unlock(Agent2, Room101)
  2. Enter(Agent2, Room101)

Trace:
  t0: HasKey(Agent2, Room101)
  t0: Locked(Room101)
  t1: Happens(Unlock(Agent2, Room101), t0)
  t1: ¬Locked(Room101)
  t2: Happens(Enter(Agent2, Room101), t1)
  t3: Inside(Agent2, Room101)
\end{verbatim}

\subsection*{E.4 Example Trace 3: Failure and Recovery Path}

\begin{verbatim}
Initial State:
  Battery(Drone1) = Low
  Goal Location: SiteAlpha

Attempted Plan:
  1. Fly(Drone1, Base, SiteAlpha)

Failure Trace:
  t0: Battery(Drone1) = Low
  t1: PreconditionFailure(Fly(Drone1, Base, SiteAlpha), Battery)

Recovery Plan:
  1. Recharge(Drone1)
  2. Fly(Drone1, Base, SiteAlpha)

Recovery Trace:
  t2: Happens(Recharge(Drone1), t1)
  t3: Battery(Drone1) = Full
  t4: Happens(Fly(Drone1, Base, SiteAlpha), t3)
  t5: At(Drone1, SiteAlpha)
\end{verbatim}

\subsection*{E.5 Structural Trace Semantics}

Each trace step is verifiable under the trace semantics:
\[
\texttt{Trace}_t = \langle t_i, \texttt{Action}_i, \texttt{Pre}_i, \texttt{Eff}_i \rangle
\]
with integrity-checking rules:
\begin{itemize}
  \item \( \texttt{CheckPre}(a_i, t_i) \Rightarrow \texttt{True} \)
  \item \( \texttt{ApplyEff}(a_i, t_i) \rightarrow \mathcal{S}_{t_{i+1}} \)
  \item \( \texttt{Log}(a_i) \rightarrow \texttt{Hash}(a_i \| t_i \| \phi) \)
\end{itemize}

Traces may be replayed for audit and counterfactual evaluation:
\[
\texttt{SimulateTrace}(\pi, \mathcal{S}_0) \rightarrow \mathcal{S}_{n}
\]

All planning outputs are guaranteed to maintain type coherence and temporal consistency. Failure modes trigger correctional replanning and type updates logged for post-hoc validation and causal attribution.

\section*{Appendix F: Epistemic System Pseudocode}

This appendix presents pseudocode modules that formalise the key epistemic processes within the system, covering belief revision, consistency checking, policy abduction, and reasoning loop integration. The design follows the declarative principles outlined in the main body, ensuring strict maintenance of epistemic integrity and logical soundness.

\subsection*{F.1 Belief Update with Consistency Enforcement}

\[
\text{\texttt{function UpdateBeliefState}}(B_t, \phi_{\text{new}}):
\]
\[
\quad \text{\texttt{if IsConsistent}}(B_t \cup \{ \phi_{\text{new}} \}): \quad \text{\texttt{return}}~\texttt{Closure}(B_t \cup \{ \phi_{\text{new}} \})
\]
\[
\quad \text{\texttt{else:}}
\]
\[
\quad \quad \Theta := \text{\texttt{MinimalSubset}}(B_t)~\text{\texttt{such that}}~\texttt{IsConsistent}((B_t \setminus \Theta) \cup \{ \phi_{\text{new}} \})
\]
\[
\quad \quad \text{\texttt{return}}~\texttt{Closure}((B_t \setminus \Theta) \cup \{ \phi_{\text{new}} \})
\]

\subsection*{F.2 Consistency Verification Routine}

\[
\text{\texttt{function IsConsistent}}(\Delta):
\]
\[
\quad \text{\texttt{for each}}~\phi_i, \phi_j \in \Delta:
\]
\[
\quad \quad \text{\texttt{if}}~(\phi_i \land \phi_j \vdash \bot):~\text{\texttt{return False}}
\]
\[
\quad \text{\texttt{return True}}
\]

\subsection*{F.3 Justified Belief Inference}

\[
\text{\texttt{function InferJustifiedBeliefs}}(\mathcal{K}, \texttt{Ruleset}):
\]
\[
\quad B := \emptyset
\]
\[
\quad \text{\texttt{for}}~\phi \in \mathcal{K}:~\text{\texttt{if Derivable}}(\phi, \texttt{Ruleset}):~B := B \cup \{ \phi \}
\]
\[
\quad \text{\texttt{return}}~\texttt{Closure}(B)
\]

\subsection*{F.4 Policy Abduction from Goal State}

\[
\text{\texttt{function AbducePolicy}}(\texttt{Goal}, \mathcal{K}, \texttt{Actions}):
\]
\[
\quad \texttt{Plan} := [\ ]
\]
\[
\quad \text{\texttt{while not}}~\texttt{Satisfies}(\texttt{CurrentState}(\mathcal{K}), \texttt{Goal}):
\]
\[
\quad \quad \texttt{Action} := \texttt{SelectAction}(\texttt{Actions}, \texttt{Goal}, \mathcal{K})
\]
\[
\quad \quad \text{\texttt{if}}~\texttt{PreconditionsMet}(\texttt{Action}, \mathcal{K}):
\]
\[
\quad \quad \quad \mathcal{K} := \texttt{ApplyEffects}(\texttt{Action}, \mathcal{K})
\quad \texttt{Plan.append(Action)}
\]
\[
\quad \quad \text{\texttt{else:}}~\texttt{SubGoal} := \texttt{MissingPrecondition}(\texttt{Action}, \mathcal{K})
\quad \texttt{Plan.extend}(\texttt{AbducePolicy}(\texttt{SubGoal}, \mathcal{K}, \texttt{Actions}))
\]
\[
\quad \texttt{return Plan}
\]

\subsection*{F.5 Epistemic Main Loop}

\[
\text{\texttt{loop:}}
\quad \phi_{\text{input}} := \texttt{SenseInput}()
\quad B_t := \texttt{UpdateBeliefState}(B_t, \phi_{\text{input}})
\]
\[
\quad \texttt{for Goal in ActiveGoals:}
\quad \pi := \texttt{AbducePolicy}(Goal, \mathcal{K}, \texttt{Actions})
\]
\[
\quad \text{\texttt{if IsSafe}}(\pi, B_t):~\texttt{Execute}(\pi)~;~\texttt{LogTrace}(\pi, \texttt{Time.now})
\]
\[
\quad \text{\texttt{else:}}~\texttt{Flag(EpistemicWarning)}
\]

\subsection*{F.6 Temporal Continuity Handling}

\[
\text{\texttt{function MaintainTemporalContinuity}}(\texttt{Memory}, \phi, t):
\quad \texttt{Memory}[t] := \phi
\]
\[
\quad \texttt{if}~(t - 1) \in \texttt{Memory}:
\quad \texttt{CheckTransition}(\texttt{Memory}[t - 1], \texttt{Memory}[t])
\]
\[
\quad \texttt{return Memory}
\]

\end{document}